# AAO OBSERVER

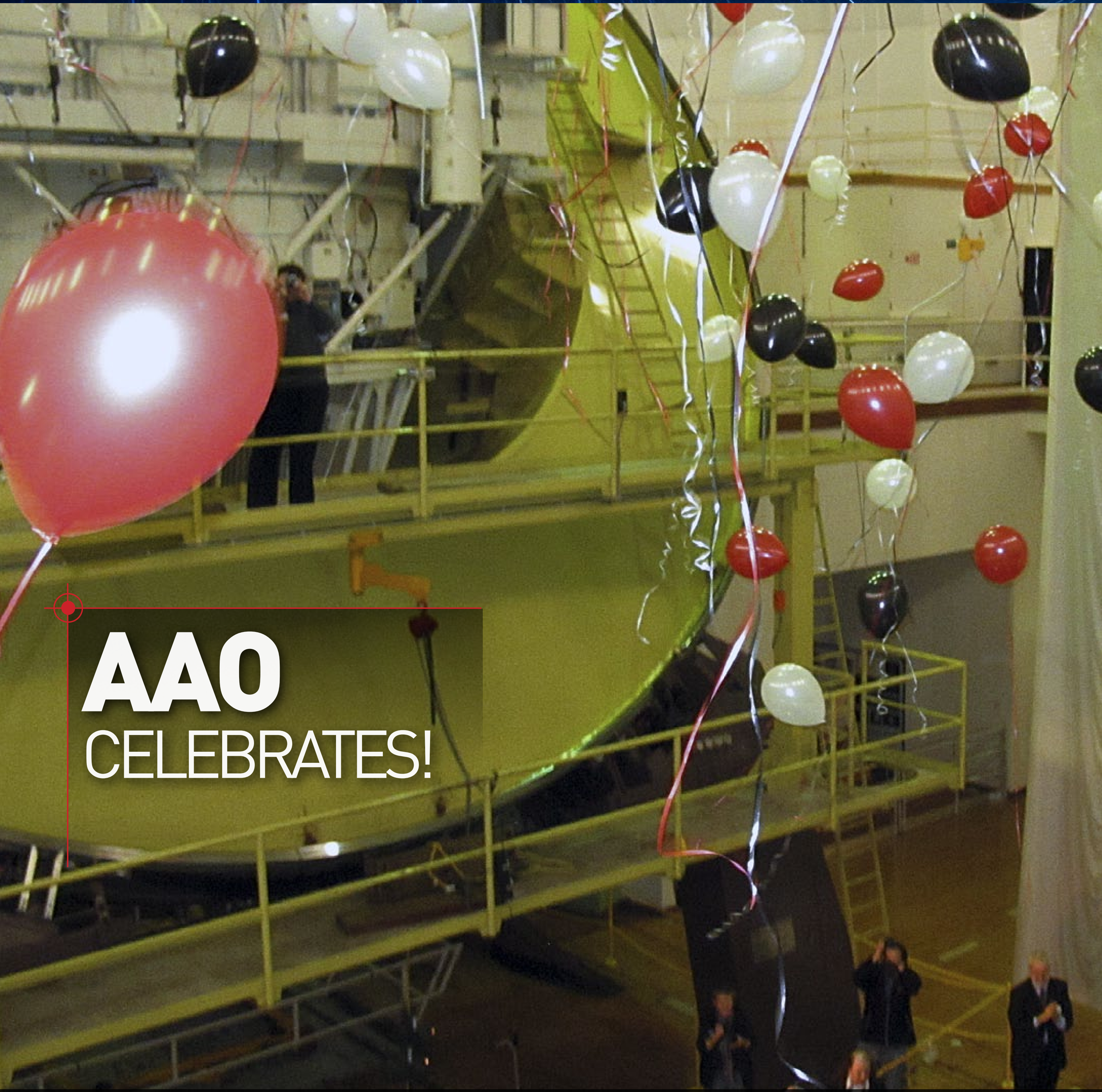

## AAO CELEBRATES!

**PCA sky subtraction for AAOmega** | **GNOSIS:** OH suppression in progress

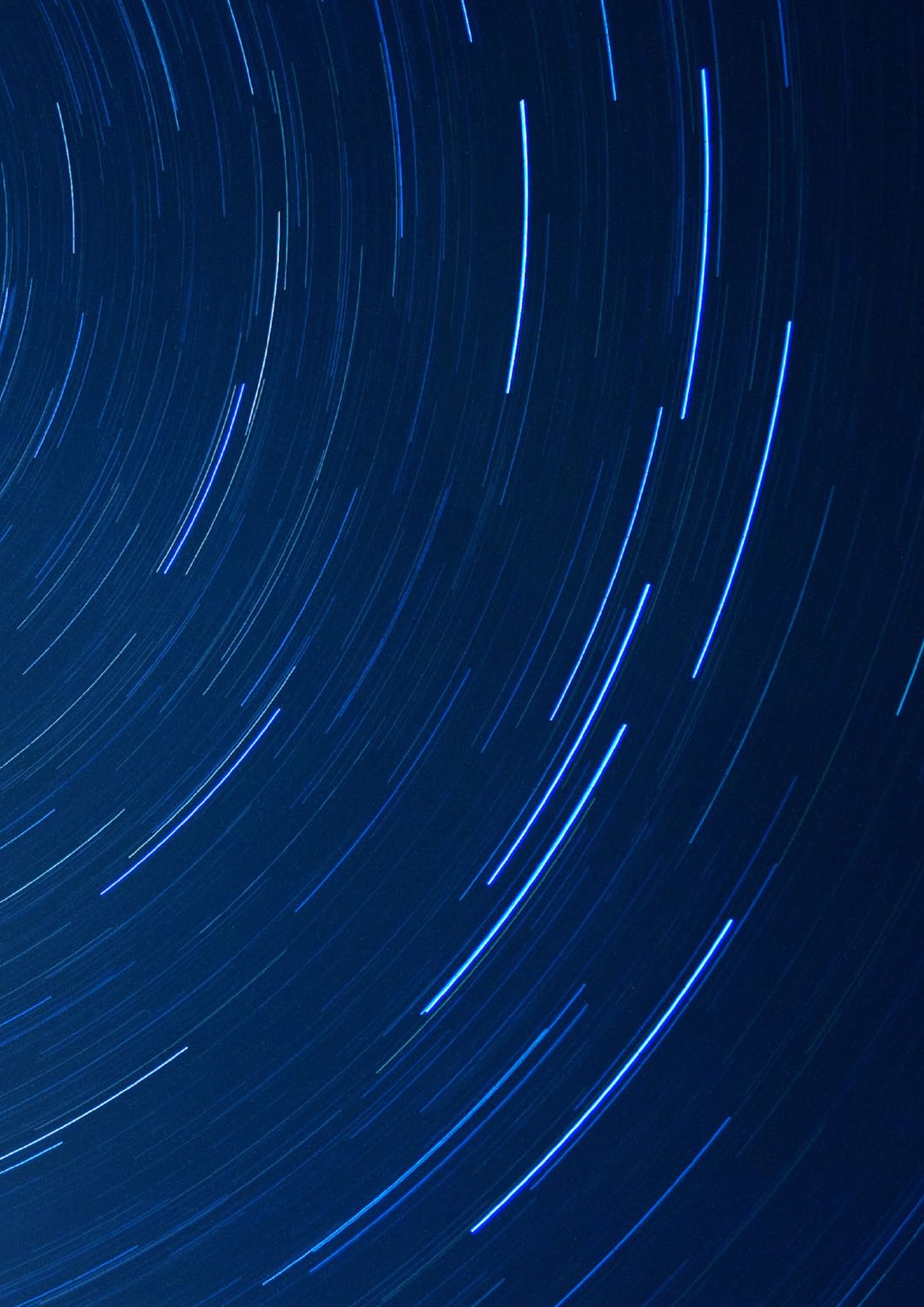

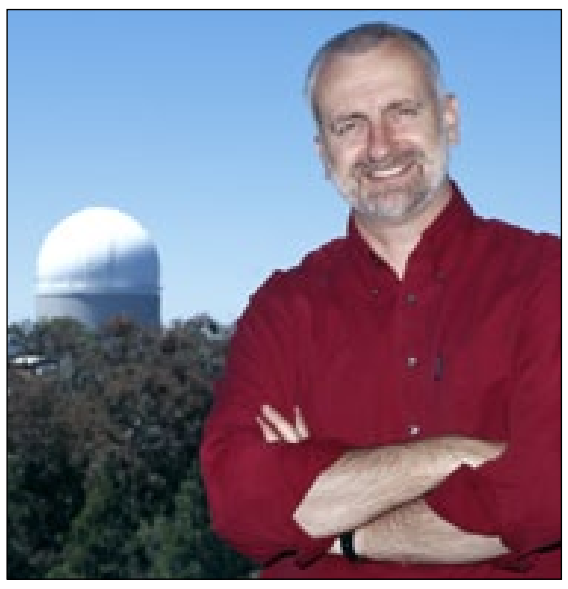

# Director's message

Matthew Colless

Welcome to the first newsletter of the Australian Astronomical Observatory! I hope you like the fresh new styling and layout; under the hood, however, you should find the same engine – the new AAO continues the mission to provide Australian astronomers with access to the world's best optical and infrared telescopes and instruments, and to support outstanding research in astronomy and astronomical technology.

The AAO celebrated the 36 prolifically productive years of the Anglo-Australian Observatory partnership with a highly enjoyable and memorable symposium held at the end of June in Coonabarabran. The meeting revived many old friendships, recalled dozens of scientific triumphs (and a few disasters), and formed a fitting memorial to one of the most outstandingly successful scientific collaborations in the two nations' histories. In a ceremony to mark the occasion, attended by representatives of both countries, hundreds of balloons were released into the AAT dome and the AAO's new name was unfurled on a giant banner – the moment captured in the picture by David Malin on the cover of this newsletter.

At this fresh new beginning for the AAO, it's appropriate to take stock of where we stand and where we plan to go from here. This is timely, as the Australian community is currently engaged in its Mid-Term Review of the Decadal Plan for astronomy. What, then, is the AAO's role over the next five to ten years? In a nutshell…

The AAO will provide cutting-edge instrumentation for the AAT to allow it at least another decade of successful, high-impact research based on its powerful existing suite of instruments and the unique new facilities currently under construction: the HERMES high-resolution multi-object spectrograph, which will probe the stellar content and formation history of the Galaxy, and the GNOSIS OH-suppression fibre feed for IRIS2, which will make the AAT the most sensitive near-infrared telescope in the world for some applications. The AAO will also continue to exploit new opportunities to upgrade existing instruments and develop new capabilities for both the AAT and the UKST.

The AAO will facilitate Australian access to the world's largest optical telescopes, currently achieved through partnership share in the Gemini 8-metre telescopes and buying time on the Magellan 6.5-metre telescopes. The AAO aims to enhance this access by developing potent new instrumentation for these telescopes and by providing value-added support for Australian users, both in proposing for time and in extracting the best science from the time they are awarded.

The AAO will plan for the longer term, as the community seeks to shift the focus of its resources from the current mix of 4-metre and 8-metre telescope access towards a greater emphasis on larger apertures, including at least the equivalent of 20% of an 8-metre telescope and 10% of an ELT. As a way to achieve these Decadal Plan goals, Australia is currently participating in the re-negotiation of the Gemini partnership and has invested in a 10% share of the construction of the Giant Magellan Telescope. The AAO will continue to play a significant role in maintaining and reviewing these arrangements in order to ensure that the requirements of the community for world-class facilities are being met.

Finally, the AAO will continue to perform high-impact research, both on its own and, more commonly, in collaboration with researchers from other Australian institutions and from overseas. The award of one Future Fellowship and four Super Science Fellowships to the AAO, and its participation in the recently announced Centre of Excellence for All-sky Astrophysics (CAASTRO), will significantly increase the AAO's research capability and productivity over the next few years. The AAO will also continue to communicate the excitement of astronomical discoveries to the public through a variety of education and outreach activities.

This outline of the way forward for the AAO assumes a reasonably predictable future. However, experience teaches that the road ahead is full of unexpected twists and turns, and that to survive and prosper, an organisation must be far-sighted, nimble, and well resourced. Thanks to the Australian government, the Australian Astronomical Observatory is appropriately funded for the foreseeable future; now it's up to us to have the vision and agility to respond to whatever that future holds.

# CONTENTS

# The Galaxy And Mass Assembly (GAMA) first data release

Andrew Hopkins (AAO) and Simon Drive (University of St. Andrews)

The GAMA survey is a Large Observing Program with the AAT, which featured in the August 2008 AAO Newsletter following an immensely successful opening set of observations (21 clear nights out of 22). GAMA was allocated 66 nights on the AAT during 2008-2010, and has now successfully completed these observations, obtaining over 130,000 spectra. The first public data release of GAMA spectra and redshifts, along with associated photometric catalogues self-consistently reprocessed using both SDSS and UKIDSS imaging, was announced on 25 June 2010 at the AAO Symposium "Celebrating the AAO", in Coonabarabran. Information on GAMA and the public survey data can be found at the GAMA website: http://www.gama-survey.org/ (see Figure 1 and back page).

GAMA has two main aims: To conduct a series of tests of the Cold Dark Matter paradigm (Driver et al. 2009, Norberg et al. 2010 in prep.), and to create a uniform homogeneous local survey for studying galaxy evolution. The GAMA survey builds on earlier surveys, the 2dFGRS and 6dFGS, and the SDSS, in a comprehensive and uniformly rigorous manner to fainter flux levels, higher redshift, higher spatial resolution, and with multi-wavelength photometric coverage spanning UV to radio wavelengths. GAMA achieves this by bringing together data from a number of world-class instruments:

- The Anglo-Australian Telescope (AAT)
- The VLT Survey Telescope (VST)
- The Visible and Infrared Survey Telescope for Astronomy (VISTA)
- The Australian Square Kilometre Array Pathfinder (ASKAP)
- Herschel
- Galaxy Evolution Explorer (GALEX)

GAMA spans three equatorial regions (centred around RA=9h, 12h and 15h) each of 4 deg x 12 deg (Figure 2). In these regions a complete sample of galaxies to $r_{Pet}<19.4$ (in the 9h and 15h regions) and to $r_{Pet}<19.8$ (in the 12h region) have been observed. GAMA has measured 135,902 spectra in total, including 134,390 spectra of 120,862 unique galaxy targets. Of these, 114,043 are reliable new galaxy redshifts, adding to the 19,000 known in these regions, to provide a highly complete sample both to the stated magnitude limits and in redshift. The sample is 98% complete in terms of sources targetted, and has a mean overall redshift completeness of 94.4 per cent (Figure 3). GAMA probes to redshift z~0.5 with a median in the redshift distribution of z~0.2 (Figures 4 and 5), and allows detection of galaxies with stellar masses as low as $10^7$ Msun, comparable to Magellanic Cloud masses.

Optical and near-infrared sample selection came from the SDSS and UKIDSS survey data. This will be complemented by imaging from two approved ESO Public Surveys, KIDS (with VST) and VIKING (with VISTA), which will also expand the photometric coverage to southern fields that GAMA plans to target. This is necessary to achieve the full survey goals, expanding the area from the existing 144 square degrees to an area of 360 square degrees. This goal is driven primarily by the need to sample a sufficient volume of the Universe, both for the analysis of the halo mass function and large scale structure, along with a comprehensive sample of counterparts to galaxies detected through their neutral hydrogen content, to be catalogued by the DINGO survey with ASKAP (PI: M. Meyer).

The GAMA survey is detailed in several publications. An overview is provided by Driver et al., (2009), the input catalogue and star-galaxy separation is given in Baldry et al., (2010), and the optimal tiling algorithm is detailed in Robotham et al., (2010). The optical and NIR photometric catalogues are described by Hill et al., (2010), with UV photometry in Seibert et al., (2010), and stellar mass estimates by Taylor et al., (2010). The survey diagnostics and first data release details are provided by Driver et al., (2010).

As future data releases are announced, the variety of data products included will be increased, with galaxy stellar masses, emission and absorption line measurements, star formation rates, metallicities, and more to be included. Initial scientific results from the GAMA collaboration that already make use of such measurements include:

- an analysis of some of the slowest forming galaxies, with masses and star formation rates as little as 1/10000 that of the Milky Way (Brough et al., 2010),
- evidence suggesting that the stellar initial mass function varies in galaxies and depends on galaxy star formation rates (Gunawardhana et al., 2010),
- a comprehensive and self-consistent analysis of galaxy obscuration curves (Wijesinghe et al., 2010),
- a 9-band photometric analysis of the full sample (Hill et al., 2010).

Further results are also being explored, with work on the evolution of the stellar mass function, and the star formation rate dependence of the mass-metallicity relationship, also underway.

The GAMA survey has been supported by the ARC with the award of 3 Super Science Fellowships to the AAO, with one position to be appointed this year, and 2 more next year. This brings the collaboration (with over 50 members) to more than 6 dedicated postdoctoral fellows and more than 10 PhD students. We are looking forward to a lot of exciting scientific results over the coming year, along with new developments as GAMA moves into a new phase with planning for the second stage of the survey underway.

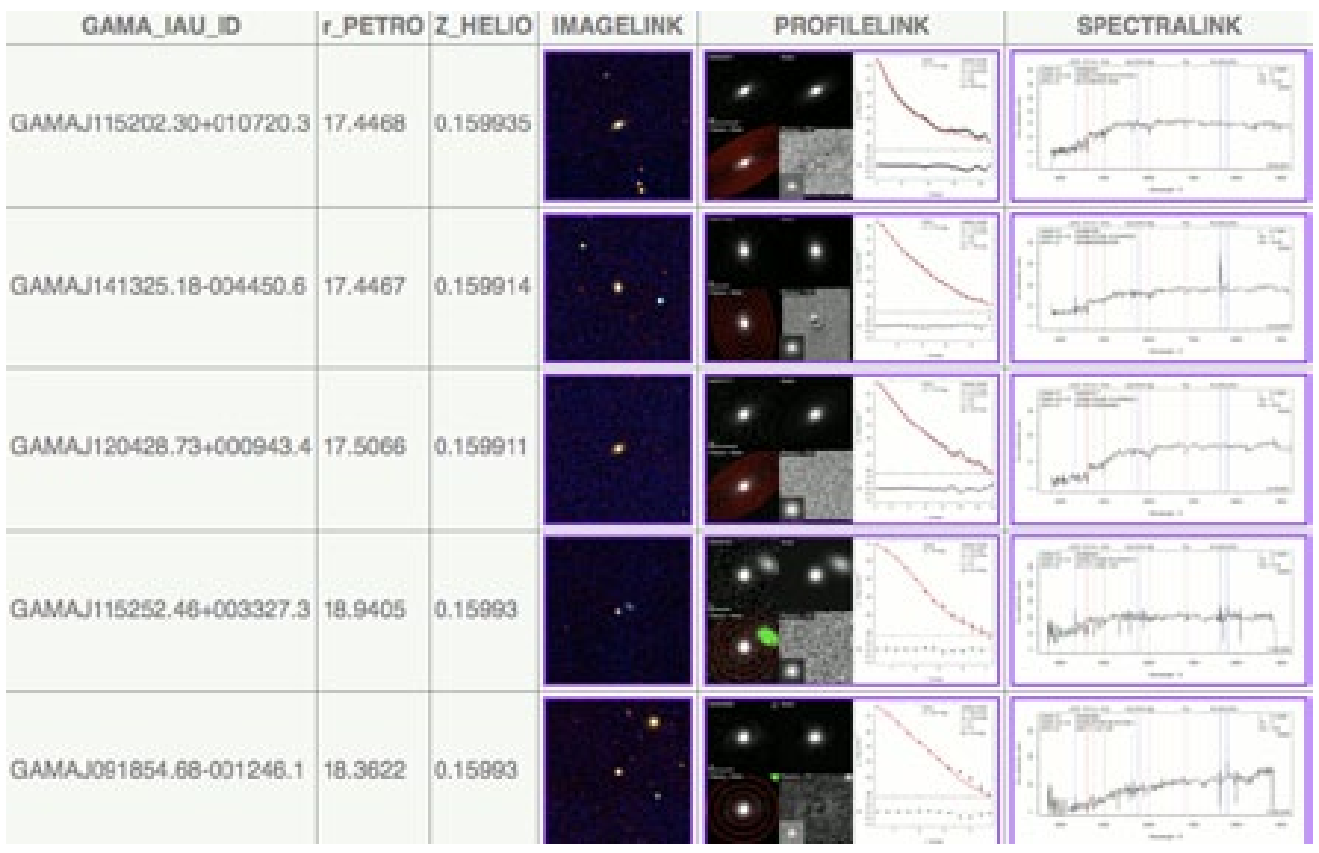

**Figure 1:** Example GAMA DR1 output illustrating photometry, light profiles, redshifts and spectra.

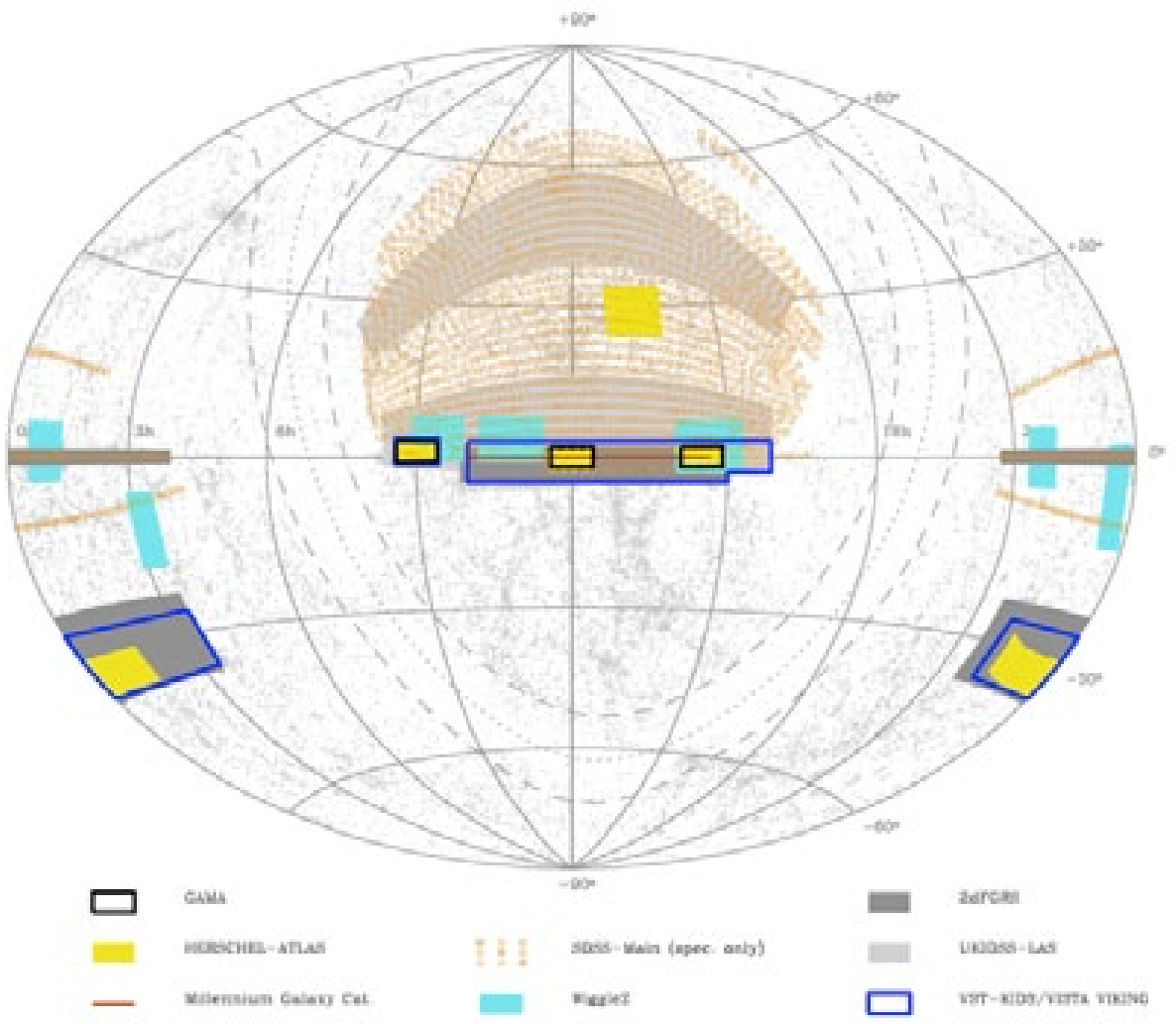

**Figure 2:** GAMA Phase I (black squares) in relation to other recent and planned surveys (see key). Also overlaid as grey dots are all known redshifts at $z < 0.1$ taken from the NASA ExtraGalactic Database.

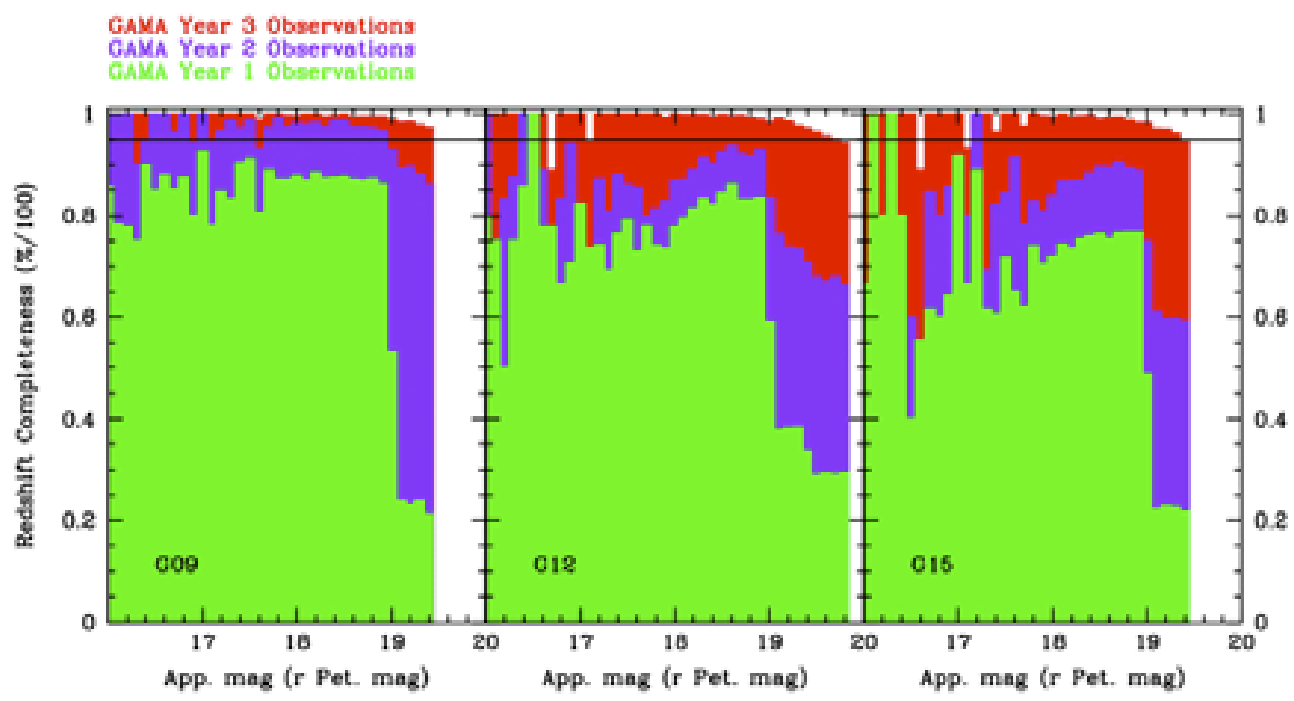

**Figure 3:** Evolution of the GAMA survey over three years of observations showing the progressive build-up towards uniform high completeness. The horizontal line denotes a uniform 95 per cent completeness.

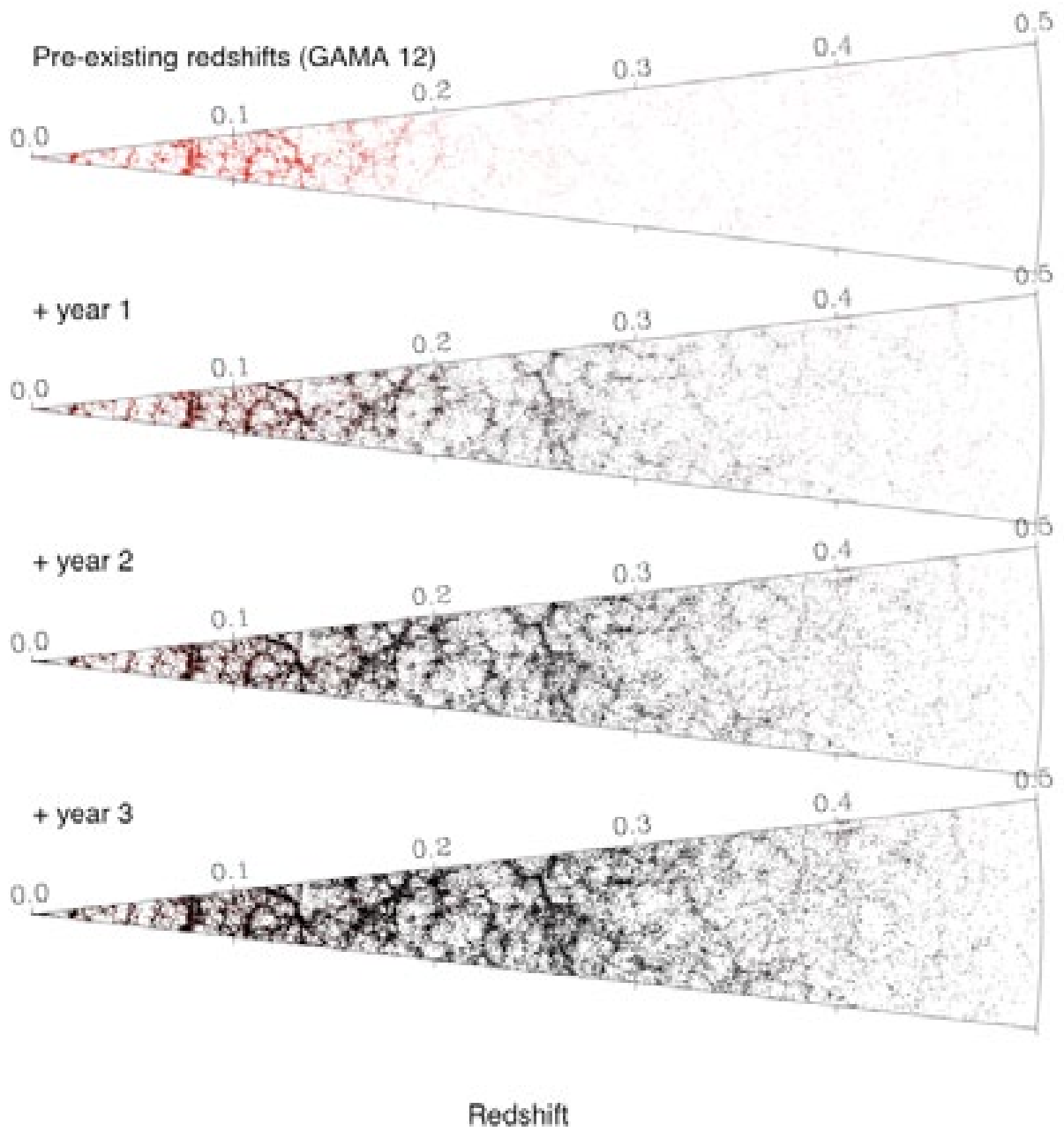

**Figure 4:** Redshift cone diagram for the GAMA 12 region showing (top to bottom), pre-existing data, year 1 data release added, year 2 data release added, year 3 data release added.

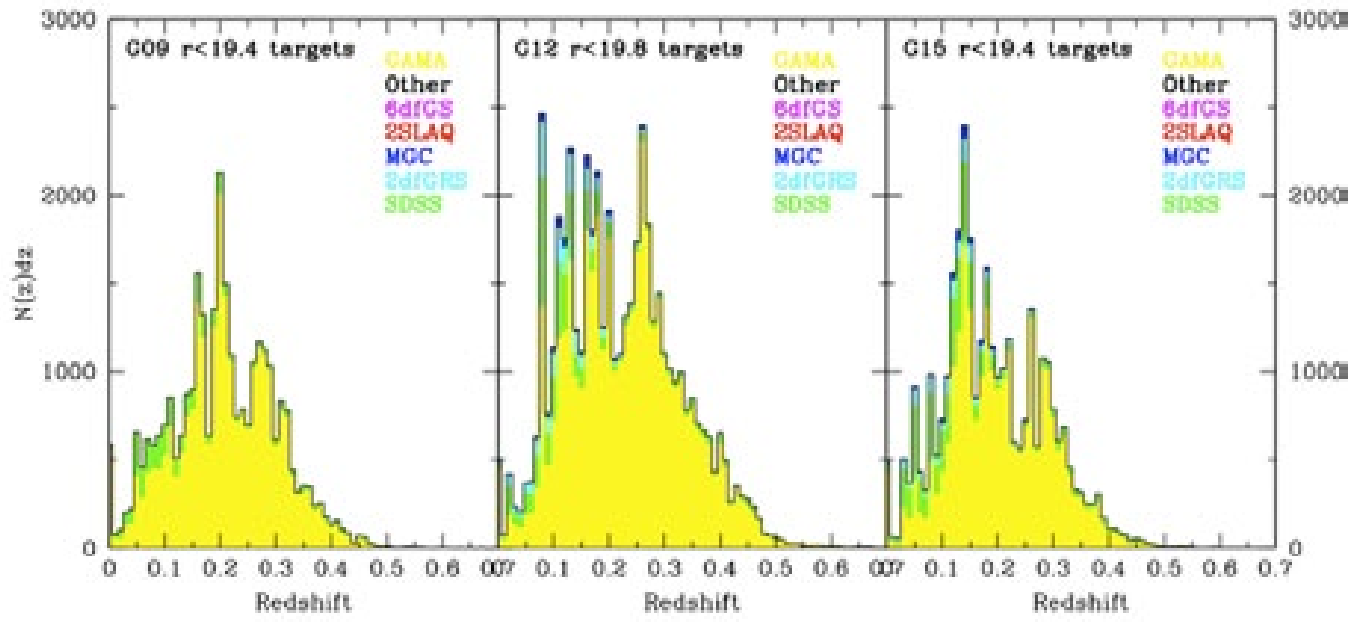

**Figure 5:** The n(z) distributions of the GAMA regions including new GAMA redshifts (shown in yellow) alongside pre-existing redshifts already in the public domain as indicated.

# PHR1315-6555: a bipolar Type I planetary nebula (PN) in the compact Hyades-age open cluster ESO 96-SC04

Q.A. Parker (Macquarie University, AAO), D.J. Frew (Macquarie University) B.Miszalski (University of Hertfordshire), A.Kovacevic (Macquarie University), P.Frinchaboy (Texas Christian University), P.Dobbie (AAO) & J.Koppen (Observatoire de Strasbourg)

**We have identified a bipolar Type I planetary nebula (PN) PHR1315-6555, in the distant, compact, intermediate-age open cluster, ESO 96-SC04. This is currently the only known "confirmed" example of a PN physically associated with a Galactic open cluster. Cluster membership is extremely important as it allows for very precise estimates of the fundamental properties of the PN as the cluster is at a known distance. Results will be of considerable interest to specialists across differing astrophysical disciplines, including PNe, white dwarfs, and open clusters. These are presented in detail in Parker et al. (2010).**

**Discovery:** The PN was found during systematic searches for new Galactic PNe for the original MASH survey (Parker et al. 2006) on the AAO/UKST Ha survey (Parker et al. 2005). The PN had been missed in earlier broadband surveys, including CCD studies of the host cluster (e.g. Carraro et al., 2005). Here we present original discovery images and CTIO 4m MOSAIC-II camera follow-up narrow-band images that reveal its bipolar morphology. We also present confirmatory spectroscopy and provide preliminary estimates of basic PN properties and abundance estimates from deeper spectra that show it to be of Type I chemistry consistent with that of the cluster and its estimated turn-off mass.

**PN cluster membership.** This rests on several key arguments and other pieces of evidence which are presented in the figures and tables.

i. Close (23") angular proximity of PN to the cluster centre well within the 32" half-light radius (see Figure1).

ii. Excellent 1kms$^{-1}$ radial velocity agreement of PN and cluster stars from our high resolution spectra (Table1). Open cluster velocity dispersions are typically only ~1kms$^{-1}$ (e.g. Mathieu, 2000 and Hole et al. 2009).

iii. Very good agreement between our independently estimated PN distance via our new surface brightness radius relation (Frew 2008) and that of the host cluster provided in the literature, to within the errors (Table 1).

iv. Independently determined cluster redenings and our own PN redening estimates are in good agreement. (important in distance estimates).

v. The PN and cluster metallicities agree and are consistent with a Hyades age cluster as inferred from previous photometric studies of the cluster.

vi. The physical PN parameters evaluated at the estimated distance, such as physical nebula extent and likely progenitor mass, are all consistent with PN values and the estimated cluster turn-off mass of ~2.5 Msun (assuming a non-binary nucleus).

vii. A Galactic |z| height and statistical likelihood argument shows that a chance alignment of an unrelated distant PN with this remote compact cluster is exceedingly unlikely (there is no Galactic warp signature at the galactic longitude of the cluster).

**Conclusion:** Taking these diverse strands of evidence together they present an extremely compelling case for membership. Now that membership is proven the astrophysical potential inherent in this rare association can be exploited. A more detailed version of this work has been submitted to MNRAS.

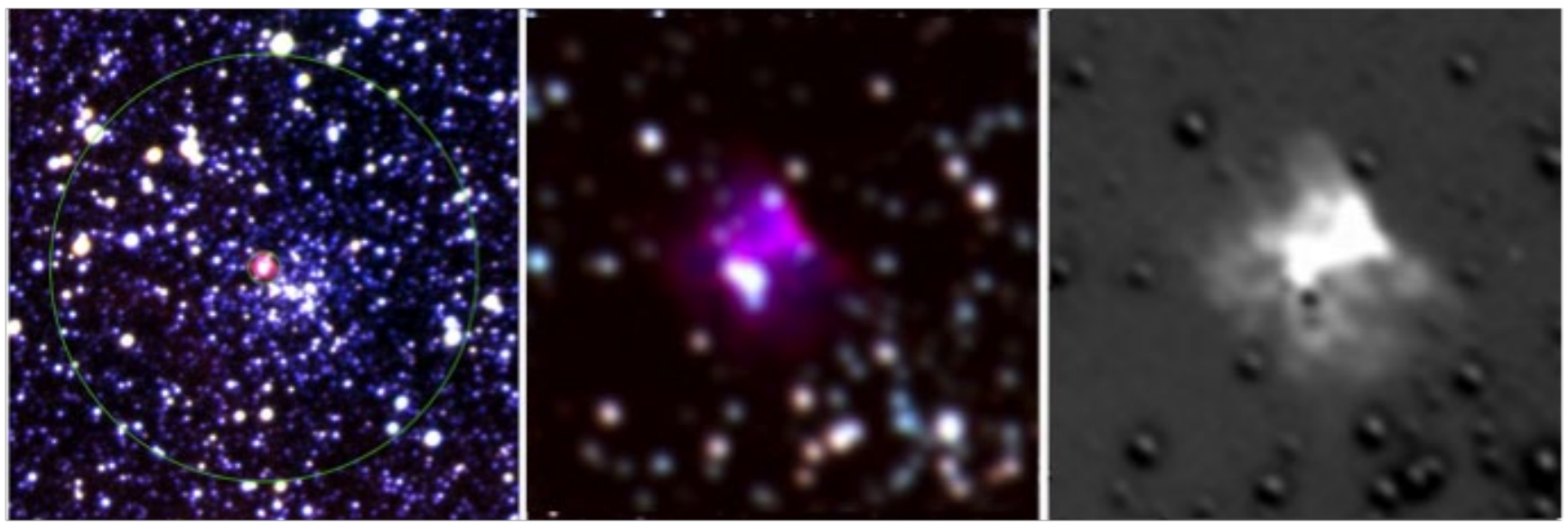

**Figure 1:** Left panel: a 5′× 5′ colour composite SuperCOSMOS discovery image with Hα (red) broad-band SR (green) and B-band (blue). The image is centred on the PN which falls well within the cluster 32" half-light radius. Middle: a 1′×1′ CTIO 4-m MOSAIC-II colour composite of the PN with its bipolar morphology evident from the Hα+[N II] (red), [O III]-off (green) and [O III] (blue) narrow-band filters. Right panel: quotient image from dividing the Hα+[N II] image by the Hα off-band image which is effective in showing the fainter outer regions of this unique bipolar PN. North-east is to the top left in all panels.

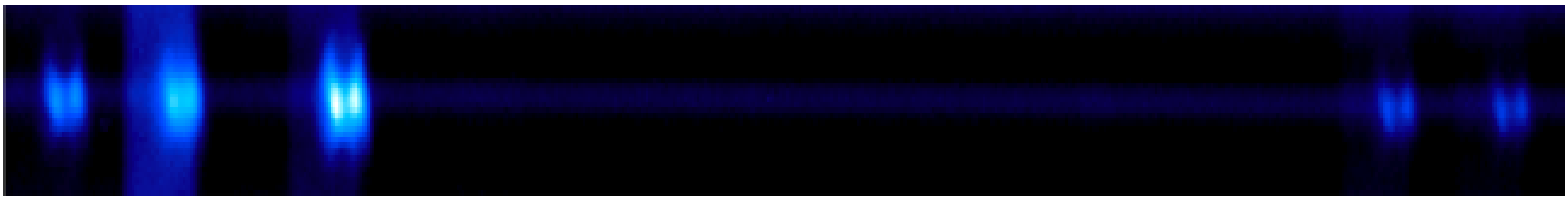

**Figure 2:** Cleaned 2-D wide-slit 2.3m DBS red spectral image of PHR1315-6555 with wavelength increasing to the right used for line flux measures. The bipolar nature of the PN is confirmed. The vertical direction is about 30" and the slit width was 19". From left to right the lines are [N II]6548Å, Hα, [N II]6584Å

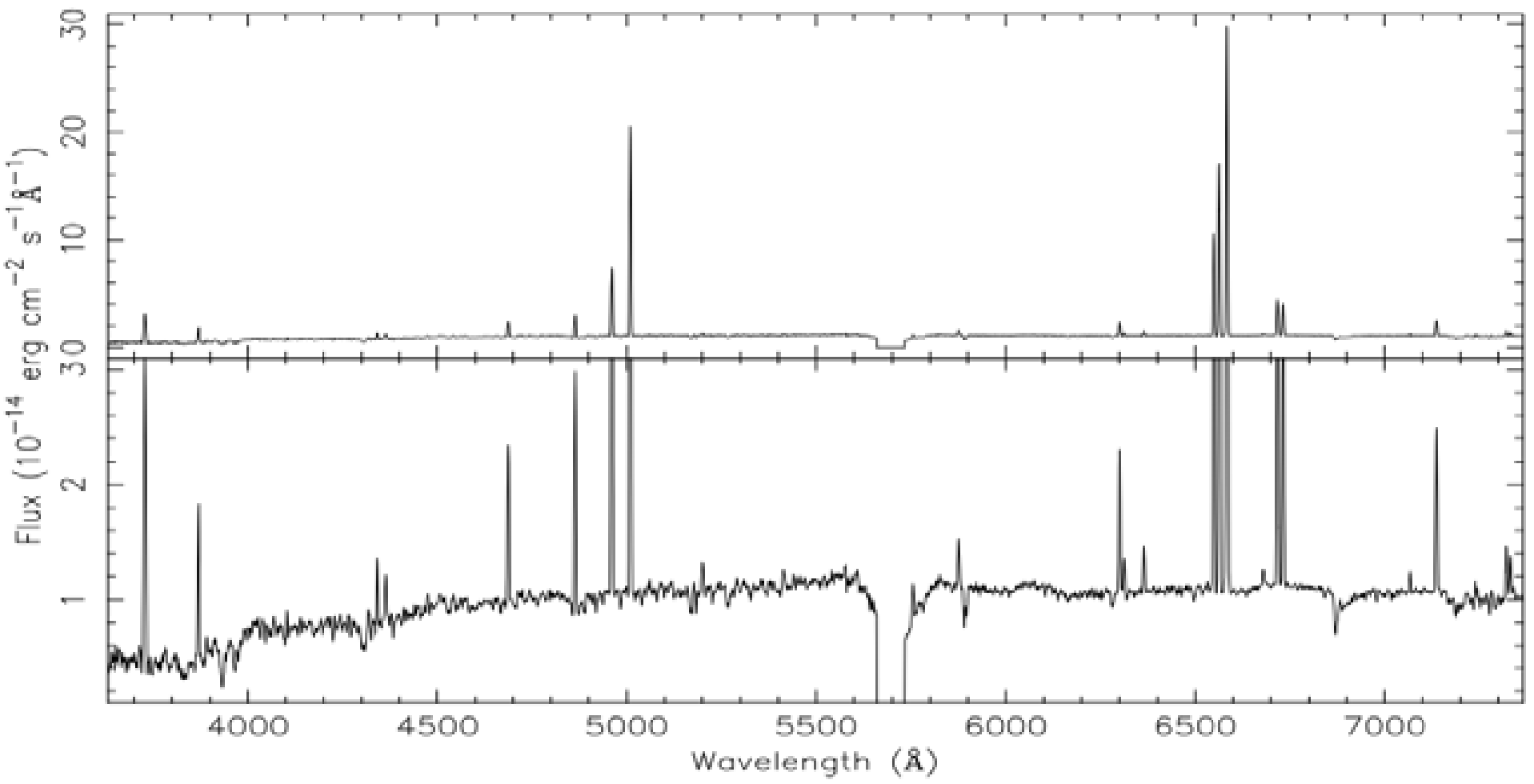

**Figure 3:** Flux calibrated, medium resolution 2.3m DBS spectrum of PHR1315-6555. The red and blue arm data have been independently flux calibrated (note gap at 5800Å). The upper panel shows the spectrum with the full range of line intensities plotted while the lower panel is re-scaled to reveal the important, fainter diagnostic lines more clearly. The [OII] doublet at 3727Å is well seen. It is clearly a PN according to its spectral signature.

| Property | PHR1315-6555 | ESO96-SC04 |
|---|---|---|
| Position RA (J2000) | 13:15:18.9 | 13:15:16.0 |
| Position Dec (J2000) | -65:55:01 | -65:55:16 |
| Position l,b | 305.368, -3.158 | 305.362, -3.162 |
| Distance (kpc) | 9.1±2.9 | 10.4±1.8 |
| Reddening E(B-V) | 0.79±0.08 | 0.72±0.02 |
| Radial Velocity | 59±2.5 kms$^{-1}$ | 57±5 kms$^{-1}$ |

**Table 1:** Main PN and cluster data comparisons

| Observation | Date | Log F Hα | Log F [OIII] |
|---|---|---|---|
| SAAO narrow-slit | May 2001 | -12.49±0.10 | -12.28±0.05 |
| DBS wide-slit | May 2008 | -12.44±0.05 | -12.32±0.05 |
| CTIO MOSAIC-II | June 2008 | -12.45±0.02 | -12.35±0.02 |

**Table 2:** Comparison of independent reddened PN line flux estimates showing excellent agreement between measures.

| Reference | Telescope | D (kpc) | *E(B-V)* | Age (Myr) |
|---|---|---|---|---|
| PJM94,JP94 | 0.9m CTIO | 7.57 | 0.72 | ... |
| CVO95 | 3.5m NTT | 11.8 | 0.75 | 700 |
| CM04 | 1.0m SAAO | 12±1 | 0.7±0.2 | 800 |
| CJE05 | 1.0m CTIO | 10.1* | 0.7* | 800 |
| Adopted values | | 10.4±1.8 | 0.72 | 800 |

**Table 3:** Fundamental parameters for cluster ESO 96SC04.

Refs: CVO95, Carraro et al. (1995); PJM94, Phelps et al. (1994); JP94, Janes & Phelps (1994); CM04, Carraro & Munari (2004); CJE05, Carraro et al. (2005).

| Characteristic | Estimated Value |
|---|---|
| Major & Minor axes | 18 x 14 arcseconds |
| Morphology | Bipolar |
| Chemistry | Type I |
| DM excitation class | 7.8 |
| Electron density ([SII]) | 240 electrons cm$^{-3}$ |
| Electron temperature, [OIII] & [NII] | 17700 & 10700 K |
| Physical radius | 0.4 pc |
| Estimated age | 16,000 years |
| Galactic z height below the plane | ~575 pc |
| Estimated ionised mass | 0.5 Msun |
| M5007 | -0.7 |
| Estimated CSPN V mag | 23.5±1 |
| CSPN temperature (cross-over) | 218,000 K |

**Table 4:** Measured & derived properties for PN PHR1315-6555. Note the derived cluster distance is used in calculation of PN parameters.

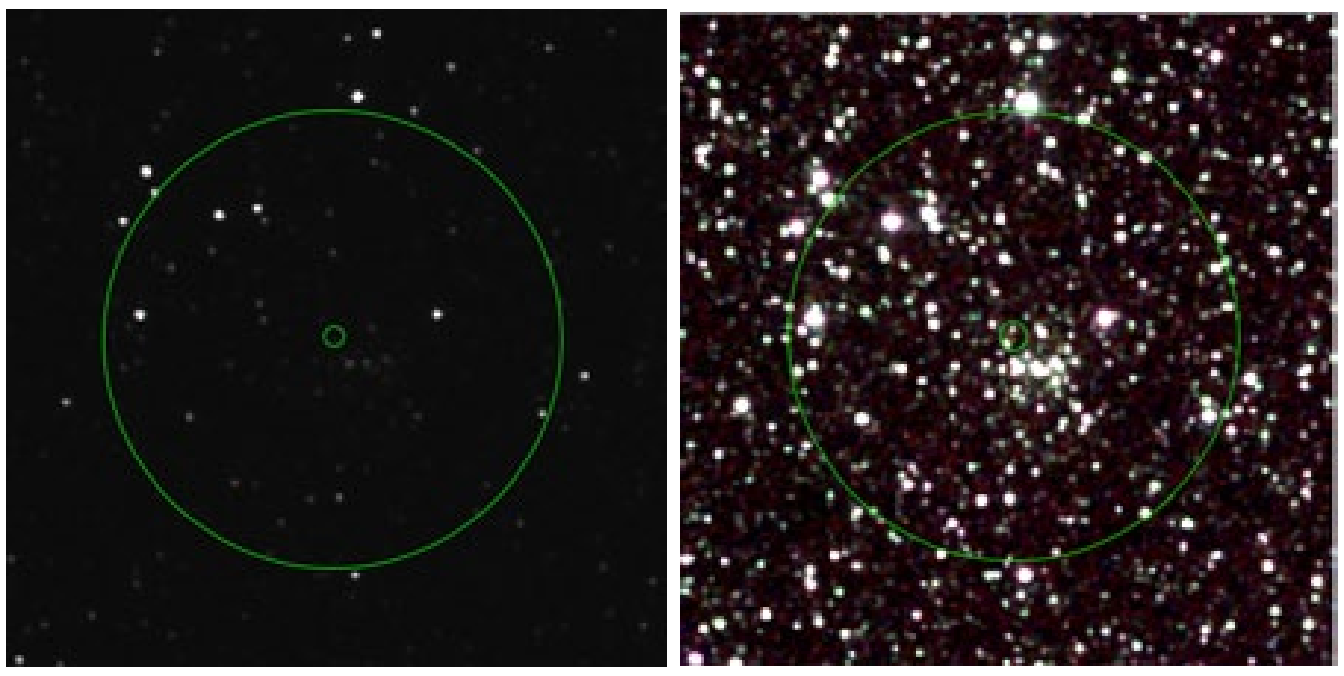

**Figure 4:** left: 2MASS K-band image centred on the PN with a 2′ radius outer circle. Note the difficulty of seeing the cluster in this standard scaled image. Right: combined 2MASS JHK cluster image at high contrast. ESO 096-SC04 is too distant for 2MASS to derive an cluster reddening from the JHK data The cluster MSTO is V = 17. Even at V-K~ 2, the top of the MS is at K = 15. The 2MASS limit is K = 14.3 and at low latitudes could be  1.5 magnitudes brighter.

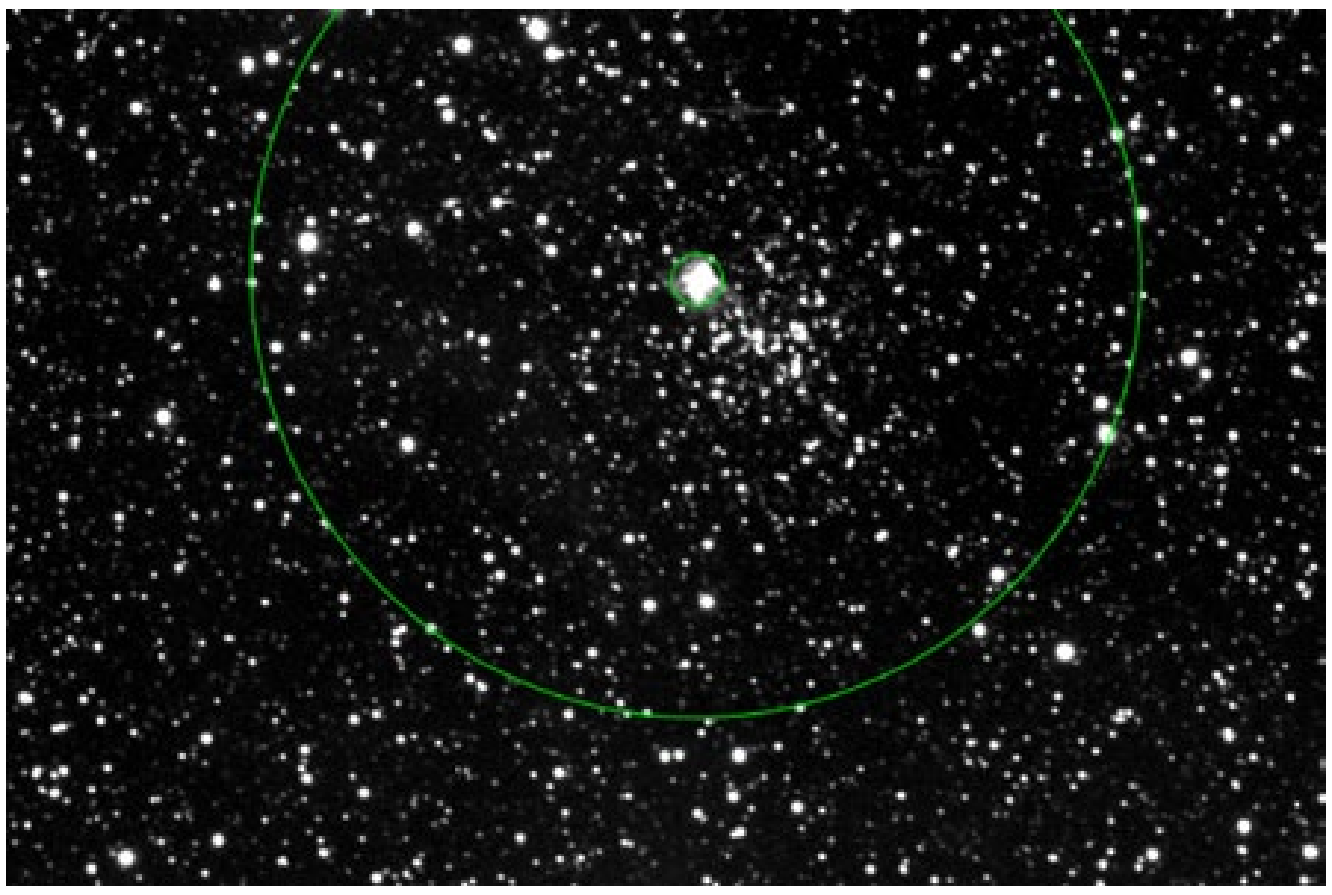

**Figure 5:** Our new, deep, high resolution CTIO 4m narrow-band [OIII] data of the PN and cluster which shows the context of PN and cluster to better effect. The outer green circle is of 4′ diameter and the small inner circle is centred on the PN.

# Dynamical mass-to-light ratios of globular clusters: analysing dark matter content and testing gravity with AAOmega

Richard Lane, Geraint Lewis & Tim Bedding (University of Sydney), László Kiss (University of Sydney & Konkoly Observatory), Rodrigo Ibata & Arnaud Siebert (Universite de Strasbourg), Péter Székely (University of Szeged), Zoltán Balog (Max-Planck-Institut für Astronomie) and Gyula M. Szabó (Konkoly Observatory)

**Using AAOmega observations we have determined the dynamical mass-to-light ratios, and masses, of ten Galactic globular clusters in an effort to determine whether weak accelerations can be explained solely using Newtonian gravitation. It has recently been claimed that globular clusters behave like large elliptical galaxies, with a flattening of their velocity dispersion profiles near their outskirts. If this could be replicated it would indicate either a large dark matter component, or a lack of understanding of weak gravitational interactions. We find that no deviation from Newtonian gravity can be detected in any of our clusters, although an interesting rise in the velocity dispersion of 47 Tuc at large radii was discovered.**

**Why globular clusters?** It has recently been claimed (e.g. Scarpa et al. 2007) that many globular clusters (GCs) have a velocity dispersion profile which flattens past about half the tidal radius ($r_t$; see Figure 1). The tidal radius of a GC is the radius at which external gravitational forces from the host galaxy are equivalent to those from the cluster itself. At this radius stars are, therefore, susceptible to being stripped from the cluster. The velocity dispersion is a measure of the range of stellar velocities within a cluster. Within the framework of Newtonian gravity, and in the absence of a significant quantity of dark matter, the velocity dispersion profile of a GC should drop off monotonically. A flattening of the velocity dispersion profile at large radii can, therefore, mean two things: either these clusters contain a large dark matter component, or we do not understand gravity at low accelerations. In either case, this would be an amazing discovery because GCs are thought to be virtually free of dark matter (see Mandushev et al. 1991; Phinney 1993; Moore 1996; Navarro et al. 1997; Odenkirchen et al. 2001; Ibata et al. 2002).

In many GCs, most stars near the tidal radius feel accelerations below $a_0$~$1.2 \times 10^{-10}$ ms$^{-2}$, the level at which either modified gravity (e.g. MOND; Milgrom 1983) or dark matter is required to reconcile the observed kinematics of elliptical galaxies with theory. This makes them an ideal testing ground for low-acceleration gravity (Sollima et al. 2010 and references therein) . Furthermore, if all GCs exhibit similar behaviour, Galactic influences cannot be the primary cause. We have studied Galactic GCs that are close enough that radial velocity information can be determined for individual stars.

**Our tests:** We chose ten GCs at varying Galactocentric distances (namely M4, M12, M22, M30, M53, M55, M68, NGC 288, NGC 6752 and 47 Tuc) and obtained a total of ~30000 stellar spectra within AAOmega fields (each of ~2 square degrees) centred on the clusters. We used a data reduction pipeline based on that from the Radial Velocity Experiment (RAVE; Steinmetz et al. 2006; Zwitter et al. 2008) to estimate radial velocities to ~1-2 kms$^{-1}$ and used various output parameters from the pipeline (such as metallicity, surface gravity and velocity) to determine cluster membership. In total, we ended up with ~5000 spectra of cluster *members*. We then used a Markov Chain Monte Carlo (MCMC) method to determine the velocity dispersions, with uncertainties, for annular bins centred on the cluster centres. We also altered the bin boundaries to see if this would alter the shape of our velocity dispersion profiles. In each case the velocity profile was unaffected by the choice of binning.

Several models exist which fit the surface brightness and velocity dispersion profiles of GCs. We chose to use the Plummer (1911) model because it is a physically motivated, fully analytic model which can be used to estimate the enclosed mass of the cluster at any radius, which is essential for determining mass-to-light ratios. Other models, such as the King (1966) models, are commonly used in GC studies, however, they are not analytic, which makes them much more cumbersome. Furthermore, our total mass estimates from our Plummer model fits agree well with literature values, most of which are estimated without the use of Plummer models, indicating the suitability of this approach. Figure 2 shows two of our velocity dispersion profiles overlaid with best fit Plummer models. The models fit the data very well indicating no deviation from the monotonic drop in velocity dispersion expected from Newtonian gravity. We did not observe any flattening of the velocity dispersion profiles in any of our clusters, except in M4 which we attribute to influence from the Galactic tidal field due to its orbit. Therefore, we could not reproduce the efforts of Scarpa et al. (2007), despite choosing two of our clusters to overlap with their studies, namely M30 and NGC 288.

We noticed a distinct rise in the velocity dispersion of 47 Tuc beyond about 30 pc from the centre (again see Figure 3), which we interpret as evidence for a recent merger between two protoclusters during the formation of this GC. Other explanations are possible, however, this seemed to be the most plausible at the time of publication (Lane et al. 2010a). Since we first detected this, Küpper et al. (2010) noted that this signature may be due to projection effects of escaped member stars and we cannot discount this possibility.

As discussed above, the lack of velocity dispersion flattening implies no significant dark matter component. To quantify this, we determined the mass-to-light ratios for each cluster. We calculated dynamical mass-to-light-ratio profiles, using the Plummer model to estimate the mass within a certain radius and using published surface brightness profiles (Trager et al. 1995) for the luminosities. We found that none of our clusters have mass-to-light ratios of ≥5 (see Figure 3) which means that visible matter (stars) dominate the mass of the cluster.

As part of this work, we extended a metallicity calibration method so that it can be used for more distant, and dimmer, clusters by using the *K* band magnitude of the Tip of the Red Giant Branch instead of the Horizontal Branch magnitude, and equivalent widths of the calcium triplet absorption lines (at 8498, 8542 and 8662Å). We measured the rotations of all ten clusters and analysed the Galactic halo using the tidal heating of the clusters by the Galaxy. All our results have recently been published in MNRAS and ApJ Letters (Lane et al. 2009; Lane et al. 2010a; Lane et al. 2010b; Lane et al. 2010c).

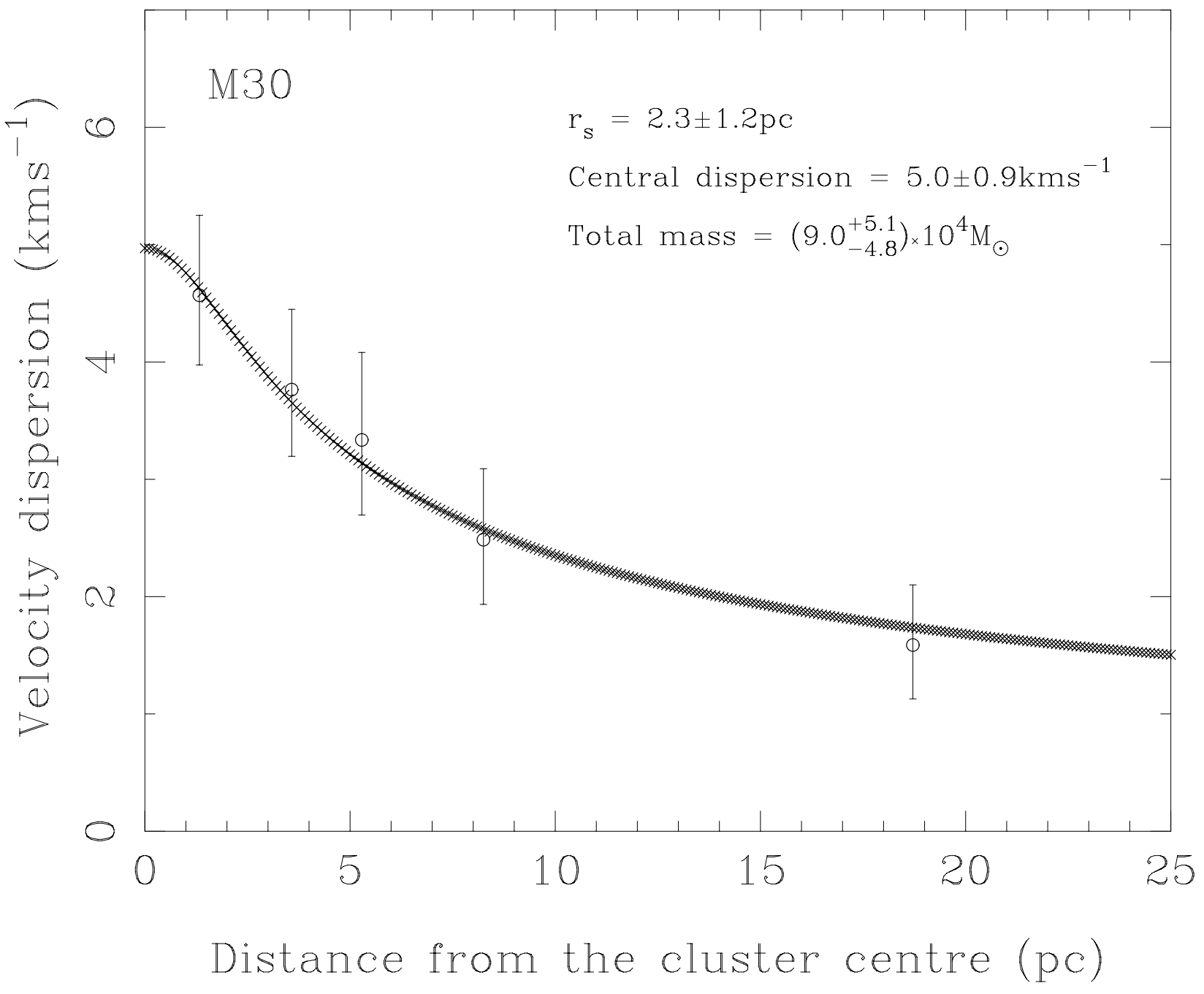

**Figure 1:** Velocity dispersion profile of NGC 7099 (M30) from Scarpa et al. (2007). An apparent flattening of the velocity dispersion profile can be seen for radii greater than ~10 pc. Note that the solid line is not a fit to the data but simply added by the authors to 'guide the eye'.

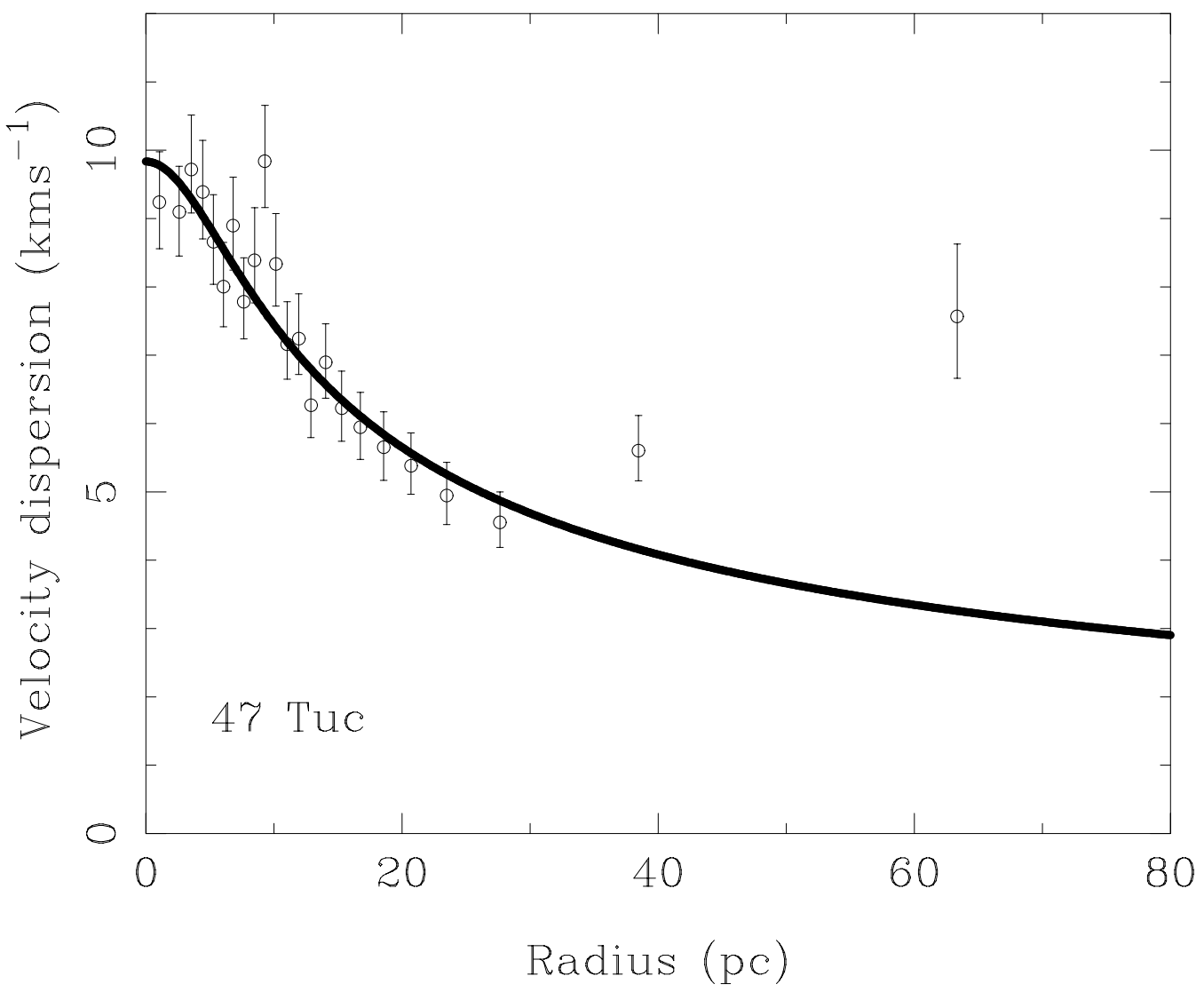

**Figure 2:** Two of the velocity dispersion profiles obtained from this study. Note the lack of flattening at large radii. M30 (NGC 7099) was one of the clusters studied by Scarpa et al. (2007). The best fit Plummer profile is overlaid onto the data and values are given for the Plummer scale radius ($r_s$; equivalent to the half-mass radius in projected Plummer models such as ours), the central velocity dispersion and the dynamical mass (all calculated from the Plummer fit). It should be noted that our Plummer model for M30 also passes through the error bars from Figure 1, despite being fitted to different data, so it is not clear that the claimed flattening in Figure 1 can be considered real. A clear rise in the velocity dispersion of 47 Tuc can be seen, which we interpret as possible evidence for the merger of two protoclusters ≤7.3±1.5 Gyr ago (note that when fitting the Plummer model to the 47 Tuc data the two outermost data points were excluded).

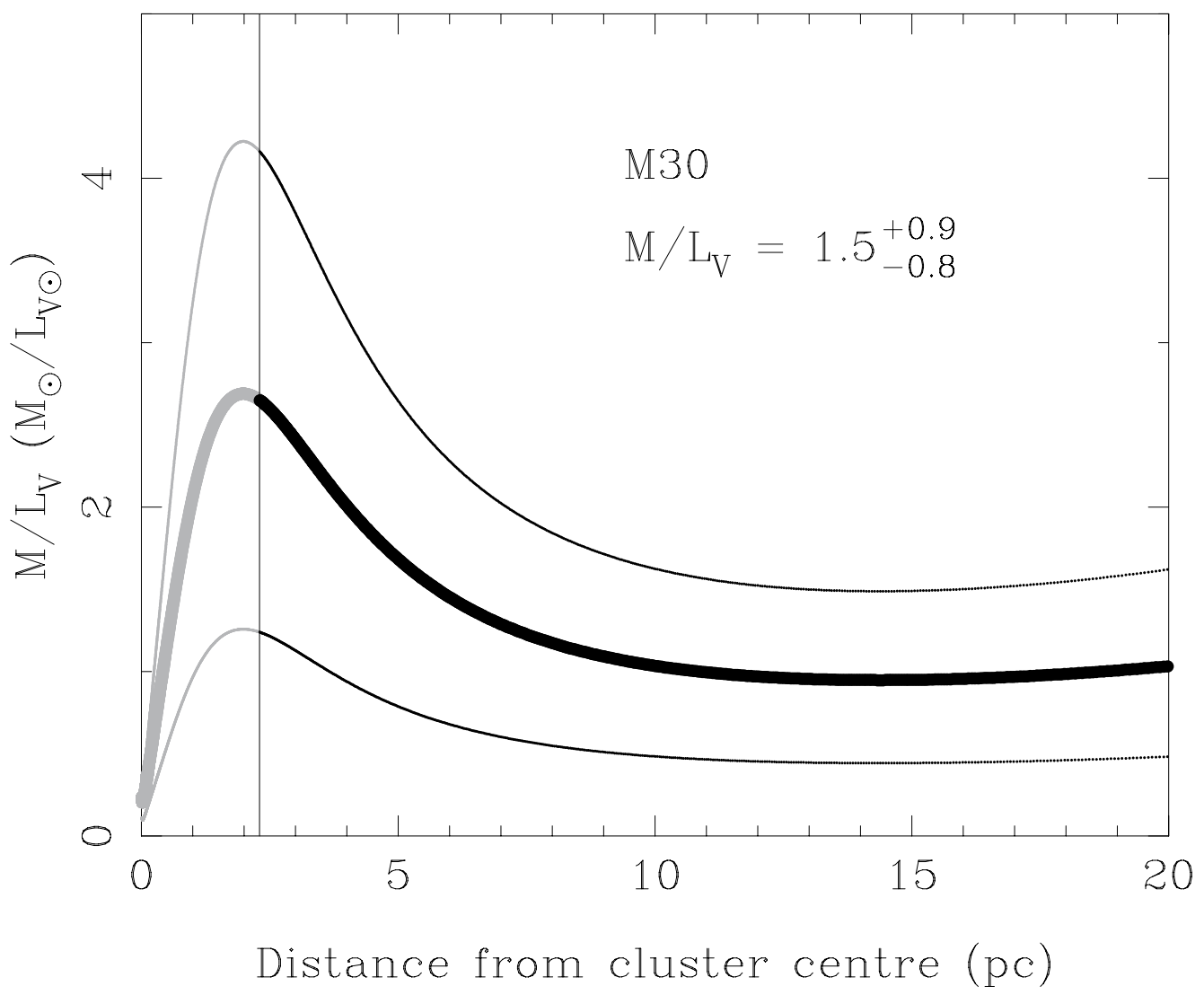

**Figure 3:** Dynamical mass-to-light ratios of M30 (NGC 7099) and 47 Tuc. In the inner region, below $r_s$ (vertical line), the mass-to-light ratio is uncertain due to crowding and confusion effects, both in velocity and luminosity, so we have only calculated mass-to-light for radii greater than $r_s$.

# The Planetary Nebula Spectrograph

Nigel Douglas (Kapetyn Institute)

The AAO has an extraordinary reputation for cutting edge instrumentation. This little item deals with a well-known instrument of which few are aware of its AAO roots. The instrument in question is the Planetary Nebula Spectrograph, or PN.S, which is celebrating its tenth year of observing from La Palma in the Canary Islands. The PN.S was designed to obtain kinematic information in early-type galaxies, which have relatively little neutral hydrogen, and therefore are difficult to investigate with radio telescopes. The idea of using planetary nebulae (PN) to obtain kinematic information in gas-poor galaxies is not at all new, and hard-working researchers had calculated kinematic data on the basis of fairly small numbers of PN long before the PN.S was built. The practical problem was that to take the required radial velocities with optical fibres was somewhat of a hit-and-miss affair leading to poor yields.

In 1994 I took sabbatical leave from the Kapteyn Institute in the Netherlands in order to work at the AAO. I was looking for a project, and became excited by an idea that had been tossed around by Keith Taylor and Ken Freeman and which aimed to improve the yield of extragalactic PN. Their idea was to use slitless spectroscopy, explaining that two slitless images, with 180 degrees between, would provide the required data to calculate the position and radial velocity of each and every PN in the field.

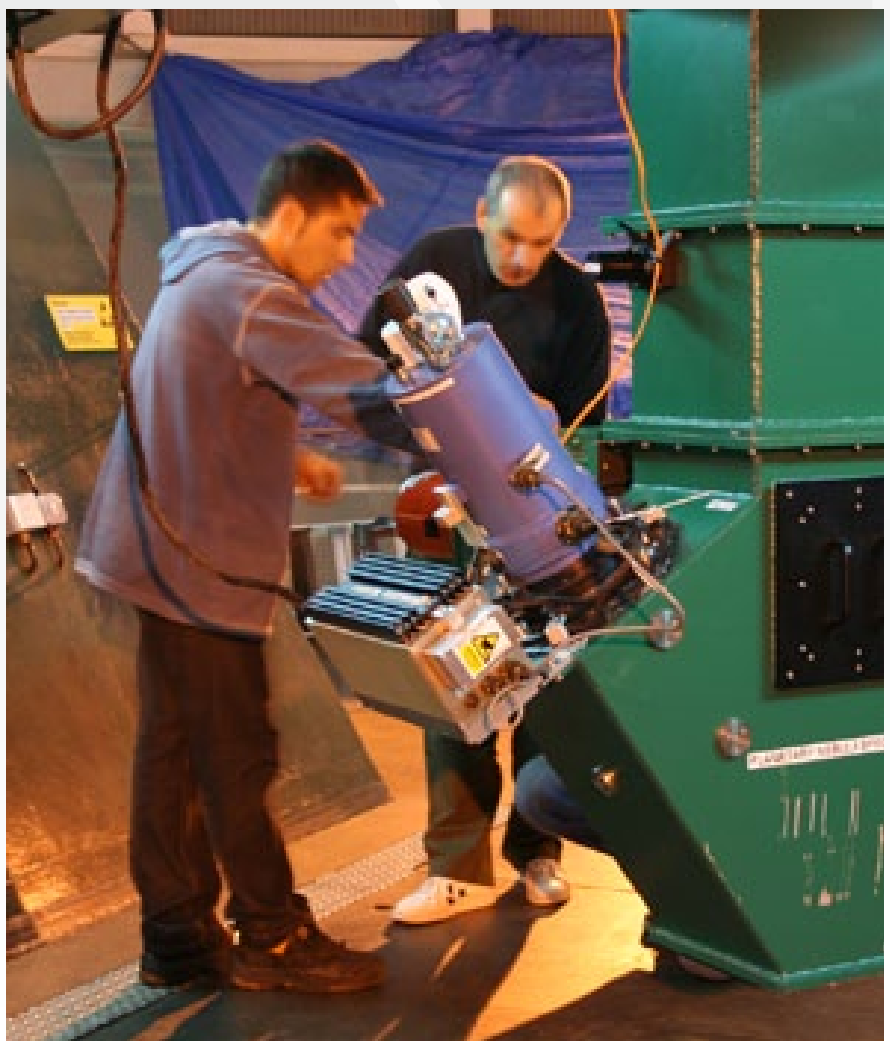

Lodovico Coccato and Nigel Douglas with the PN. S

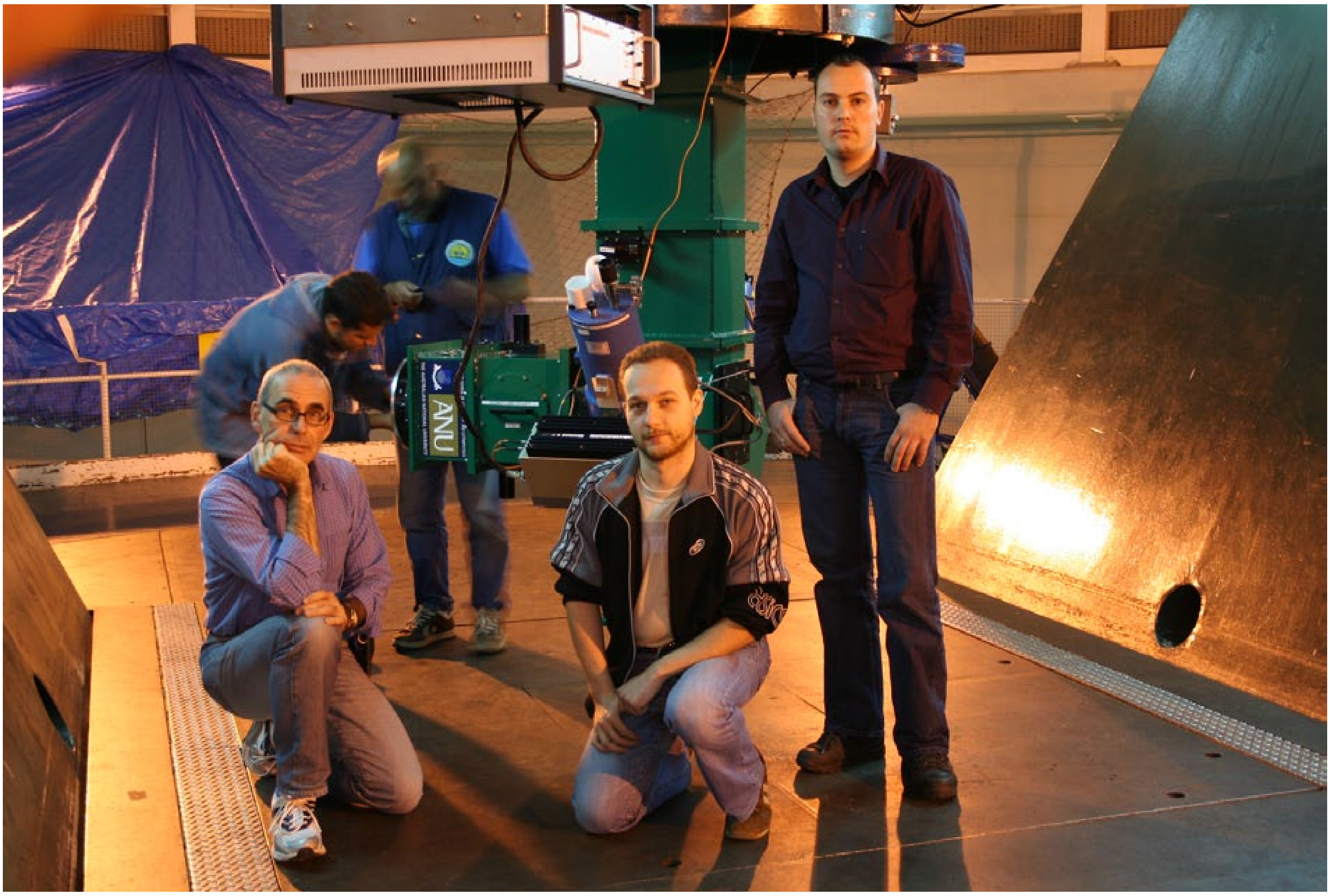

Technical staff preparing the PN. S

(Later, we were somewhat embarrassed when G. Monnet told us at a conference that we had re-invented Ferhenbach's 1947 objective prism technique.)

During my sabbatical I used the RGO spectrograph and the AAT to verify that the idea would work. While doing so, I realized that the rotating of the spectrograph could lead to flexure and that, in fact, one didn't need to do it anyway. I sketched a version of what we were then calling the PN Seeker in which the collimated beam is split into two arms by replacing the rotating grating with two fixed gratings in a symmetric arrangement that split the light into two separate cameras. This was the final configuration of the PN.S. and we called the arrangement "Counter-dispersed imaging" (CDI).

Back in the Netherlands, I started looking for funds for the design and building of the instrument, the first injections coming from the Observatory of Capadimonte via Massimo Cappaccioli, the Kapteyn Institute, and the Australian National University (ANU) Major Equipment Fund, via Ken Freeman.

The AAO was unable to build the PN.S because their books were full at that time, but luckily the ANU took it on, thus keeping the project in mostly in Australian hands, which nicely rounded off my sabbatical. I must mention the very substantial roles that were subsequently played by Damian Jones of Prime Optics (optical design), and the ANU staff John Hart, Julia Hu, and Gabe Bloxham, who turned the concept into an opto-mechanical design. Heavy components were manufactured by ASTRON in the Netherlands.

When shipping the completed instrument the ANU fell foul of the fact that any place in the Canary Islands larger than a corner shop includes the name "La Palma" and after desperate midnight phone calls realized that the PN.S had been sent to the wrong island. This was the first setback for the PN.S, and the last.

During the weekend of July 14-15 2001, Aaron Romanowsky and I put the instrument together, and on July 16 we had mounted the spectrograph on the 4.2m WHT; ING staff had hooked up the left and right detectors, and we had fired up the simple DOS commands I had written for control of shutters (those were the days!). Seeing was poor during the first-light observations of M31 and NGC 7457 but lingering doubts about whether the principle of the PN.S would work was blown away as the counter-dispersed images appeared before us. Rough estimates of the radial velocities of PN could be made straight of the screen on the basis of the separation of the dual images made by each PN.

As mentioned, the PN.S is in its tenth year of fruitful operation, with highlights such as the catalogue of nearly 200 PN in NGC 3379, the analysis of which leading to a controversial "dearth" of dark matter and an entry in Science, and 255 PN in NGC 4494 providing kinematic data over 7 effective radii. In a collaborative survey of M31 PN.S data was combined with fibre data taken by a team headed by D.Carter and in 12 nights of observing the PN.S found 3,300 emission objects, of which 2,615 we designated as planetary nebulae.

In the meantime PN.S has been equipped with a third camera, providing direct images in Hα, and the DOS machine has been replaced by a laptop brilliantly programmed by Electronx in the Netherlands – ready for many more years of service – and it all started at the AAO!

# An analysis of the component principle of sky subtraction

Rob Sharp (AAO)

Barring the deployment of mirror extensions, a cloud suppressor or the world's most extreme wide-field AO system, the primary factor limiting the depth to which AAOmega can be used to accurately record the spectra of astronomical sources is the accuracy with which the atmospheric OH-airglow lines can be subtracted from long-duration exposures. Conventional wisdom for sky subtraction has been to use the *mean sky* method whereby 5-10% of the available fibres are allocated to blank sky positions in each fibre configuration (and you did carefully check each position was really blank didn't you). Combining these blank sky spectra creates a *master sky* spectrum for the observation which is essentially *noise free* by virtue of the enhanced signal-to-noise ratio from stacking a number of spectra.

Scaling and subtracting this *noise free* spectrum from each and every science observation should allow one to achieve a residual noise level of between 1% and 3% of the sky signal (based on the assumption of a Poisson process of arrival of photons for an hours exposure, see Figure1). However, systematic artefacts in the sky subtraction across each frame limit the accuracy to which results can be obtained. For short exposures, these artefacts are typically at a level similar to the Poisson limit, but for longer exposures on fainter sources they can result in significant reduction in overall sensitivity (this typically resulting in half crazed and near blinded astronomers desperately trying to identify those missing 20% of redshifts from that last oh-so-important survey field). What is needed is: firstly, a proper understanding of the problems limiting the accuracy on the sky subtraction; and secondly, a way to fix it!

The problem is not to dissimilar to that faced by slit mask spectroscopy and we are all now familiar with the elegant solution Karl Glazebrook (1998) identified to the problem; the *nod-and-shuffle* observing technique (Glazebrook & Bland-Hawthorn 2001, Abraham et al. 2004). There are however a number of intrinsic inefficiencies with the nod-and-shuffle technique (see Sharp & Parkinson 2010). Even with the advent of improved fibre extraction for AAOmega (Sharp & Birchall 2010) which has allowed the realisation of the mini-shuffle technique (see Sharp 2009 AAO newsletter article discussing the technique) observations still suffer a reduced efficiency at the root-two level with respect to the *noise free* mean sky approach.

An alternative approach to sky subtraction was proposed some time ago by Kurtz & Mink (2000). They advocate the use of a Principal Components Analysis (PCA) to identify and remove systematic residual in the sky-subtracted data. This idea was developed further by Wild & Hewitt (2005) who applied the approach to archival SDSS data in order to improved systematic residual which trigger false positives in work such as searches for the spectra of background lensed galaxies.

Encouraged by these earlier works, both the WiggleZ (Drinkwater et al. 2009) and GAMA (Driver et al. 2009) large AAOmega projects implemented the PCA process to enhance the survey data. Credit for these Herculean undertakings must be given to Swinburne University PhD student Emily Wisnioski (an Honours student at the time of her investigations) and Edinburgh University PhD Hannah Parkinson. While both implementations have been working exceptionally well over the last two years of operation, ultimately the approach developed for the GAMA project appears to provide the most reliable results and therefore it is this approach which has been adopted for inclusion with the 2dfdr data processing software used with AAOmega observations.

The details of the process and the results achieved are detailed in Sharp & Parkinson (2010). In this paper we reproduce a selection of these results in order to demonstrate the potential power of this approach to sky subtraction for faint astronomical sources, particularly when coupled with the planned upgrade of the AAOmega red CCD to a new high-resistivity devices which will boost performance in the range 850nm-1100nm by a factor of two-to-three.

Future applications of the technique include new approaches to the fibre-optic system on ELTs, such as the MANIFEST concept proposed by the AAO for GMT. The use of PCA allows sensitive measurement of faint sources "without all that tedious mucking about with nod-and-shuffle", although there are likely still applications where the superb pixel-to-pixel fidelity of spectra which nod-and-shuffle provides is still essential. The process cannot be directly applied to observations of stellar sources, such as the current generation of AAOmega stellar programs or those planned for the HERMES instrument. This is primarily because data with significantly similar spectral features across many input spectra will contaminate the PCA with real spectral features. A modified approach, where a larger number of fibres than usual are allocated to blank sky, might provide an interesting solution that still results in improved sky subtraction.

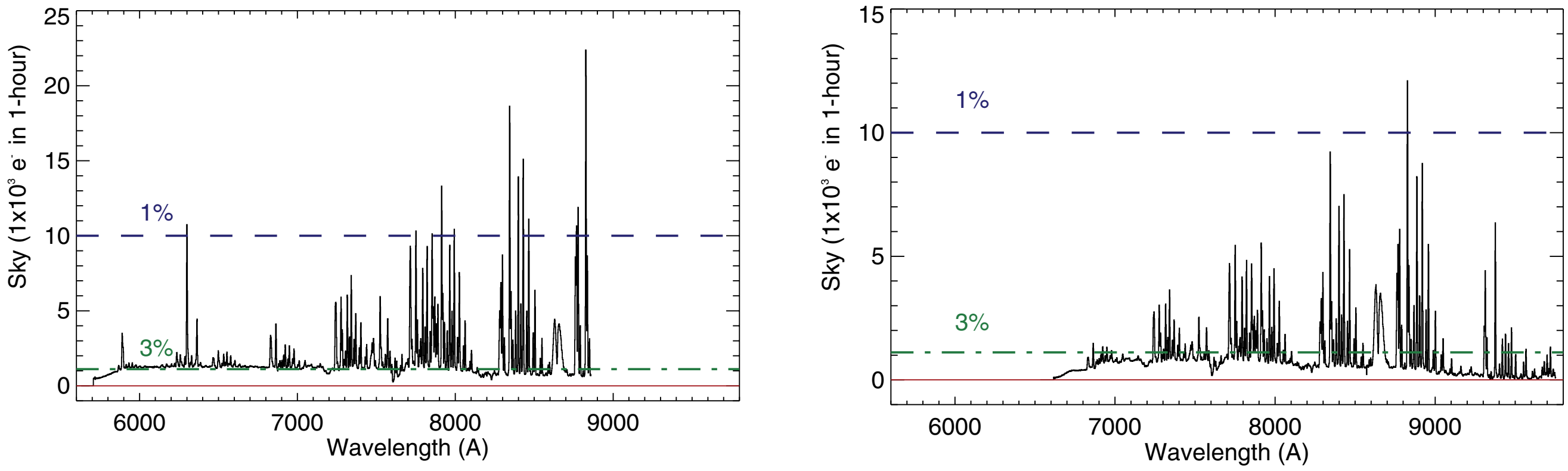

**Figure1:** Example sky spectra (converted to electrons per hour after gain correction) using the observing configurations for the GAMA and WiggleZ programs in the red-arm of AAOmega. The spectral regions in which 1% and 3% sky subtraction accuracy are possible are marked (i.e. the regions in which the residual Poisson error from the sky-line count rate is of the order 1% or 3% of the sky signal).

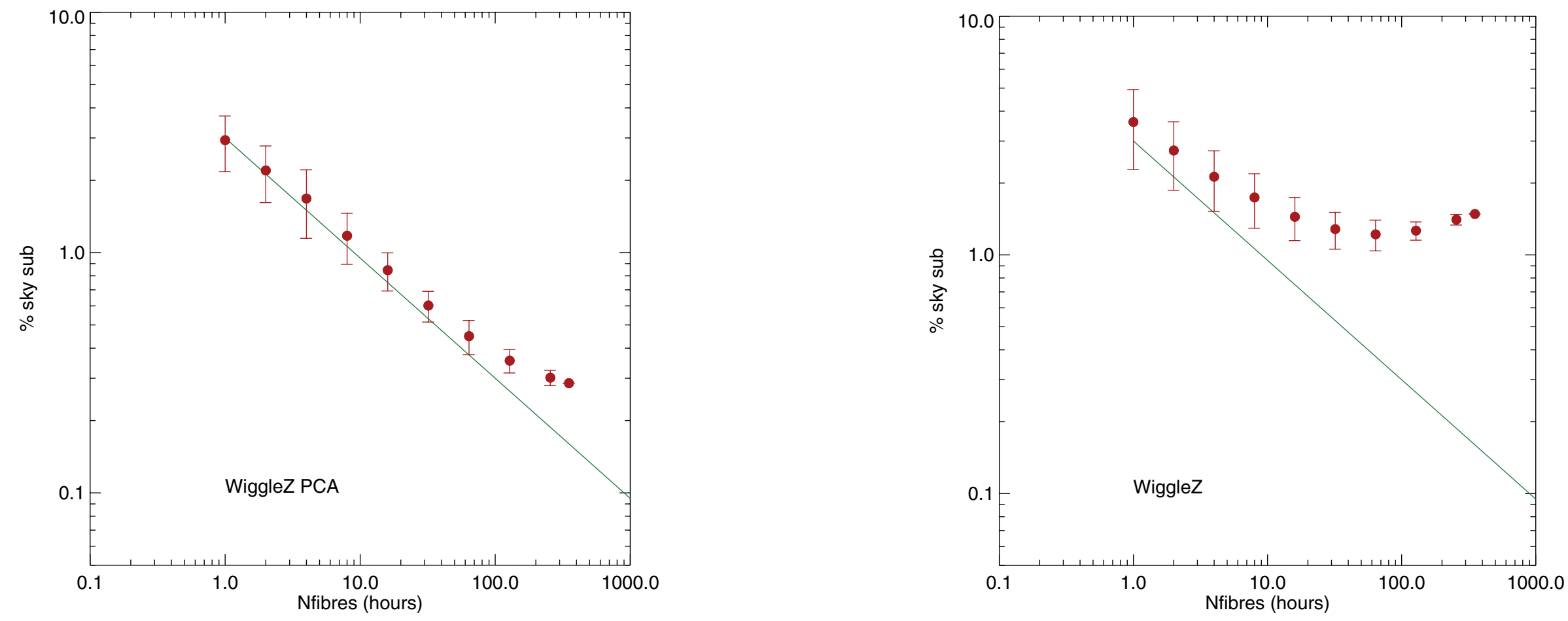

**Figure 2:** Using a large number of frames from the WiggleZ survey project, one is able to simulate long on sky exposures. WiggleZ targets are high EW emission line sources and so have little source continuum contributing Poisson noise to the majority of the spectral pixels. Using the 2009 WiggleZ survey data set the sky subtraction accuracy achievable in a very long duration exposure can be simulated. Here, the rms scatter in a stacked spectrum is shown as a percentage of the sky level as a function of the effective length of the exposure. We see that without the PCA sky subtraction, there is a significant departure from the square-root-of-time suppression of the sky residual (indicated by the green line). With the application of the PCA process, a rate much closer to the Poisson rate is recovered. The departure for very long durations is consistent with the residual signal level from low signal-to-noise continuum emission in the WiggleZ spectra.

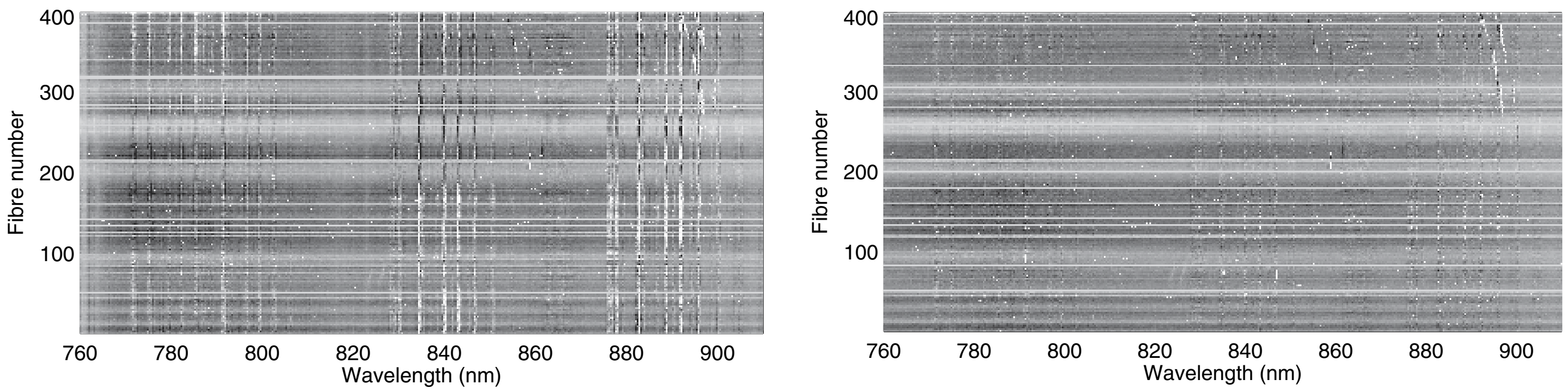

**Figure 3:** A graphic illustration of the effect of using the PCA sky subtraction is seen in stacking thirty hours of data from the WiggleZ survey, with and without the PCA sky subtraction. The images are centred the strong OH bands around 840nm. Without the PCA processing, Strong systematic residuals are present in the sky subtraction. On application of the PCA algorithm to the individual frames prior to stacking, the result is consistent with the expected Poisson residual. Note, the curved features are the result of a small number of bad columns on the red AAOmega CCD in this wavelength range.

# GNOSIS: an OH suppression unit for existing near-infrared spectrographs

**S. C. Ellis** (University of Sydney), **J. Bland-Hawthorn** (University of Sydney), **J. S. Lawrence** (AAO, Macquarie University), **J. Bryant** (University of Sydney), **R. Haynes** (Astrophysikalisches Institut Potsdam), **A. Horton** (AAO), **S. Lee** (AAO), **S. Leon-Saval** (University of Sydney), **Hans-Gerd Löhmannsröben** (Astrophysikalisches Institut Potsdam), **J. Mladenoff** (AAO), **J. O'Byrne** (University of Sydney), **W. Rambold** (Astrophysikalisches Institut Potsdam), **M. Roth** (Astrophysikalisches Institut Potsdam), **C. Trinh** (University of Sydney)

**GNOSIS is an instrument designed to filter atmospheric OH emission lines from light being delivered to near-infrared spectrographs. This is achieved using aperiodic fibre Bragg gratings: extremely complex filters printed into optical fibres. The consequent reduction in background will result in the darkest near-infrared background of any ground-based observatory.**

Near-infrared spectroscopy suffers from a very bright and variable background (see figure 1). This background severely hinders accurate sky subtraction; the brightness introduces high levels of Poissonian noise and the variability adds high levels of systematic noise. This situation is unfortunate since deep near-infrared spectroscopy is desirable for many areas of observational astronomy and crucial for some.

For example, deep near-infrared spectroscopy would allow us to measure the star-formation rate at redshifts $0.7 \le z \le 1.7$ using the redshifted Hα emission line. This would be very beneficial since this is the epoch during which the star-formation rate is thought to peak, and when galaxies were being assembled: an important era in the history of the universe. Furthermore measuring the star-formation rate from the Hα line would allow direct comparison with the star-formation rates measured in the same manner in the nearby universe.

Similarly, identifying and characterising low-mass stars is essential for a complete understanding of star-formation and planet-formation, but low-mass stars emit most of their light in the near-infrared. Deep near-infrared spectroscopy would enable the measurement of the temperatures, surface gravities and atmospheric composition of low mass stars.

Indeed, deep near-infrared spectroscopy is so desirable that NASA, the European Space Agency and the Canadian Space Agency are together planning on launching an infrared space telescope in 2014 (the James Webb Space Telescope) into an orbit 1.5 million km from Earth, at a cost of $4.5B, in order to circumvent the infrared background (Gardner et al. 2006).

GNOSIS is an OH suppression unit designed to achieve this goal from the ground. This is made possible by a novel technology developed in collaboration between the University of Sydney, the Anglo-Australian Observatory, the Astrophysikalisches Institut Potsdam and industrial partners. This technology is based on two recent advances in photonics: aperiodic fibre Bragg gratings, and photonic lanterns. The first makes it possible to construct extremely complex optical filters. The second allows very efficient conversion between multi-mode and single-mode fibres. This technology is at the heart of GNOSIS, and we discuss its behaviour and performance below, followed by a description of its planned implementation in GNOSIS. First though, we review the properties of the near-infrared background.

## The near-infrared background

The reason that the near-infrared sky is so bright and variable is due to the de-excitation of OH radicals in the upper atmosphere (see Ellis & Bland-Hawthorn 2008 for a review of the near-infrared background). At an altitude of 87km, in a layer about 11km thick a reaction takes place between hydrogen and ozone giving rise to an OH molecule, viz.,

$H + O_3 \rightarrow OH^* + O_2$.

The OH molecules are excited by sunlight throughout the day, and then rotationally and vibrationally de-excite throughout the night giving rise to a very bright forest of emission lines (see figure 2). The brightness of the lines declines by about half throughout the course of a night, but density waves in the atmosphere cause variations in intensity of around 10% on the timescales of minutes (Ramsay, Mountain & Geballe 1992; Frey et al. 2000).

In addition to the OH lines other major sources of background are the zodiacal scattered light (sunlight reflected of dust in the plane of the solar system) and thermal emission from the telescope and instrument optics. The thermal emission starts to become dominant at $\lambda > 1.8$ μm for telescopes at room temperature.

Although the OH lines are very bright, they are intrinsically very narrow (FWHM $\approx 5 \times 10^{-6}$ μm), and significant gaps exist between the lines, in which the intrinsic background will be dominated by zodiacal scattered light. Therefore if the OH lines can be selectively filtered, whilst leaving the interline regions unaffected the near-infrared night background could be made very dark indeed.

However it is very important to realise that whilst the intrinsic interline background is very dark, in practice this region is dominated by scattered OH light due to the spectrograph optics. This scattering is present in all spectrographs and is very difficult to control. For this reason the noise from the OH light cannot be removed simply, e.g. using software subtraction, or by observing at high spectral resolution. Therefore the OH light must be removed prior to entering the spectrograph.

## OH suppression

The OH suppression in GNOSIS is achieved using fibre Bragg gratings. These are optical fibres which have a variation in the refractive index of their cores. As light propagates down the fibre it undergoes Fresnel reflection at each variation in refractive index. The variations in refractive index are small, and so the reflections at each change are also small, but by having many such variations along the length of the fibre very strong reflections can be built up at a particular wavelength.

Fibre Bragg gratings were originally developed for the telecommunications industry and to make them useful for astronomy two adaptations must be made. Firstly, many notches must be printed in a single fibre (i.e. there must be many wavelengths which are reflected; telecommunications gratings have only a single notch), and at very precise wavelengths. Secondly they must be made to work with multi-mode fibres. Fibre Bragg gratings are printed in single-mode fibres, which have very narrow cores (10μm diameter), but in order to gather a sufficient amount of light from an astronomical object we must use fibres with a larger diameter (e.g. 140μm for AAOmega), and which are therefore multi-moded.

The first problem was solved by Bland-Hawthorn et al. (2004). This technology has now matured to the point where the GNOSIS fibre Bragg gratings will suppress 103 of the brightest OH doublets between 1.47 and 1.70μm, using two gratings in series.

The performance of these gratings is very good. The wavelengths of the notches are printed to an accuracy of $11 \times 10^{-6}$ μm and are $< 2 \times 10^{-4}$ μm wide. The depth of the notches scales with the brightness of the lines to be suppressed, and the deepest notches remove 99.999% of the light. Furthermore the profile of the notches is extremely square, which maximises the amount of interline light that makes it through to the spectrograph. The interline throughput of the fibres is >90%. Altogether the fibre Bragg gratings provide very efficient and effective filtering of the OH spectrum.

The problem of multi-mode to single-mode conversion was solved by Leon-Saval et al. (2005), using fibre tapers to convert between a multi-mode fibre and a parallel array of single-mode fibres and back again. Fibre Bragg gratings can then be connected to the single-mode fibres. These devices, called photonic lanterns, are the key to unlocking the potential of many photonic technologies which require single-mode fibres for astronomical applications. Noordegraaf et al. (2009) have shown that it is possible to make efficient photonic lanterns by matching the number of modes in the multi-mode fibres to the number of single-mode fibres. The losses in the MMF-SMF-MMF conversion, including splicing are < 20%.

The performance of our OH suppression technology was demonstrated in on-sky tests carried out at the AAT in December 2008. These tests showed that good OH suppression can be achieved using fibre Bragg gratings and our results reported by Bland-Hawthorn et al. (2009).

## GNOSIS

GNOSIS is an OH suppression unit designed to exploit the technology described above. The primary requirement of GNOSIS is to demonstrate the potential of fibre Bragg grating OH suppression. This will be done by measuring the interline background and through the demonstration of OH suppressed observations of astronomical targets.

GNOSIS is fully funded by an Australian Research Council Linkage Infrastructure, Equipment and Facilities grant led by the University of Sydney, with further contributions from the Astrophysikalisches Insitut Potsdam. GNOSIS will be built by the Australian Astronomical Observatory. The project is currently in the preliminary design phase, and is expected to see first light by early 2011. In its first implementation GNOSIS will feed IRIS2 on the AAT with OH suppressed light over two-thirds of the H band. A J band suppression unit is a proposed upgrade. Thereafter it is planned to redeploy elements of GNOSIS to feed the GNIRS spectrograph on Gemini.

The rapid time frame for the design and build of GNOSIS is possible because GNOSIS will feed existing spectrographs. Therefore GNOSIS will consist of interfaces between the telescope, the OH suppressing photonic lanterns, and the spectrograph. The preliminary design is as follows.

### Fore-optics

GNOSIS will be mounted at the f/8 Cassegrain focus of the AAT. A fore-optics unit will take light from the telescope and feed it to an integral field unit. A beam splitter will divert some fraction of the incoming light to the acquisition and guide-camera. A deployable mirror will allow a Xe arc lamp to illuminate IRIS2 for wavelength calibration. A schematic drawing of the fore-optics is shown in figure 3.

### Integral field unit

GNOSIS is required to permit spectroscopy of a single source at a time. This will be achieved with a seven element hexagonally packed lenslet array. Each lenslet will feed a 50 μm core diameter fibre. This fibre diameter is chosen to match the number of modes to a 1×19 photonic lantern; larger fibres would require larger lanterns, which would be more expensive to produce. Therefore in order to increase the field of view to accommodate the seeing an IFU is adopted.

The median seeing at Siding Spring is ≈1.2'' at 1.6μm. The field of view of GNOSIS will be 1.4'' which maximises signal to noise in the aperture whilst minimising losses due to focal ratio degradation when feeding the lanterns.

### Grating unit

Each fibre from the IFU will be fusion spliced to a 1 × 19 photonic lantern. The single-mode fibres in the lantern will be fusion spliced to the two fibre Bragg gratings in series. This will be followed by the reverse transition back to multi-mode fibres. Thus GNOSIS will contain 14 photonic lanterns (an input and output for each of the seven IFU channels) and 133×2 fibre Bragg gratings (i.e. two gratings in series for each single-mode fibre). These will be housed together in box in the Cassegrain cage. The fibre Bragg gratings are provided in a temperature compensating tube, which maintains the grating performance from -10 to +25 degrees Celsius. Thus there is no need to temperature control the grating box unit.

### Relay optics

GNOSIS will feed IRIS2 (Tinney et al. 2004). The output fibres from the grating unit will be arranged into a slit block. A pair of relay lenses and a fold mirror reimages the fibre array onto the entrance slit of IRIS2, and provides the necessary magnification to reimage the fibre cores to the appropriate slit size. A sketch of the relay optics is shown in figure 4.

The interface assembly also provides a light source to back illuminate the slit block, via a deployable fold mirror, to allow acquisition.

### Expected performance

The major contributions to the GNOSIS background are the OH lines, the telescope and instrument thermal background and the zodiacal scattered light. These are shown in the top panel of figure 2. The OH emission is by far the dominant component. The estimated throughput of the different system components is shown in the middle panel of figure 2, giving rise to the final background spectrum at the detector as shown by the bottom panel.

The GNOSIS background shown in figure 2 has been used to estimate the 5σ limiting sensitivities in 1 hr. These are shown in figure 5.

We further illustrate the performance of GNOSIS with some simulated observations. These follow the simulations developed by Ellis & Bland-Hawthorn (2008), which include systematic variation in the OH lines as a function of time, and systematic errors in wavelength calibration, both of which lead to a realistic simulation of sky-subtraction errors.

Figure 6 shows a simulated 6hr exposure of a z=1.4, H=20 mag emission line galaxy. The galaxy spectrum was taken from the SDSS spectral templates, and then scaled according to the redshift and magnitude. The simulation shows the detection of the Hα emission line at λ=1.575 μm. Such observations are highly desirable; measuring Hα emission line strengths in redshifted galaxies allows a direct comparison of the instantaneous star-formation rate with that of nearby galaxies. Measuring the star-formation rate in redshifted galaxies currently requires comparison of different signatures making calibrations between the different signatures necessary. Furthermore GNOSIS will be able to measure Hα over the redshift range $1.2 \leq z \leq 1.6$, sampling the cosmologically important epoch at which galaxy assembly and star-formation are thought to be at a maximum.

Measuring the number density of low-mass stars as a function of mass is essential for a full understanding of star-formation. To achieve this it is necessary to determine the temperatures, surface gravities and atmospheric composition of low mass stars. These observations require near-infrared spectroscopy, such as the simulation shown in figure 7 of a T5 dwarf with J=20 mag. The methane and water absorption features are easily identifiable, allowing accurate spectral typing of the star. Thus GNOSIS will allow spectroscopy of very faint low mass stars, increasing the search volume and number of objects able to be properly characterised.

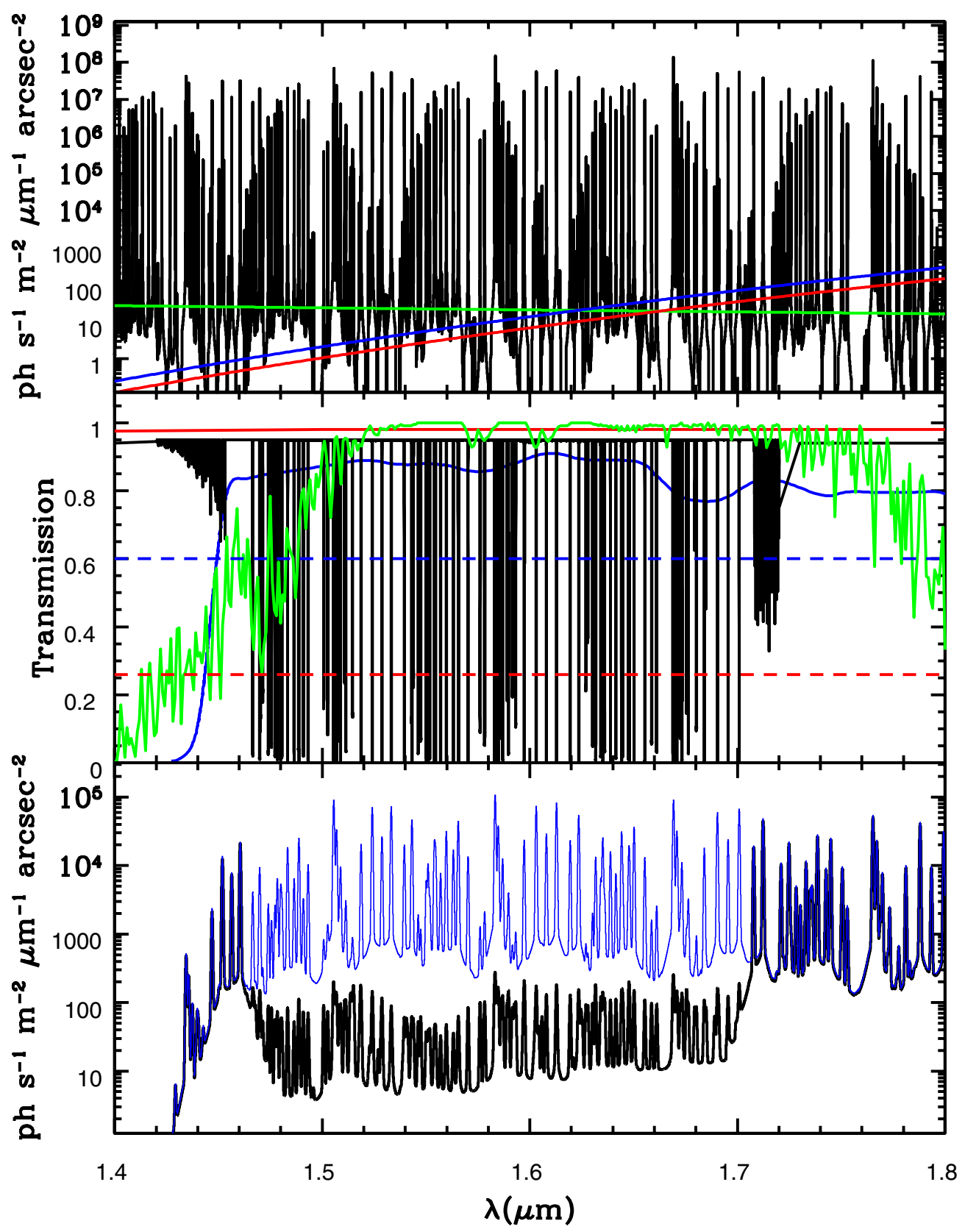

**Figure 1:** The night sky surface brightness as a function of wavelength at a good observing site. The near-infrared night sky is more than a thousand times brighter than in the visible.

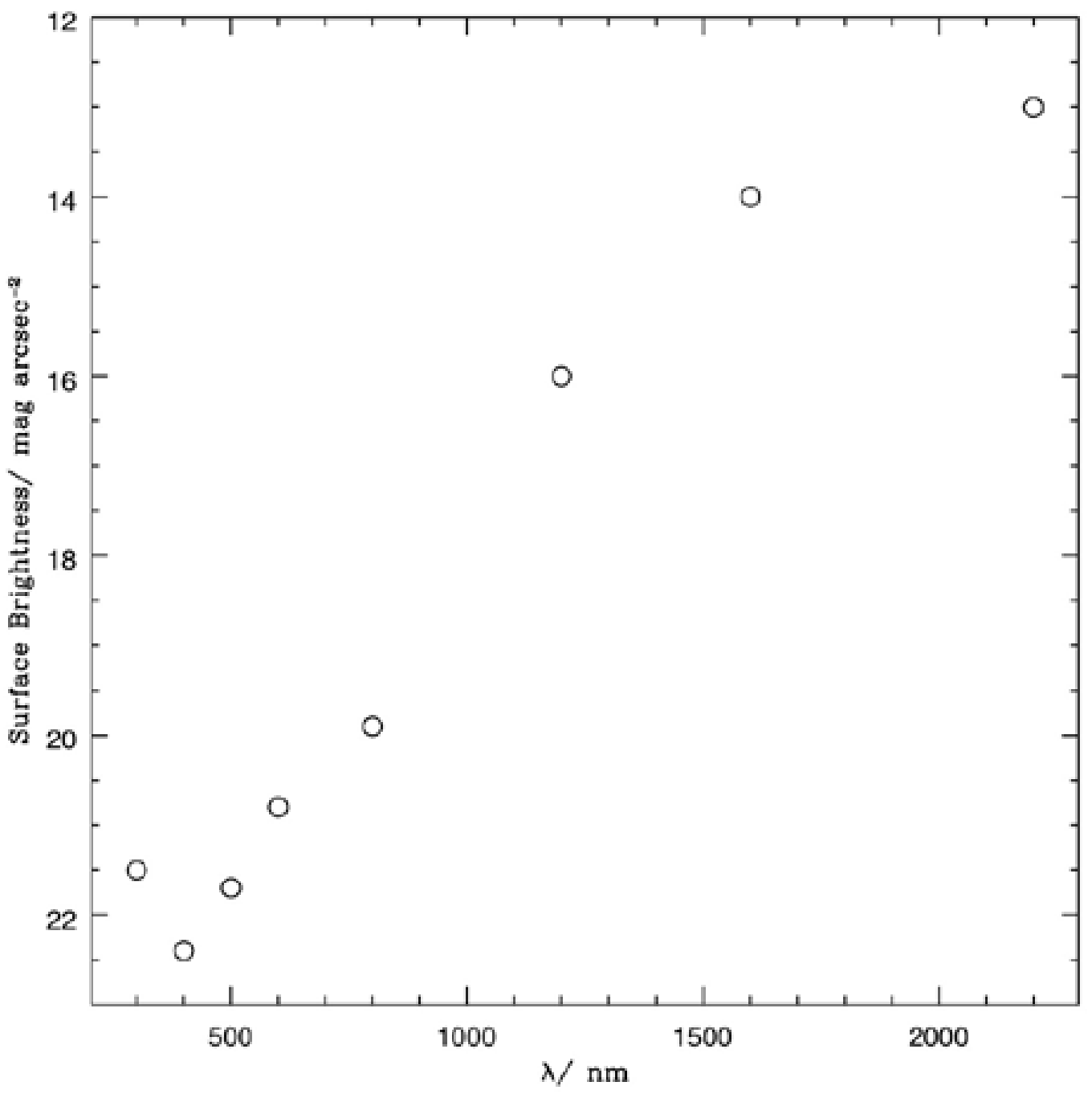

**Figure 2:** The GNOSIS background. The top panel shows a model of the OH emission line spectrum (black), the telescope thermal background (red) and the instrument thermal background (blue). The middle panel shows the system throughput, including the fibre Bragg gratings (black), atmosphere (green), the telescope reflectivity (red), the GNOSIS optics (dashed blue), the IRIS2 throughput (dashed red) and the Hs filter response (blue). The bottom panel shows the resulting background spectrum, at the resolution of IRIS2 (R=2400) incident on the detector (black), and for comparison the spectrum of an identical system without fibre Bragg gratings.

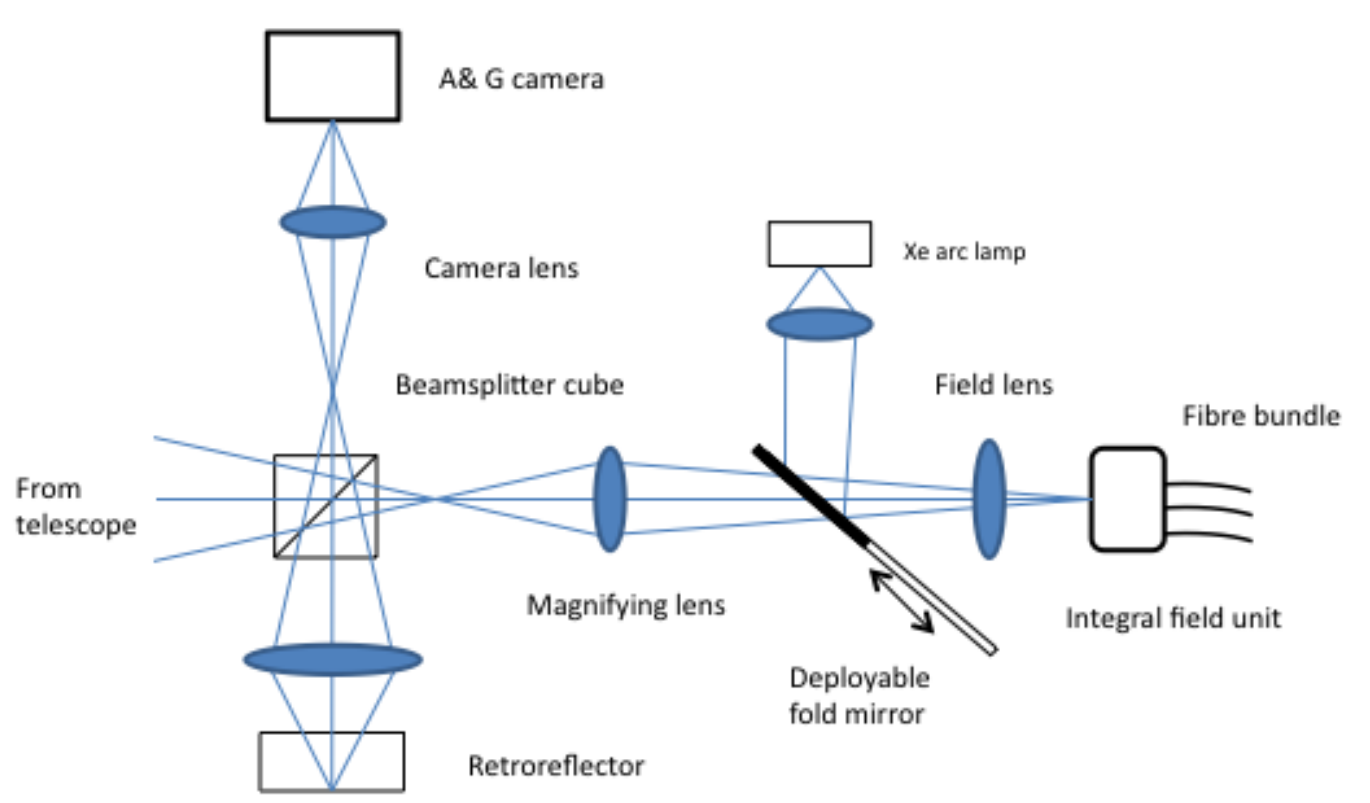

**Figure 3:** Schematic diagram of the GNOSIS fore-optics.

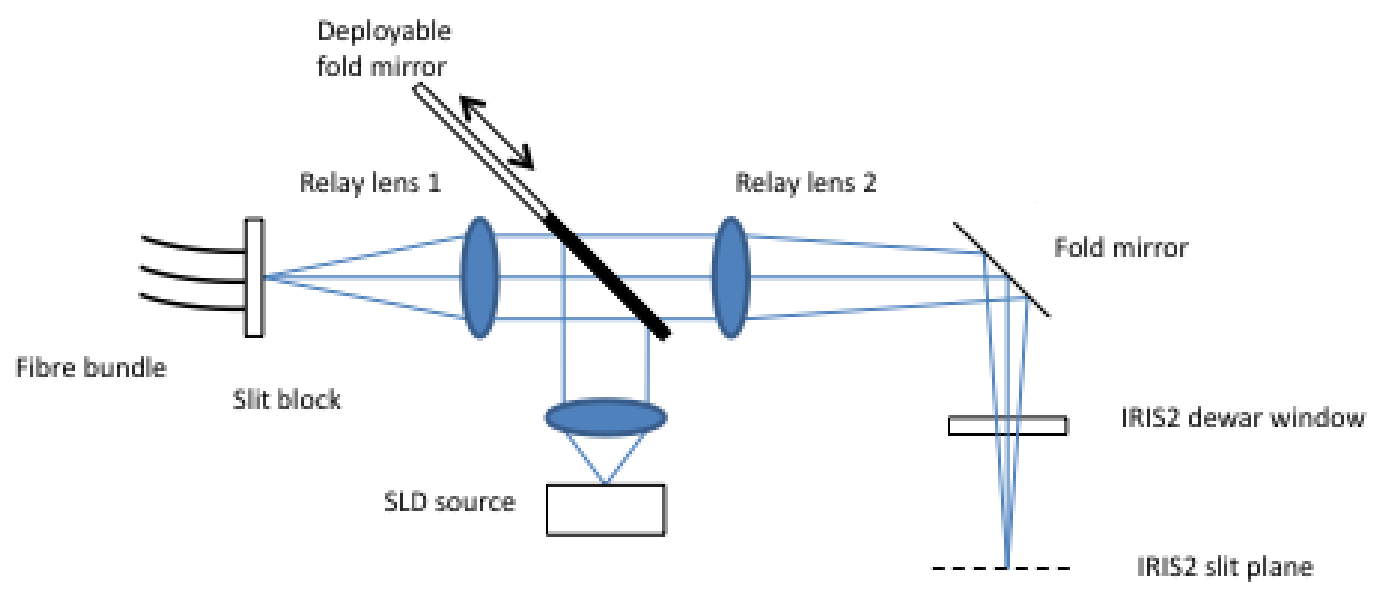

**Figure 4:** Schematic diagram of the GNOSIS relay-optics.

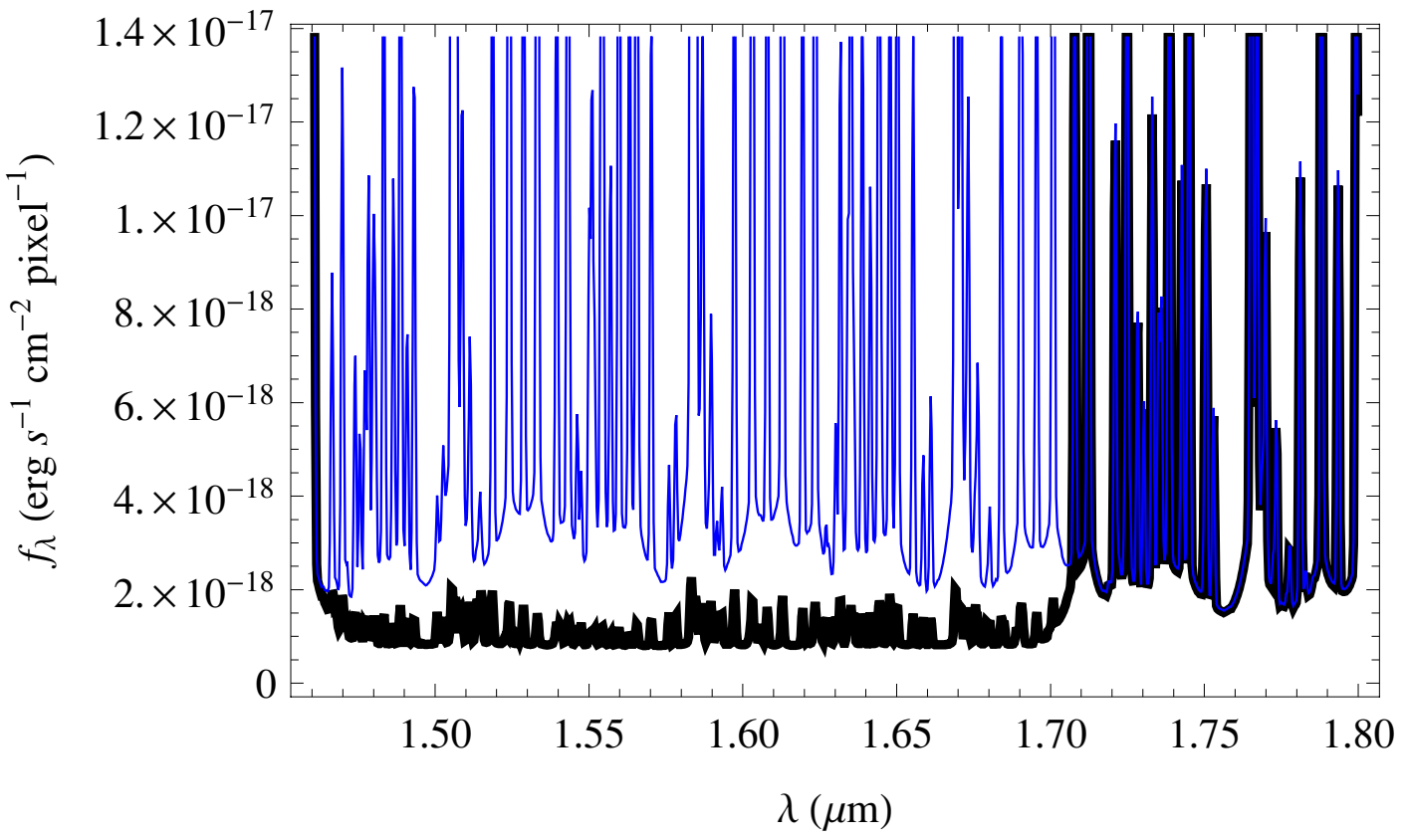

**Figure 5:** The 5σ limiting sensitivities in 1hr for GNOSIS (black) and for and identical system without fibre Bragg gratings (blue). There is a small decrease in sensitivity at the location of residual OH emission lines.

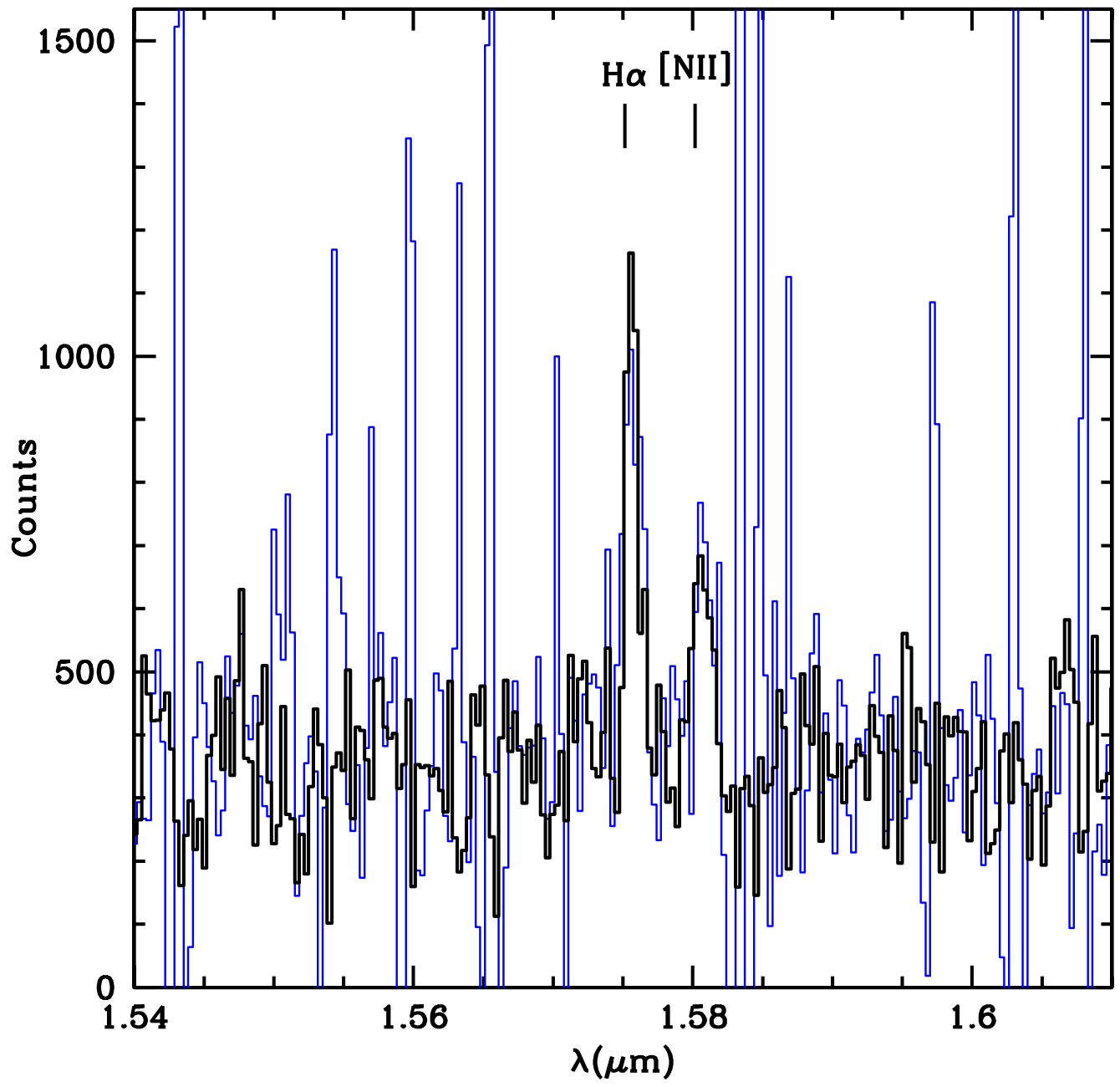

**Figure 6:** A simulated 6hr GNOSIS observation of an H=20 mag emission line galaxy at z=1.4 (black) shows that Hα is detectable. Also clearly visible is [NII]6584Å. The blue spectrum shows simulations for an identical system without fibre Bragg gratings, in which the Hα line is lost in the residual sky lines.

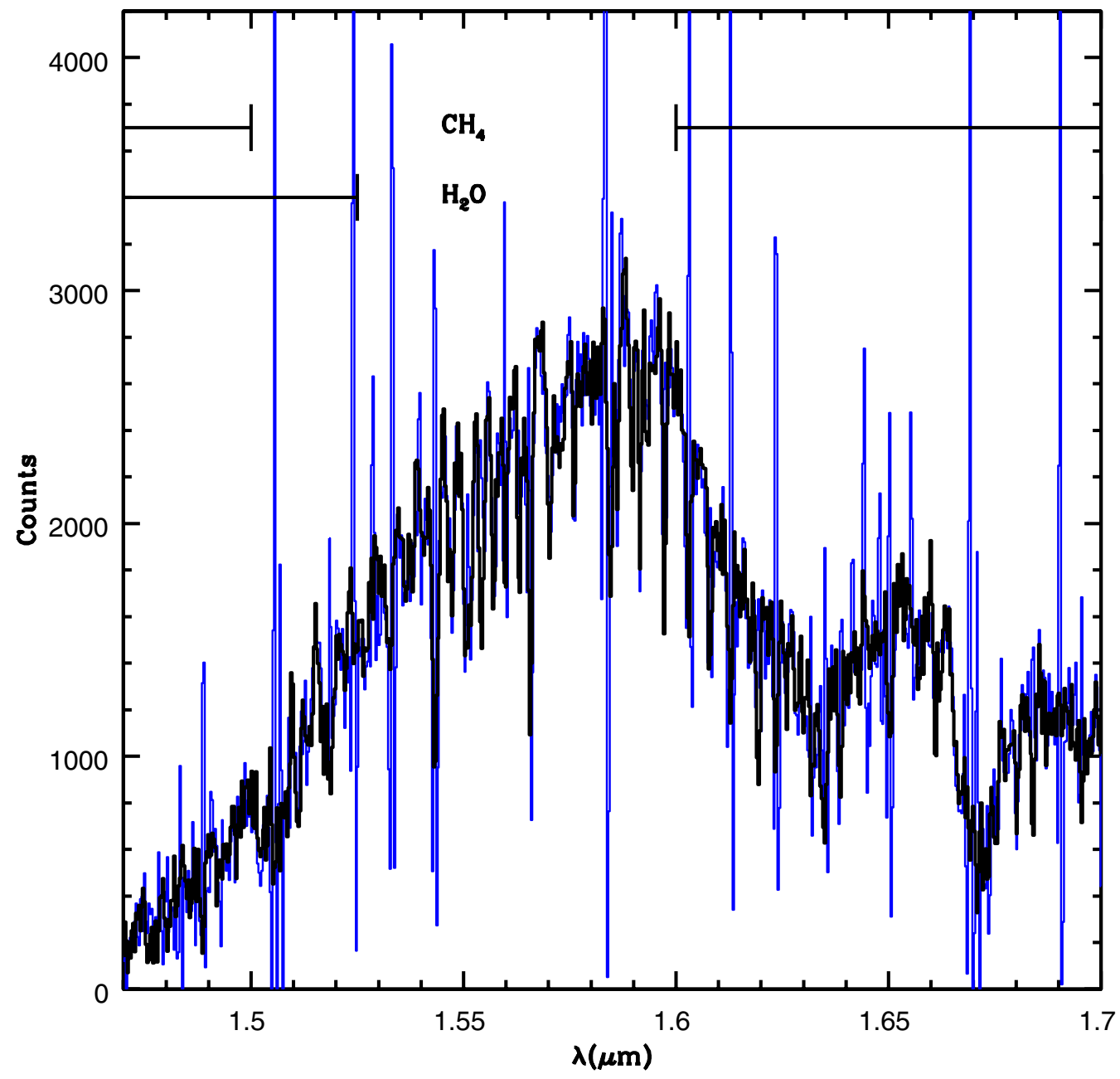

**Figure 7:** A simulated 6hr GNOSIS observation of a J=20 mag T5 dwarf star (black). The blue spectrum shows simulations for an identical system without fibre Bragg gratings.

# AusGO corner

Stuart Ryder (Australian Gemini Office, AAO)

## Semester 2010B

In this semester we received a total of 26 Gemini proposals, of which 16 were for time on Gemini North, 2 were for exchange time on Keck or Subaru, and 8 were for time on Gemini South. While total demand for Gemini North and exchange time in 2010B was down 20% on 2010A, demand for Gemini South time almost doubled. A 20% increase in the number of hours available to ATAC in 2010B (as an accumulated imbalance in partner share improves) partly accounts for the oversubscription factor for ATAC time on Gemini North (including Keck and Subaru exchange time) dropping from last semester's 3.59 to 2.74, while the oversubscription for Gemini South went up from 1.24 to a more healthy 1.70. At the ITAC meeting Australia was able to schedule 17 programs into Bands 1–3, 11 of which involved joint allocations with other Gemini partners.

For Magellan we received 6 proposals, well down on the record number of 11 received in 2010A, but Magellan oversubscription was still a healthy 2.43. In addition to the new optical wide-field mosaic camera Megacam, the near-IR imager and multi-object spectrograph MMIRS, and the Planet Finding Spectrograph PFS all commissioned during 2010A, two more instruments are expected to be commissioned on Magellan during 2010B. The Folded port InfraRed Echellette (FIRE) provides R~6000 cross-dispersed spectroscopy covering the entire near-IR, as well as a lower-resolution (R=2500 in *J*, R=1300 in *H*, R=900 in *K*), higher-throughput mode. FourStar is a wide-field (11'x11' at 0.16" per pixel) near-infrared imager using four HAWAII-2RG array detectors, equipped initially with *J*, *H*, *K*s, and a set of methane filters.

## Completion rates

In Semester 2009B, all ATAC queue programs with the exception of one Band 3 program were completed. Both of the classical/exchange programs were successful, and one PhD student program in the Poor Weather queue received 27 hours for which Australia is not charged. This excellent result is testament to the efforts of the Gemini summit observers and queue planners, as well as the flexibility of Australian Principal Investigators (PIs) in being prepared to relax their observing condition constraints (particularly in Band 3) in order to maximise their chances of obtaining useful data. Maintaining this high level of queue efficiency will be particularly challenging in the tight budgetary environment brought on by the UK's intention to withdraw from Gemini post-2012, but AusGO is working with the new Head of Science Operations Dr Andy Adamson to identify areas where we can take more responsibility, and reduce the burden on Gemini staff.

## Joint proposals database

One such recent AusGO initiative was a proposal to coordinate the technical assessment of joint proposals (those requesting allocations from more than one Gemini partner). Traditionally, joint proposals receive separate technical assessments from each National Gemini Office (NGO) involved (resulting in unnecessary duplication of effort) which are not necessarily consistent, leading to confusion for the PI. The proposed solution prepared by Deputy Gemini Scientist Dr Christopher Onken at RSAA is for AusGO to host a Joint Proposals Database (JPD) built with DokuWiki, that lists all joint proposals submitted each semester and designates the Primary and Secondary Technical Assessment NGOs. The Primary Assessment NGO is responsible for completing a standardised Technical Assessment which, together with the PI's response to any queries raised, is then shared with the other NGOs involved. Secondary Technical Assessments can also be contributed by other partners. Proportionate adjustments to the amount of time required from each partner can also be displayed. Semester 2010B was the first time the JPD was in operation, and all indications are that it has improved both the quality and consistency of technical assessment across the partnership, while reducing the overall workload for each NGO. While the operation of the JPD is effectively transparent to the PIs, we hope that they too will notice an overall improvement.

## Gemini school astronomy contest

Following the success of the International Year of Astronomy Gemini School Astronomy Contest last year, AusGO decided to run a similar contest again this year for Australian high schools to win one hour of time on Gemini South to image their favourite target in multiple filters with GMOS. The winning entry for 2010 came from the astronomy club at Sydney Girls High School, supervised by Jeff Stanger. They proposed to observe the merging galaxy system NGC 6872. For the second year running, one of the runners-up was submitted by the Forest Lake College astronomy club, who proposed to image another merging system, IC 5250. Indeed, merging galaxies appear to be "flavour of the month", since the other runner-up submitted by Benjamin Graham from Whitefriars College was to observe Arp 256, another pair of galaxies in the process of merging.

The Phase 2 program to observe the winner's target has now been prepared and sits in the Band 1 queue, so observations are likely to take place in the next couple of months. In the meantime all three groups will get to participate in a "Live from Gemini" interactive event from the Gemini control room, where a Gemini staff member will introduce them to the Gemini telescopes and observing process. Once the winning target has been observed, the data will be processed by AusGO staff and the multiple filters combined to yield a spectacular full-colour version to be unveiled at Sydney Girls High School later in the year. To follow the action of this event please visit **http://ausgo.aao.gov.au/contest/news.html**, or look for us on Facebook.

## AGUSS

Each year since 2006 AusGO has offered talented undergraduate students enrolled at an Australian university the opportunity to spend 10 weeks over summer working at the Gemini South observatory in La Serena, Chile on a research project with Gemini staff. The Australian Gemini Undergraduate Summer Studentship (AGUSS) program is kindly sponsored by Astronomy Australia Ltd (AAL). In addition to offering studentships at the Gemini Observatory, this year we also hope to be able to offer at least one studentship to work at the adjacent Las Campanas Observatory, which operates the twin 6.5m Magellan telescopes to which Australia has access. Applications for the 2010/11 program close on 31 August 2010, so if you know of any students in Australia who might be interested then please draw their attention to http://ausgo.aao.gov.au/aguss.html

### Changes in advisory committee structures

Prior to March 2010, the Australian Research Council appointed, and sought advice from members of the Australian Gemini Steering Committee (AGSC) on matters relating to Australian membership of the International Gemini Partnership; access to other 8m-class telescopes (including Magellan); and the operation of the Australian Gemini Office. In April 2010, a new advisory committee was established by AAL which merged the functions of AGSC and the Australian Giant Magellan Telescope Advisory Committee (AGMTAC). The Optical Telescopes Advisory Committee (OTAC) reports to the AAL Board on the operational performance of all Australia's national optical/infra-red telescopes, particularly those projects funded by AAL. The Terms of Reference and current membership of OTAC are available from http://astronomyaustralia.org.au/otac.html.

With the transition to a wholly Australian-funded AAO from 1 July 2010, it has been agreed that the new Australian Astronomical Observatory Users Committee (AAOUC) will take on the role of providing user feedback to the AAO and OTAC on the performance of the Australian Gemini Office, and issues to do with the quality of Australian data collected at Gemini and Magellan. This allows OTAC to focus on the "big picture" of Australia's ongoing usage of, and access to 8m-class telescopes. Australian users of Gemini and Magellan who have concerns about the level of support received from AusGO, Gemini Observatory staff, or the Magellan Fellows should bring these to the attention of the Australian Gemini Scientist (ausgo@aao.gov.au) in the first instance. They are also free however to bring these concerns (or to relay praise!) to their local AAOUC member or the AAOUC Chair. The Terms of Reference and current membership of the AAOUC are available from http://www.aao.gov.au/about/aaouc.html.

# Science meets parliament

Simon O'Toole (AAO)

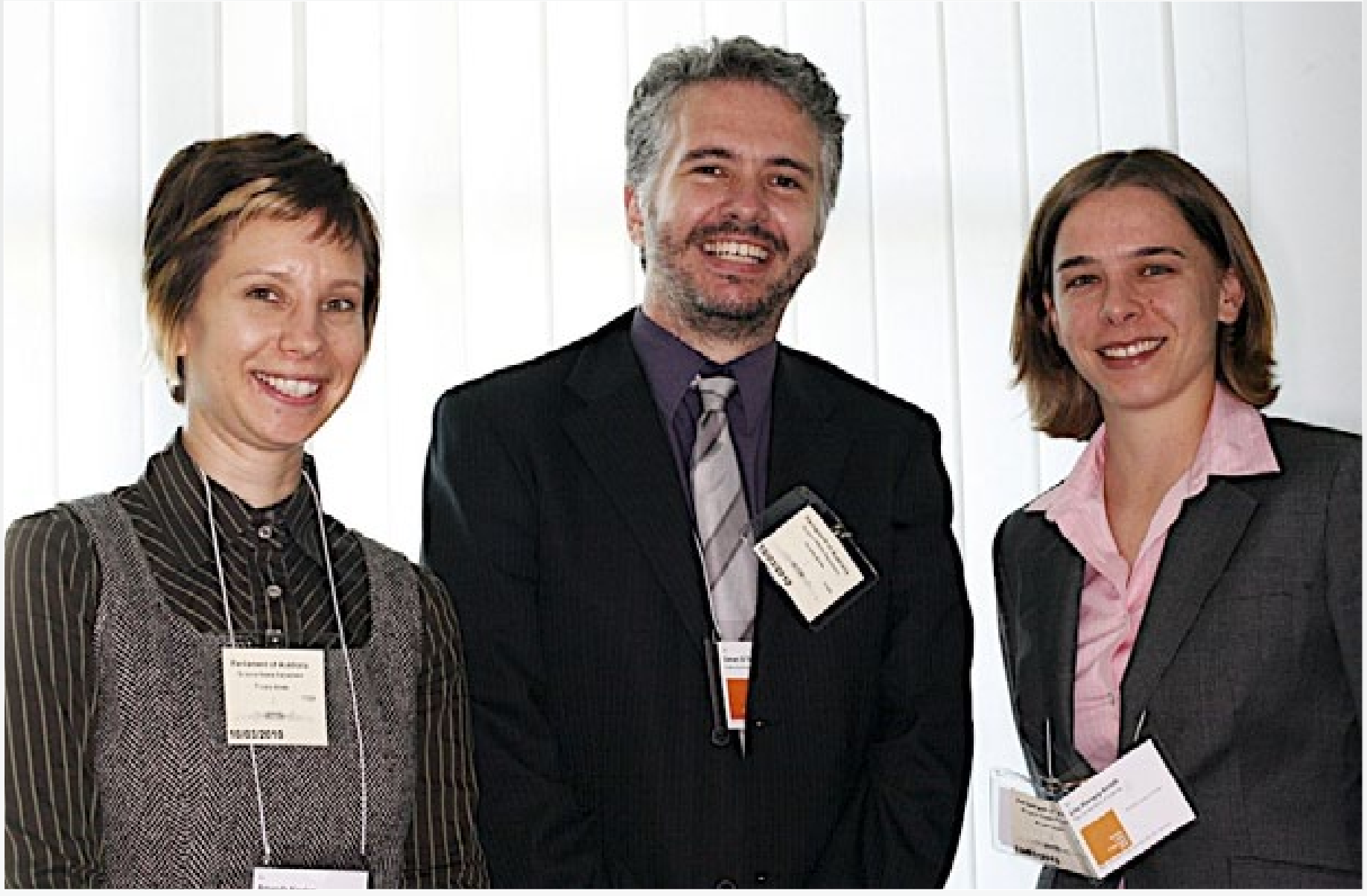

**Figure 1: (Image credit: Lorna Sim / FASTS):**
From left to right: Amanda Karakas (Australian National University) Lisa Harvey-Smith (University of Sydney), Simon O'Toole (AAO)

I was lucky enough to attend Science Meets Parliament down in Canberra in March this year, along with several other astronomers from around Australia. This event is organised by the Federation of Australian Scientific and Technological Societies (FASTS), and is designed to give scientists a chance to meet with Federal politicians to discuss science issues. More broadly, it also presents ideas for communicating science with the media and the general public. The concept of the meeting has always interested me, especially after hearing several reports from people who had attended in the past. As someone who deals regularly with the media, but mainly with science journalists, it was very interesting to hear perspectives on what makes science stories interesting to non-science journalists. Also very useful was the discussion on communicating your science to politicians and their advisors. Below are a few of my thoughts and impressions on more specific parts of the meeting.

The first day was made up of a series of seminars and forums on communicating science to a general audience. The first session included a welcome from FASTS president Cathy Foley and Executive Director Anna-Maria Arabia. We were given an exercise on the various levels of influence amongst politicians, the media, lobby groups, and the general public. This turned out to be very informative, as it showed how complex our society is, especially when you are trying to influence policy – whether it be science-based or otherwise.

Next up was a great session on conveying your message to the mass media, i.e. not just science journalists! At first I'll admit I though that having non-science journalists here was a bad thing, but I soon realised that if you want to reach a truly broad audience, then you have to speak to more than just New Scientist, etc. Obviously science journalism has an important role to play, but for scientists to reach out to people who don't go looking for science stories (most people), we need to know how.

In the afternoon, there were two sessions that over-lapped in some sense: the first was about conveying your message to politicians, while the second tried to develop the ability to be as clear and concise as possible. The highlight for me (perhaps of the whole meeting) was the presentation by the Prime Minister's speechwriter, Tim Dixon. He spoke very well, and gave us six key points to consider when dealing with politicians. Some of these had been covered earlier (e.g. finding a key message or hook), but others, such as considering issues that are on the public policy agenda and framing your message within the politician's agenda, not your own, opened my eyes somewhat. It was very good preparation for our actual meeting with politicians the next day.

In the second part of the afternoon, we practiced delivering our message in the time it takes for a sparkler to burn, or at least we got to think about it, as there wasn't enough time for everyone to have a go. This really highlighted to me that short sound-bites – often lambasted – are actually quite important in political communication, given the limited amount of time politicians have to discuss all the issues.

The conference dinner was held in the Great Hall of Parliament House, and along with lunch, morning and afternoon tea, gave us a great chance to meet other scientists, especially those in very different fields. I was very happy to meet some climate scientists in particular. Peter Yates of the Australian Science Media Centre gave the dinner address, and I must admit I wasn't that impressed by his speech overall, although I do appreciate his great enthusiasm and support for science.

Before meeting two parliamentarians, I attended a breakfast briefing dealing with open access to research articles in Australia, which was very informative. There was a lot of debate about the best model for open access among the non-astronomers in attendance. This slightly bemused the astronomers in the audience, as the model used in astronomy and astrophysics is well tried and tested. In my opinion it's possibly the best one out there!

After this, I was lucky enough to meet with two politicians (the coincidental visit by the Indonesian President caused plenty of disruption), one from each side of the political divide: a Liberal backbench MP and an ALP senator. Both seemed very interested in astronomy in Australia and one even expressed interest in visiting Mt Stromlo Observatory just outside of Canberra. Unfortunately, the Senate called for a division, so my second meeting was cut short, but I still managed to get my message across. Luckily I was very well prepared by the previous day's seminars! I was also very impressed with the general awareness and even enthusiasm of all the politicians I met (there was one at my table at dinner as well) with perhaps the largest astronomical project on the horizon, the Square Kilometre Array.

Finally, we were fortunate to be able to go to the National Press Club to hear author and science communicator Chris Mooney speak on the issues for scientists surrounding the communication of climate change information, along with what he called the US Republican's "war on science". It was a great way to cap off the two days for me, and showed how science can be hijacked, by what Richard Dennis of the Australia Institute had earlier called "articulate dumb loud people", and also what we can try to do about it.

I am very grateful to the CSIRO Staff Association and the CPSU for sponsoring me to attend Science Meets Parliament 2010. Thanks also to FASTS for organising the event. I certainly recommend it to anyone interested in science communication!

## Large observing programs on the AAT

### Request for proposals semester 11A

Heath Jones (AAO)

The AAO aims to provide opportunities for Australian and international astronomers to make effective use of the Anglo-Australian Telescope's unique capabilities to address major scientific questions through large observing programs. These Large Programs may use any instrument, or combination of instruments, on the AAT.

Until now the Large Program Requests for Proposals have been made in the B semester each year. However for at least the next two years (i.e. 11A-12B) the AAO anticipates making a call for Large Programs every semester. This will allow the Australian Time Allocation Committee (ATAC) to respond to various changes occurring over this period, including the end of current dark-time Large Programs in 10B, a prolonged AAOmega downtime during HERMES installation (currently planned for 12B), and the advent of HERMES itself (currently planned for 13A).The AAO is therefore issuing this Request for Proposals (RfP) for Large Programs using part or all of the period from semester 11A to 12B (i.e. from February 2011 to February 2013, inclusive). All proposals will be evaluated by ATAC. Ambitious projects are encouraged, and the AAO expects Large Programs to be awarded in total at least 25% of the available time on the AAT; in some past semesters Large Programs have been allocated almost 50% of the available time.

Proposers are encouraged to form broad collaborations across the Australian and international communities in support of their programs. The PIs for Large Programs will generally be expected to commit to the project as the main focus of their research over the program's duration. Proposers should also familiarise themselves with the method of time accounting on the AAT (**http://www.aao.gov.au/AAO/astro/newformula.html**). The web page **http://www.aao.gov.au/AAO/astro/apply/longterm.html** gives the status of current Large Programs.

Proposals for Large Programs should be submitted to ATAC by the standard proposal deadline of 15 September 2010.

Proposers should consult the Request for Proposals for more information about the process: **http://www.aao.gov.au/AAO/astro/Large_Programs_RfP_11A.pdf**

Anyone considering submitting a large program proposal should contact the AAO Director (**director@aao.gov.au**) in advance to discuss their plans.

# HERMES news

Gayandhi De Silva, Anthony Heng, Keith Shortridge (AAO)

**HERMES, is the High Efficiency and Resolution Multi-Element Spectrograph currently being built at the AAO. It will provide a unique and powerful new facility for multi-object astronomy.**

The HERMES system is built upon the AAT's existing two-degree field (2dF) optical fibre positioner (http://www.aao.gov.au/AAO/HERMES/2df_facility.html), which can collect the light from ~ 400 stars at a time. The positioner feeds a powerful new spectrograph (http://www.aao.gov.au/AAO/HERMES/spectrograph.html) which covers four optical bands simultaneously at a spectral resolution of ~28,000.

The HERMES project has been ongoing since early 2008. The Preliminary Design Review was completed in February 2010. Commissioning of the instrument is expected to occur early 2013 with the start of science operations to follow soon after. Below we highlight the progress of the project since the last AAO Newsletter article (2010, 117, 7)

## Science update

Gayandhi de Silva was appointed as HERMES Project Scientist in March this year. The new HERMES webpages are online at http://www.aao.gov.au/HERMES. They contain an up-to-date description of the design and capabilities, as well as information about the project history, the science requirements, and the people involved.

The primary science driver for HERMES is the 'Galactic Archaeology' (GA; http://www.aao.gov.au/AAO/HERMES/science_case_req.html) Survey, which aims to reconstruct the history of our Galaxy's formation from precise multi-element abundances of 1 million stars derived from HERMES spectra, utilizing the chemical tagging technique (Freeman & Bland-Hawthorn, 2002; De Silva et al., 2007,)

The wavelength channels have been designed to meet the requirements for the GA survey as given in the table below. These wavelength regions were chosen to be the optimum bands in which a large range of suitable elemental lines are observed. The target elements include those from the several nucleosynthesis processes such as light elements, alpha-elements, odd-Z elements, iron-peak elements, light and heavy s-process elements and r-process elements.

| Channel | Wavelength range (Å) |
|---|---|
| Blue | 4708 – 4893 |
| Green | 5649 – 5873 |
| Red | 6481 – 6739 |
| IR | 7590 – 7890 |

HERMES is designed to suit many other high resolution multi-object spectroscopic studies, beyond the GA survey. The individual channels are therefore designed to handle much wider wavelength bands. The usable bands for each channel and those defined for the GA survey are shown in the figure below.

## Science with HERMES workshop

The Science with HERMES workshop will be held on 28-29th September 2010, in the ATNF lecture theatre in Epping. The purpose of the workshop is to present the design and capabilities of the HERMES instrument to the astronomy community and to explore the scientific uses of HERMES, including the planned GA Survey and its synergies with the SkyMapper survey and the GAIA mission. This will also be an opportunity to generate ideas for future upgrades or additional features required to meet the community's science goals. We look forward to hearing about the many scientific programs utilising the unique capabilities of HERMES.

If you would like to attend, registration is now open at http://www.aao.gov.au/HERMES/ScienceWorkshop/HermesWorkshop.html. There is no registration fee but for catering purposes please register before the 17th September 2010.

## Project update

The financial year 2009/2010 has been a very busy year for the HERMES project. HERMES has continued to progress since the Configuration Design Review. Performance specifications for the collimator, camera optics, fold mirrors, slit, test cryostats, prototype gratings, beam splitters, CCD detectors, the spectrograph base and data reduction have been developed and reviewed. The HERMES science requirements have been signed off. A mini design review for the test cryostat was carried out internally. The following analyses and studies for HERMES were carried out:

- Ghosting and image quality for the instrument,
- Floor vibration analysis for the Coude West room (the HERMES instrument room),
- Ghosting and image quality,
- Dual fibre cables vs fibre connector tests,

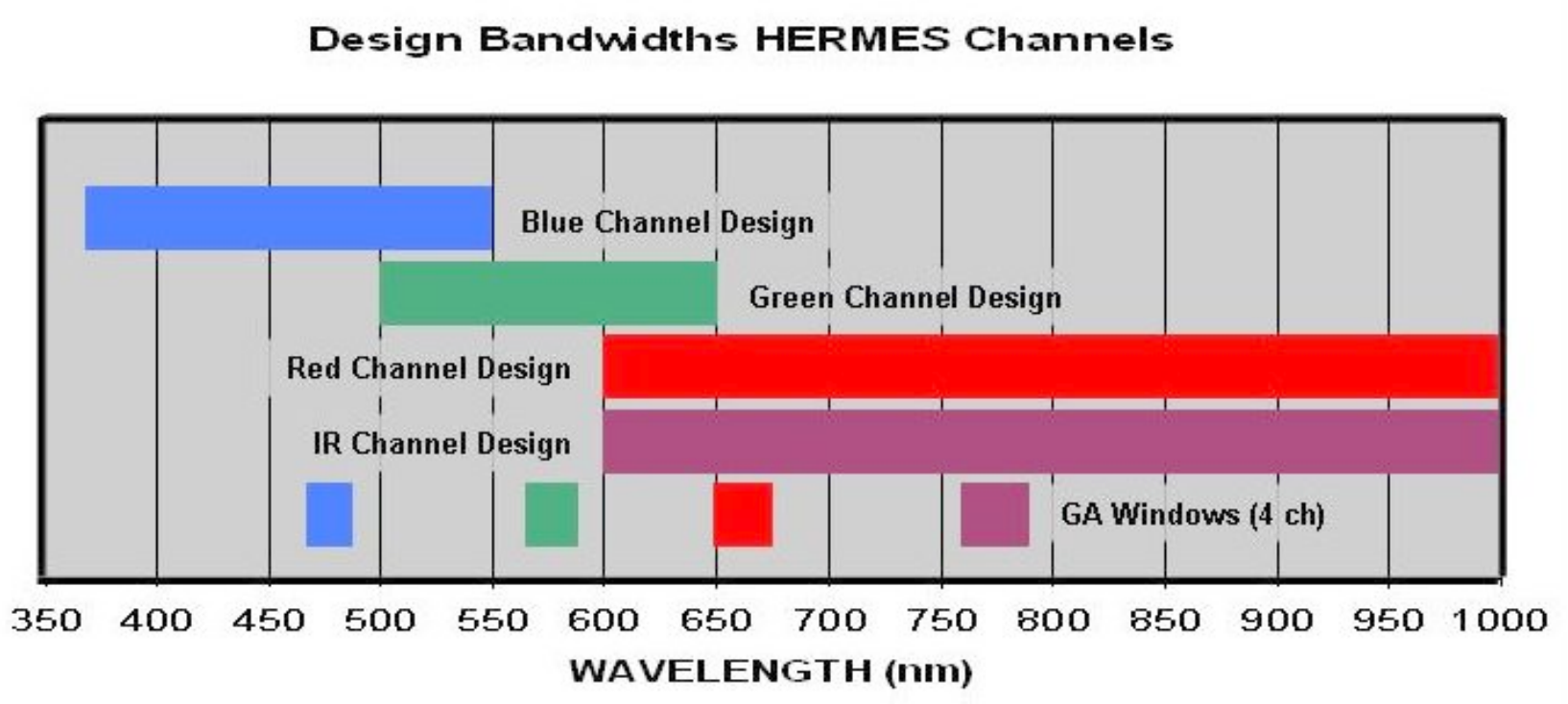

**Figure 1:** Wavelength bands for the HERMES channels and GA spectral windows (Barden et al, 2010).

- Spectrograph base vs spectrograph frame.
- Thermal and tolerance analysis for collimator and camera optics.

Data simulation for the whole instrument was developed and implemented (Goodwin et al, 2010). A product breakdown structure for HERMES was developed. The Assembly, Integration and Testing (AIT) plan was developed. A project safety management plan was developed. HERMES project personnel were trained in safe working practices. Planning for the HERMES staging area was developed and approved. The final instrument location for HERMES was also planned and developed. The full scale model of HERMES is also developed by the AAT team. The model will be used to confirm the position of HERMES in the Coude West room.

The location of the staging area for the HERMES instrument has been confirmed and will utilise part of the existing workshop spaces of AAO in Epping. The required size and the proposed room position have been confirmed. We envisage that the staging area for HERMES will be completely established at the end of year 2010 in Epping.

All parts for the test cryostat have been received. Expressions of interest and formal quotations for the following items have been ordered:

- Collimator,
- Camera optics,
- Cryostats and Cryocoolers
- Slit lens,
- Shutters,
- Beam splitters,
- Fold mirrors,
- Detectors,
- Fibre cable,
- Instrument cryostats and cryocoolers.

The clean room for the staging area, and the crane will follow the open tender process of the Department of Innovation, Industry Science and Research. The room design for the HERMES instrument has commenced. A permanent partition wall is to be erected in July 2010 in the coude west room. This will separate the AAOmega and HERMES instruments.

A management review was held in October 2010 to confirm the health of the project. Several months later, a preliminary design review was carried out in February 2010.

HERMES received positive feedback from the external reviewers, with some minor shortcomings pointed out. The installation of two brand new fibre cables (800 each) has been chosen over the fibre connector option. The fibre cable will be installed and tested. AAO also received confirmation of funding from AAL for the 4th channel (the Infra Red channel).

The challenge ahead for HERMES is to test the slitlet and grating prototypes. Two prototype gratings have been received. These tests are scheduled to be completed by the end of December 2010. The adhesives required for the fibre cable, slit and prism are still in the process of being determined, particularly for the v-groove plate in the retractor, as it has the specific requirement that it be easy to remove. The adhesive for the button to ferrule joint will be Vitralit 4280 however, and Epotek 302-1 is the leading candidate for the slit glue and the ferrule to fibre glue.

## Research thesis excellence award for new AAO research fellow

Max Spolaor, who joined in June the AAO as a new Research Fellow, has been awarded the Research Thesis Excellence Award 2010 for best PhD thesis in the Faculty of ICT at Swinburne University of Technology. The thesis work of Max, titled "Radial Gradients in Elliptical Galaxies", features the discovery (Spolaor et al. 2009, ApJL, 691, 138) of a new relationship between galaxy mass and metallicity gradients of their stellar populations, which are interpreted as fossil record of a galaxy merging history.

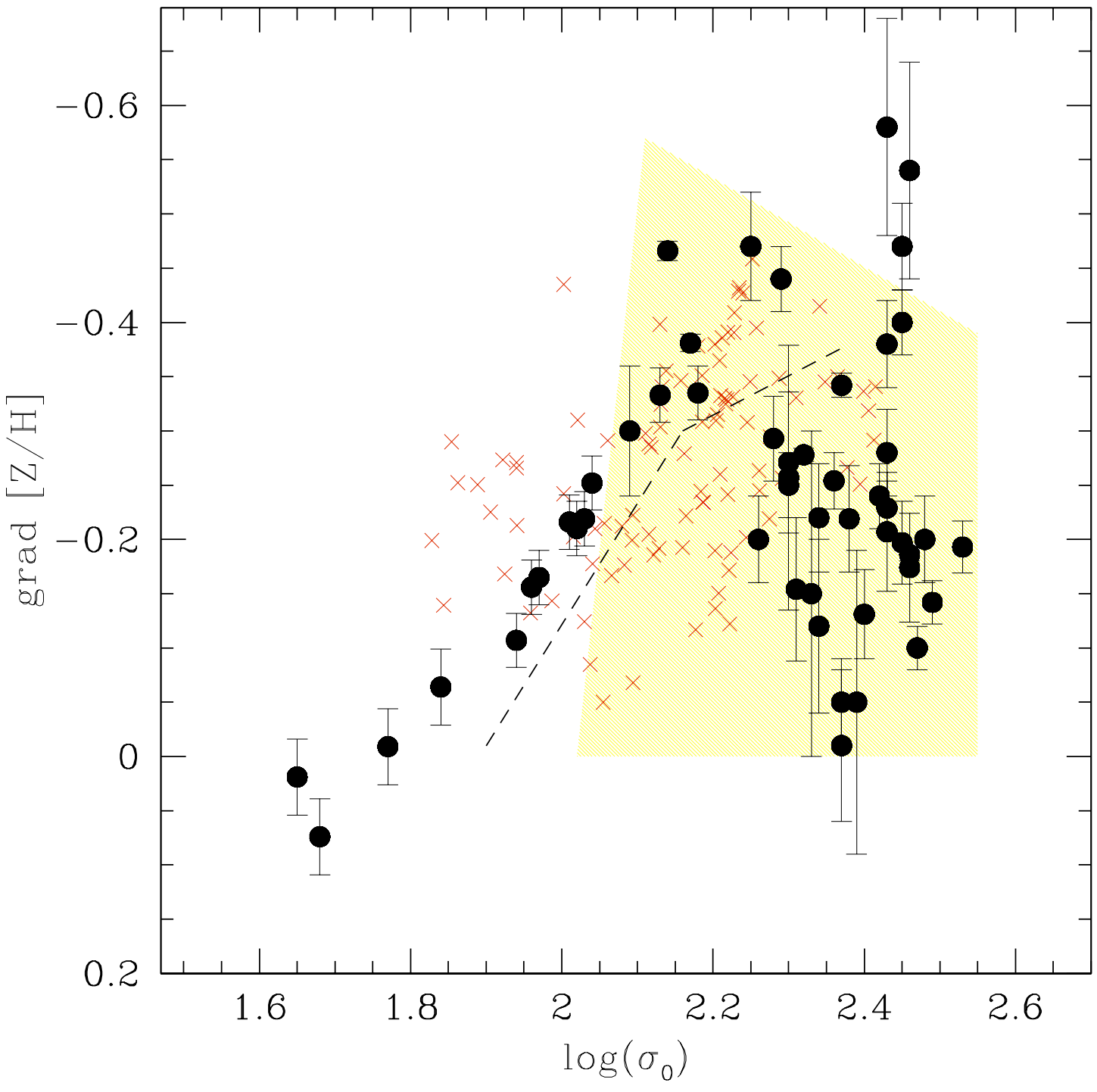

Metallicity gradients as a function of galaxy central velocity dispersion (proxy of galaxy mass). Black points indicate the observations reported in Spolaor et al. (2010, in press). The predictions of an early star-forming collapse of Kawata & Gibson (2003) are shown as black dashed lines. The region occupied by the remnants of major mergers between gas-rich disc galaxies, as simulated by Hopkins et al. (2009), is shown by the light yellow shading. The red crosses indicate merger remnants simulated by Kobayashi (2004).

# Celebrating the AAO: past, present, and future

Sarah Brough (AAO)

A group of astronomers, old and young, international and local, gathered in Coonabarabran's Shire Hall from 21 - 25 June this year to mark the transition between being a UK/Australian collaboration and becoming wholly Australian funded. We also marked the 36th year of the Anglo-Australian Telescope and celebrated the past, present and future of what is now the Australian Astronomical Observatory.

As one of the organizers of this event I'd like to share my rather tongue-in-cheek perspective of some of the highlights. For those of you wanting more serious content, please see the conference webpage (http://www.aao.gov.au/AAO2010/) where talks will be put up in pdf format. We also plan to provide a record in the form of a book, capturing the history and current state of the AAO for posterity. Those who gave presentations at the Symposium should download the template from the website for inclusion in these proceedings. The book will be published by our new governing body, the Australian Department of Innovation, Industry, Science and Research (DIISR), and will be distributed free to all registered participants, as well as to major libraries and institutions.

We started well on the evening of Sunday 20 June with a few drinks and nibbles and catching up with old friends at the Acacia Lodge in Coonabarabran, remarkably having not lost anyone being transported up by bus from Sydney. First thing Monday morning Local Aboriginal Elder, Bill Robinson, gave the Welcome to Country speech and accepted a plaque recognizing the original owners of the land to be placed in the AAT's visitor's gallery. Martin Gascoigne then opened the meeting with a moving memorial to his father, Ben Gascoigne (Astronomical Advisor to the AAT Project Office 1967–1974 and AAT Commissioning Astronomer 1974–1975), who recently passed away. Martin accepted the plaque that will mark the "Gascoigne Room", the recently refurbished AAT Meeting Room on the first floor of the telescope building.

The conference then formally began with everyone happily ensconced in the Shire Hall with wireless access and hot drinks (two things no astronomer should be deprived of). Former AAT Board Chairman Bob Frater indicated the need to take into account all the various skills utilized in the observatory and its research, rather than just publication statistics, in future hiring decisions. The AAO's former Chief Executive Officer, Joan Wilcox, wittily reminded the gathered astronomers that she saw them mostly as children, with a few grown-ups scattered through to keep things running smoothly (I will cross my fingers in aspiration). That evening several of us nearly ended up in Narrabri as our bus driver missed the turning for dinner in the dark and continued heading North. However, we finally made it back and up to a rather grand property and winery for a very interesting 7-course degustation dinner that was to set the standard for meals throughout the conference. However, it was a late night.

On Tuesday, Keith Shortridge compared an analogue phone to an iphone as a simile for the old and new telescope control systems: they have the same basic function, with a little difference in memory and extra abilities. John Storey shared his entries and mischievous poems in the Infrared Photometer/Spectrometer (IRPS) comments book. The book seems to have been surreptitiously hidden, probably as a result of John's entries. That evening we rushed through a multi-course Chinese dinner in order to be at the Shire Hall in time for David Malin's public talk. It allowed some of us more recent arrivals to understand a little bit more about astronomy with photographic plates, and

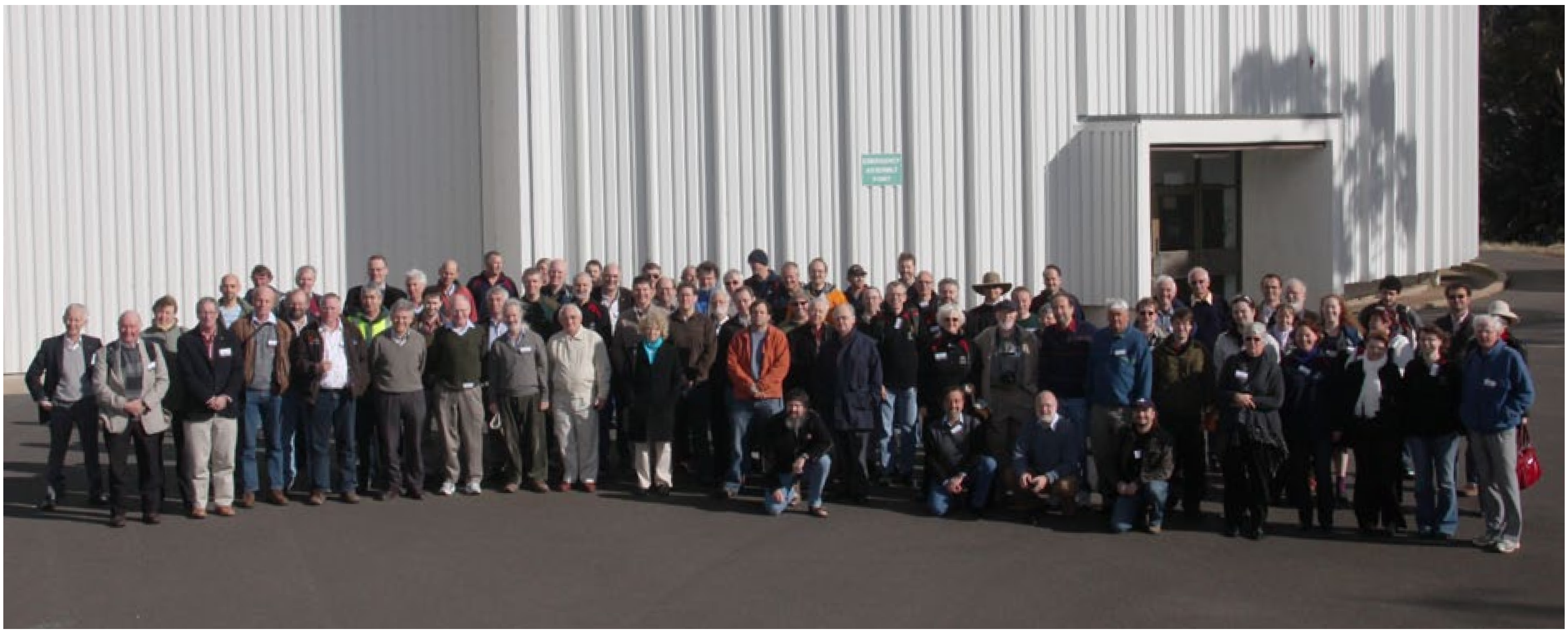

**Figure1:** Group photo of conference attendees outside AAT. (Image credit: Steve Lee)

where he actually sat in the telescope to acquire those images – and some of the more delicate personal moments that occurred if the telescope broke while he was still up there.

Wednesday morning started with Michael Burton being corrected by Jeremy Mould in his belief that he'd been involved in the first infrared camera at the AAO. Peter Gray noted his wish to jump off the AAT dome, with a paraglider! An interesting idea, but maybe no longer possible (or advisable) in this Occupational Health and Safety-conscious world. Russell Cannon morphed into Keith Taylor and then Damien Jones to tell us that Melbourne is at the end of the known Universe and about the early days of building the 2dF instrument. Sam Barden walked us through the dark days of the AAO's instrumentation group when WFMOS funding was pulled and into the current lighter days with many exciting projects funded and planned over the next few years.

On the Wednesday afternoon everyone was taken up to the AAT for the group photo (Figure 1), a tour of the AAT itself, the UK Schmidt Telescope, the new Skymapper telescope and the ANU 2.3m. The tour certainly shocked some of the old Directors (Figure 2) when they saw what we'd done to the control room with it's new lounge and coffee area.

That evening the conference dinner was prefaced by a spectacular 3-screen sound and light extravaganza from David Malin and Steve Lee, showcasing some astronomical and historical photographs. The dinner itself was MC'd by Julie McCrossin who pulled some likely faces out of the crowd to share their (mostly amusing) AAO experiences (Figure 3).

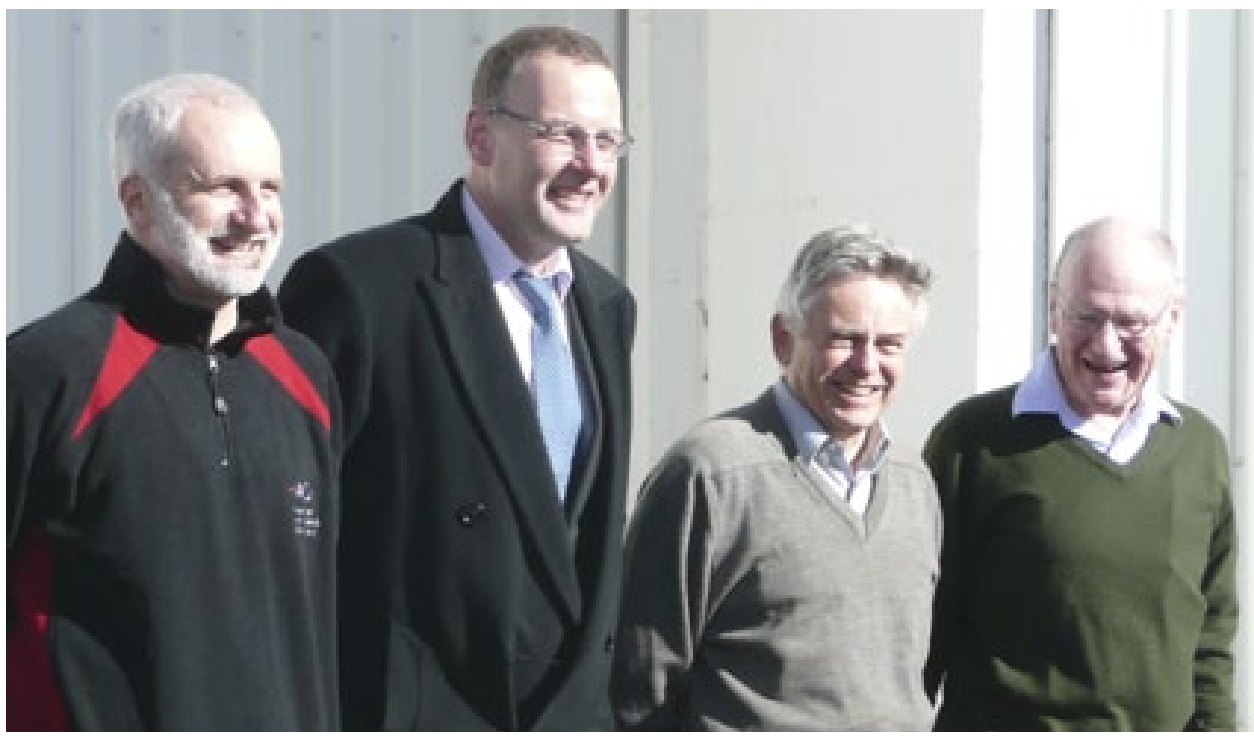

**Figure 2:** The current Director with three out of four of his predecessors

One of the surprises of this meeting was discovering how many people met their partners through the AAO in one way or another. Maybe we should start a dating agency, perhaps Julian Dating?

Malcolm Longair suggested betting on the horses as a means of running an observatory,. As Malcolm's talk ended on the Friday morning, so did Kevin Rudd's time as Australian Prime Minister as the first female Australian Prime Minister, Julia Gillard, came into power. Karl Glazebrook kindly if reluctantly performed the nod and shuffle two-step for us all, which will certainly make future nod and shuffle observations pass with a smile. Will Saunders morphed seamlessly into Rob Sharp for the AAOmega presentation. Chris Tinney taught us the difference between astronomers and astrophysicists ... anyone considering themselves an astrophysicist may not wish to know the answer. Pat Roche reminded us of benefits to AAT visits outside of the telescope itself, bringing back memories of staying in the lodge (still a reality for some of us – although the legendary monkey-bum sandwiches do seem to be a thing of the past) and of walks in the National Park.

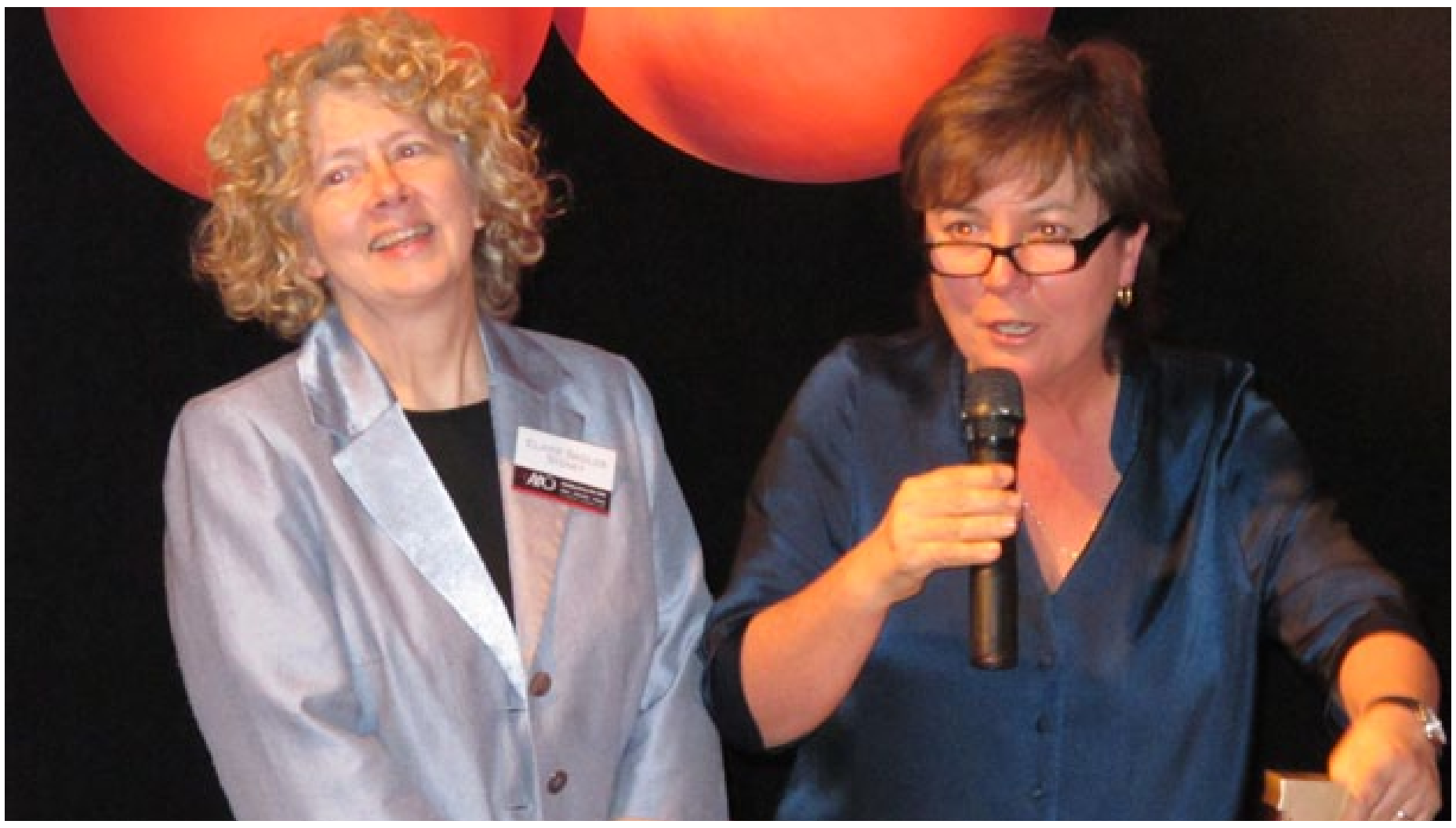

**Figure 3:** Elaine Sadler being interviewed by Julie McCrossin. (Image Credit: David Malin)

On the final day we had significant media coverage with a team from ABC Western Plains radio doing an all-morning outside broadcast and WIN and Prime coming by to film and interview some of us – to the noise of heavy trucks. With a huge sigh of relief, a massive organizational effort came together successfully to deliver us several VIPs from Canberra, despite the previous days political upheaval: DIISR's Parliamentary Secretary, Richard Marles MP, Department Secretary Patricia Kelly (DIISR), and the Deputy High Commissioner to the UK, Jolyon Welsh. The attendees and VIPs all headed up to the telescope for a grand event to mark the handover between Australia and the UK. Everyone was ushered onto the fourth floor of the AAT where there were chairs and, unexpectedly, hundreds of helium-filled multicoloured balloons attached to them, giving the place a remarkably festive appearance. Patrick Healy from Questacon MC'd and directed a rather spectacular show, exploding several balloons filled with hydrogen. Happily these were not tethered to chairs. As the speeches ended we were invited to untie the balloons and let them loose into the dome as a 10m-high Australian Astronomical Observatory banner was unveiled (Front Page Image). The balloons remained trapped inside the dome and the face of that night's observer, Paul Butler, was said to be a picture!

Everyone returned to the Shire Hall after lunch to begin the process of winding up the meeting. Not before rushing through a multi-course Malaysian dinner in order to make Malcolm Longair's public talk in which Malcolm's enthusiasm for astronomy was spread far and wide.

The week was filled with activity and occasionally a good deal of hilarity, but the new AAO was launched in style with much goodwill all round.

Obviously many thanks are due in organizing such an event, mine go out to the rest of the Organizing Committee (Matthew Colless, Andrew Hopkins, Russell Cannon, Steve Lee, David Malin, Chris Springob, Helen Sim, and Stuart Ryder), Marnie Ogg (our supremely professional conference organizer), Aileen Bell (team leader - regional development) and many other people in Coonabarabran for making us all very welcome.

# Letter from Coonabarabran

Rhonda Martin (AAO)

After a warm winter so far we have been thrown into a time of unbelievable frosts when the air almost crackles with the cold. The kangaroos have taken to wearing mufflers and sleeping in clumps to keep warm, which at least keeps them off the road. My horses, at over 30 years of age, tend to feel that at their venerable age they should come inside and sleep in front of the fire! I was looking at trees yesterday, over 40' high, on which every leaf was a block of ice and the water pipes of staff members froze. This is unusual for Australia. Only two days ago the daytime temperature of Siding Spring Mountain struggled up to 4 degrees Celsius and the wind was off the snow far to the south. Ouch! The mountain road has been sheathed in ice which even in the afternoon had not completely melted.

The cooling weather did not effect the commissioning of CYCLOPS, however. Chris Tinney, once part of the AAO and always welcome since his departure, worked on this new fibre feed from Cassegrain to UCLES. Yet another weapon in the AAO's armory of super instruments.

To celebrate the Past, Present and Future of the AAO a symposium was held in Coonabarabran where astronomers from all over the world got together to celebrate the excellent work done in the past and to look forward to a productive future under the auspices of the Australian Government as the Australian Astronomical Observatory. By all accounts the event was a roaring success thanks to the hard work done by a number of people from both Epping and Coonabarabran. It was certainly good to meet people again not seen for many years.

For those of you who may remember Tom Cragg, who worked for the AAO for a number of years and has been retired now for quite some time, we wish to tell you that Tom's health has not been good for a while now. He has suffered a couple of small strokes which are keeping him in hospital but have in no way impaired the sharpness of his mind. Tom, I remember, studied sun spots every day and had almost a lifetime of spot records. He was very fond of variable stars as well.

The Letter from Coonabarabran will in future be written by Katrina Harley, the new Administration Officer of the AAT, as I have now retired. I would just like to say that I have had a wonderful twenty years at the AAT and have met some great people. I have learnt a great deal, was inspired to build my own telescope, and being in the forefront of a fascinating science, albeit only from an administration side, was a heady ride. My thanks to you all and I am sure that you will extend to Katrina the same courtesy and fun that you did to me.

# Epping news

Sarah Brough (AAO)

As we move into our new position as Public Servants and part of the Department of Innovation, Industry, Science and Research (DIISR) we welcome many new people and sadly farewell others.

Dr Gayandhi De Silva joined the AAO as a Research Astronomer in March 2010. Gayandi was born in Sri Lanka and later her family migrated to Melbourne in the early 90's. Gayandi completed her undergraduate studies at Monash University, and her PhD at Mt Stromlo Observatory, ANU. The aim of her thesis was to test the viability of the 'Chemical tagging' technique, which aims to reconstruct dispersed groups stars in our Galaxy by looking for common chemical signatures. This involved studying high-resolution stellar spectra in order to measure precise abundances of a large range of elements, and Gayandi continues to work in this field to date.

In 2006, immediately after submitting her thesis, Gayandi relocated to Chile to take up an ESO fellowship. There she supported the science operations at Paranal observatory. In 2008 she transferred my fellowship to ESO headquarters in Germany, working for the user support department. Now at the AAO Gayandi is the Project Scientist for the upcoming HERMES instrument (see article on HERMES). She is very happy to be back in Australia and looks forward to the exciting prospects that HERMES will offer.

Dr Max Spolaor started his AAO Research Fellow position at the end of June 2010, on the opening day of the AAO Symposium "Celebrating the AAO: Past, Present, and Future". Max obtained a PhD in Astrophysics in February 2010 from the Swinburne University of Technology in Melbourne with a thesis on stellar populations and chemo-dynamical properties at large galactocentric radii of early-type galaxies. Max's interest in Astronomy began in high school when he joined the CCAF cultural club for Astronomy, which led him to study Physics at the University of Trieste in Italy, where he graduated in 2006. Max's scientific research at the AAO will focus on extending the understanding of star formation, chemical enrichment, and merging history of galaxies at intermediate and high-redshift by deriving ages and chemical abundance ratios of their stellar populations. I have a strong interest in instrumentation which I will be able to foster at the AAO by supporting 2dF and the AAOmega spectrograph as instrument scientist and collaborating in the many upcoming new instrument designs. I am very much looking forward to the research opportunities that working at the AAO offers."

Dr Chris Lidman was awarded a prestigious four year 'Future Fellowship' from the ARC to work at the AAO. Before moving back to Australia and to his new position at the AAO, Chris spent 14 years working as a support astronomer at the European Southern Observatory (ESO). The first 4 of those years were spent at La Silla. The remainder were spent at Paranal. Chris saw many changes over those 14 years, the most significant of which, from the point of view of a support astronomer, was the change from direct support of visiting astronomers to a much more remote form of support based on service observing.

Another significant change over that time has been the power and complexity of instruments. At ESO, Chris specialised in near-IR instruments and in adaptive optics (AO). At one time or another, he supported nearly all of the near-IR instruments and AO systems at ESO. He started with IRAC1 and IRAC2 on the ESO 2.2m, then moved onto SOFI on the NTT, then onto ISAAC on Antu, and finally onto NOAS-CONICA on Yepun. The change in the power and complexity of these instruments is extraordinary. One only has to compare images of the Galactic Centre taken with IRAC2 with those taken with NAOS-CONICA.

Chris is now back in Australia and working at the AAO. While he will spend most of his time pursuing high redshift galaxy clusters and distant supernova, Chris also looks forward to working with the excellent staff here at the AAO. No doubt, the first instruments on the next generation of extremely large telescopes will be more complex than the current generation of instruments on the 8m telescopes. We will need to learn new ways of how to build, operate and manage them.

Two of our Distinguished Visitors for 2010, Prof. David Koo and Prof. Kim-Vy Tran, have both arrived in Australia. David is from Santa-Cruz, and will be spending the next few weeks working with Andrew Hopkins on a complementary analysis of David's high-redshift galaxy survey (AEGIS) and Andrew's low-redshift one (GAMA). Vy is from Texas A&M and will be working with Sarah Brough on integral field spectroscopy of Brightest Cluster Galaxies. In the next few weeks we will also be joined by Prof. Andy Connolly from the University of Washington. Andy will be working with Andrew Hopkins on large galaxy survey strategies, learning from Andy's experience with the SDSS and LSST. Prof. Richard Ellis will visit from Caltech in September 2010. Richard will engage with Andrew Hopkins in a program of identifying which multi-wavelength photometric and spectroscopic observables most strongly constrain the shape of the stellar initial mass function for star clusters, galaxies, and galaxy populations.

Jon Lawrence has been acting in the role of Head of Instrument Science since Roger Haynes departure in Dec 2009. However, his position also had a 50% commitment to working within the Department of Physics and Astronomy at Macquarie University. Since July, he is now employed by the AAO in this role full-time. Jon is looking forward to leading the R&D effort of the expanding AAO Instrument Science Group on a range of new and exciting astrophotonics projects, and working closely with our collaborators at the University of Sydney, Macquarie University, and the Astrophysics Institute in Potsdam.

Luke Gers joined us in July to start his job as our new optical engineer within the instrumentation group. His primary priority will be ensuring that the optics for HERMES are up to spec. He will also provide guidance and consulting for other projects involving optical components. Luke spent the previous ~4 years in Maui designing and building telescopes for use in Space Situational Awareness (SSA). He has also been fortunate to work on the Keck Interferometer for JPL building up the metrology system as a co-op student. Luke obtained his BSE and MS in Optical Science and Engineering from the University of Arizona College of Optics in 2004 and 2006. His work and research has been primarily in optical design, testing, alignment, and stray light analysis. On the lighter side Luke loves any sport activity that has to do with the water and is addicted to surfing.

Scott Case has joined the Instrumentation Group as our new Fibre Optics Engineer. Scott comes to us from Finisar where they fabricate fibre optic wavelength division multiplexers for the telecom industry.

A former summer student, Talini Jayawardena, has rejoined the Instrumentation Group for the next 4 months to assist in HERMES fiber assemblies, grating evaluation, and MANIFEST/Starbug efforts.

Jamie Gilbert has returned to the AAO after his summer studentship to continue his work on Starbug fibre positioners. Starbugs are an integral part of MANIFEST, a facility instrument concept for the Giant Magellan Telescope. With the design study for MANIFEST approaching, Jamie is aiming to demonstrate the power of Starbugs by producing a fully functional laboratory model, with multiple bugs walking underneath a curved glass field plate.

Murat Gedik has started in July as a mechanical designer to work on HERMES. He will work on the structural frame of HERMES for at least the next 6 months.

Sadly people have also decided to leave the AAO. We wish him all the best in their new endeavours!

Our head of instrumentation for the past 7 years, Sam Barden, has announced his impending departure from the AAO. After long consideration, Sam has accepted employment with the National Solar Observatory near Alamogordo, New Mexico to be program manager of the adaptive optics and wavefront sensors for their new telescope project (ATST http://atst.nso.edu/). Sam stated that the AAO has been a great place to work and that he believes he is leaving it on the upswing with HERMES and several other projects underway. He has appreciated all the hard work and dedication of his staff and the rest of the observatory. Sam will be around until late July to help with the transition to DIISR.

Rhonda Martin has decided to retire from the AAO after 20 years of keeping everyone at the AAO in Coonabarabran (visitors included) in order. We will miss her calling us 'chicken', keeping us in order and her 'Letter from Coonabarabran'. Katrina Harley has accepted the position of site administrator and started on the 21st June.

Stephen Marsden has accepted a two-year lecturer position at James Cook University in Townsville, starting in November. Before then he is going to Germany to undertake a 3-month fellowship at the Kiepenheuer-Institut für Sonnenphysik (KIS) in Freiburg, to work with Professor Svetlana Berdyugina on a Doppler imaging project.

Ian Saunders has decided to move on from the AAO to be nearer to family and friends located in Perth. In his short time here at the AAO, we have had the opportunity to benefit from his experiences/expertise in both project management and in systems engineering. Among his numerous activities, his major contributions were the development of the MANIFEST proposal and current Starbugs effort, enhancement of a systems perspective for HERMES, and development of a space-frame concept for mounting HERMES.

Our Personnel Officer Suzanne Tritton has decided to move on having seen us safely into our new position in DIISR. Rajni Prasad commenced as the AAO's new Personnel Officer in July for an initial 6 month period.

Sadly some former AAO staff have recently passed away. An obituary for Ben Gascoigne, follows after this article. Walter Stibbs - an astronomer who spent some time at the AAO a number of years ago – also passed away in April. Walter was 91 years of age. Walter's academic career began at the University of Sydney. He then became a member of the staff at Mt Stromlo shortly after the commencement of the second world war and carried out innovative work on optical munitions. He also lectured on Defence Science at the University of New England. Following the war he returned to Mt Stromlo and collaborated with Richard van der Riet Woolley on a classic text on stellar atmospheres: The Outer Layers of a Star. Also, during this period, Walter published seminal papers on the variability of magnetic rotating stars. Walter travelled to the UK in 1951, obtained a doctorate from Oxford in 1954 and was subsequently appointed, in 1959, Napier Professor of Astronomy at the University of St. Andrews -a position which he held for 30 years. Following his retirement, Walter returned to the ANU and held a Visiting Fellow appointment in the School of Mathematical Sciences and at Mt Stromlo. Amongst other activities, Walter busied himself reorganising the Mathematics Library and resurrecting the Tour de Stromlo bicycle race which had been instituted during his early career at Mt Stromlo.

## AAO PhD scholarship scheme recipients

We welcome 3 new students into the AAO PhD scholarship scheme and look forward to spending time with them here in the next 3 years.
They join Caroline Foster, Minnie Mao and Christina Magoulas as our ongoing PhD Scholarship

Florian Beutler, UWA (AAO supervisors: Jones)

PhD title: "Analysis of large scale structure in the universe using the 6dFG survey and ASKAP simulations"

Madusha Gunawardhana, USyd (AAO supervisor: Hopkins)

PhD title: "The stellar initial mass function. Is it universal?"

Timothy White, USyd (AAO supervisor: O'Toole)

PhD title: "Asteroseismology of stars from the Kepler mission and the Anglo-Australian Planet Search"

Madusha is also the winner of the Astronomical Society of Australia's 2010 Bok prize for the best Honours thesis.

## AAO Summer students

The AAO runs a twice-yearly fellowship programme to enable undergraduate students to gain 10-12 weeks first-hand experience of astronomical-related research. The current crop of students are from the Northern hemisphere and are enjoying spending their summer in our winter.

Jessica Davis comes to us from the currently warm and sunny campus of the California Institute of Technology, having just completed her second year of studies toward a B.S. in Astrophysics. Working under the supervision of Stuart Ryder, Jessica will be processing a series of images (spanning 2 years, taken by the Gemini North 8m telescope) from the Luminous Infrared Galaxy UGC 8387, combining the different epochs to achieve the deepest and most detailed image yet obtained for this LIRG. After figuring out how to preserve the best quality for this final image, as well as creating a "true colour" image of UGC 8387, she will ultimately try and answer such questions as "Why is UGC 8387 in a LIRG phase?", "How far beyond the galactic nucleus does its star formation extend?", and "How long has this burst of star formation been going on?" Until gets the data she needs, however, Jessica is attempting to make friends with IRAF.

Maya Petkova is a student from Bulgaria who has just finished the second year of a masters undergraduate degree in Astrophysics at the University of St Andrews, UK. She is working with Matthew Colless and Darren Croton analyzing data generated by a semi-analytic model of galaxy evolution. The aim of Maya's project is to observe how different galaxies accumulate mass over time, and hopes to compare these models against real data.

# Ben Gascoigne – a "founding father" of the AAO

Russell Cannon, Peter Gillingham and David Malin (AAO)

During the AAO2010 Symposium in Coonabarabran in June there was a small ceremony to mark the naming of the recently refurbished AAT Meeting Room as "The Gascoigne Room". A photograph in the room, (Figure 1), bears the inscription "This room has been named in memory of Sidney Charles Bartholomew 'Ben' Gascoigne AO FAA, 1915-2010, Astronomical Advisor to the AAT Project Office 1967-1974, Commissioning Astronomer 1974-1975. Ben's many contributions were fundamental to the success of the AAT, and to its reputation as one of the best and most productive large optical telescopes in the world". This now joins the infamous "Gascoigne's Leap" plaque in the dome.

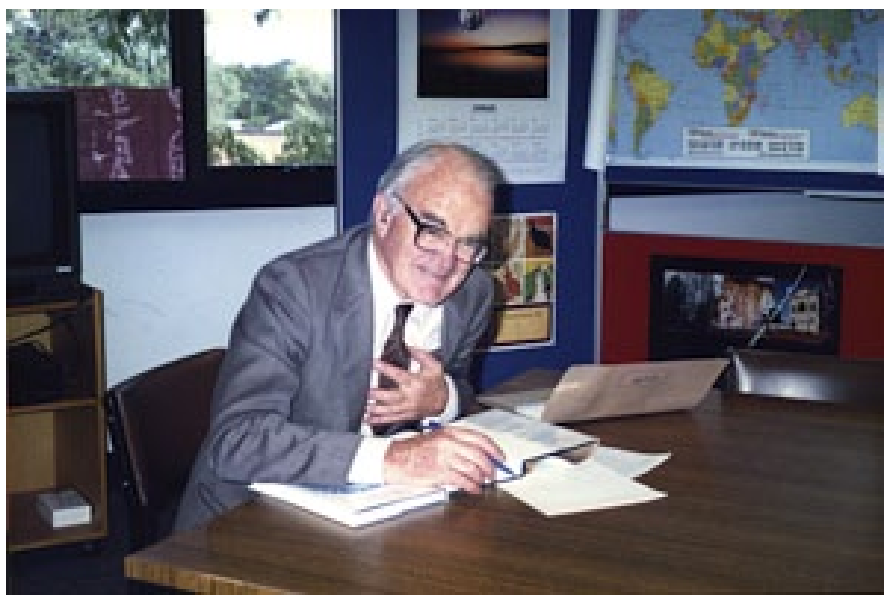

**Figure 1:** Ben Gascoigne signing copies of "The Creation of the AAO" in the AAO Epping Library, April 1991 (Robyn Shobbrook).

How did one man come to have his name commemorated twice in the otherwise austere setting of a telescope building? Several obituaries have been published in newspapers and other publications, outlining Ben's work as an astronomer and his role as husband of the famous Australian artist the late Rosalie Gascoigne. Here we will concentrate on his contributions to the AAT as an advisor on optics, commissioning astronomer and lead author of the book "The Creation of the AAO".

When the giant Anglo-Australian 150-inch Telescope (AAT) was being planned in the late 1960s, Ben was appointed to advise the design team of engineers at the Project Office in Canberra. He was uniquely qualified for this role. Following a degree in maths and science from Auckland University College, he completed a PhD in optics at the University of Bristol before returning to New Zealand in 1940, early in World War II, to work on optical munitions. He soon moved to the then Commonwealth Solar Observatory on Mount Stromlo, making gun sights and other optical devices. His first connection with astronomy involved the establishment of a new Australian Time Service in 1945. He went on to play a major role in commissioning a series of optical telescopes which established Mount Stromlo Observatory as an internationally recognized research centre: the re-built 48-inch Great Melbourne Telescope (originally commissioned in 1869) in the early 1950s, then the 74-inch Grubb Parsons reflector in 1955 and finally the first telescope on Siding Spring Mountain, the Boller and Chivens 40-inch, in 1964. During this period Ben invented what became known world-wide as the 'Gascoigne Corrector', a lens which gave Ritchie-Chrétien telescopes a much wider field of good definition. Some years after its application to the 40 inch telescope, this design of corrector was used with the Irénée DuPont 2.5-metre telescope in Chile. It has also been used on a number of other major telescopes. Ironically, such a corrector was not adopted for the AAT.

Ben himself used the Mount Stromlo telescopes for important observational projects. He pushed the techniques for optical and near-infrared photometry to new limits of faintness and precision, and soon became internationally known for the reliability of his observations of the so-called 'Cepheid' variable stars, a key link in determining stellar distances. Writing about this period in an autobiographical note for the Academy of Science, Ben says: "I'll never forget when all this suddenly fell into place how triumphant I felt, an extraordinary feeling of elation, here was one part of astronomy I really understood and what was more I had become one of them, I had joined the professionals." Later Ben turned his attention to the star clusters in the Magellanic Clouds. By 1960 he was photographing these with the new 74-inch telescope at Mount Stromlo and calibrating the stellar magnitudes using a two-channel photo-electric photometer on the 50-inch telescope. He discovered that the LMC and SMC contained some relatively young globular clusters, in contrast to the population of very ancient globulars in our Galaxy: evidence for very different histories of star formation.

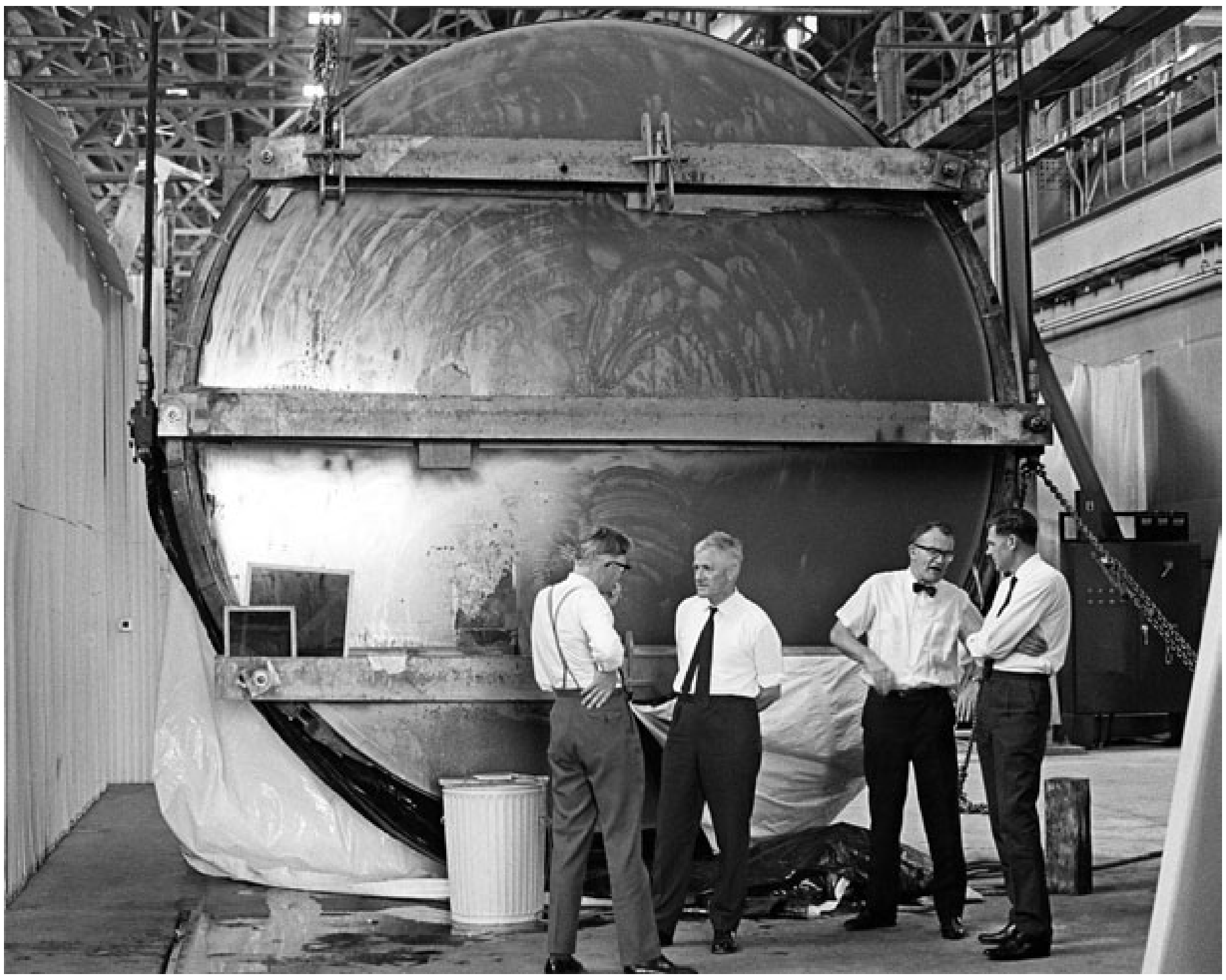

**Figure 2:** L-R: David Brown (Grubb Parsons), R.O. Redman, Ben Gascoigne and John Pope in front of the newly-cast AAT mirror blank at the Owens-Illinois glassworks, May 1969.

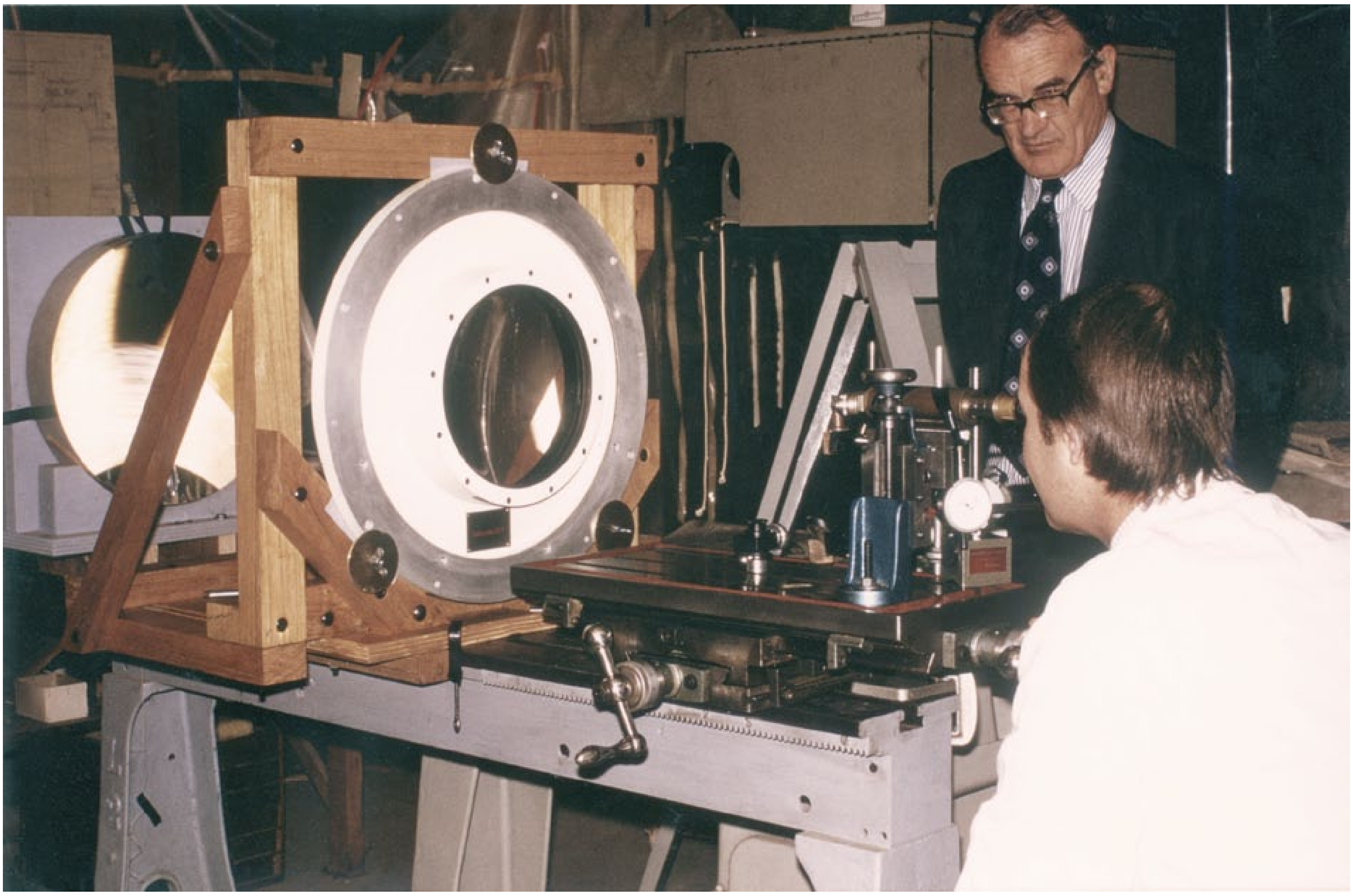

**Figure 3:** Ben Gascoigne (standing) and Bill James testing AAT PF corrector lenses, June 1973.

For the AAT, Ben's earliest work included determining the optical specification of the telescope mirrors. Although the AAT design was based on that of the Kitt Peak and Cerro Tololo 150-inch telescopes, many details were changed, including a longer focal length for the primary mirror. Ben was so content with the mirror specifications that he maintained he'd be happy to have the conic constant of the primary, 1.1717, engraved on his headstone. Ben's work on the design team involved visits to the Owens-Illinois glass works in the USA, to check and eventually do the acceptance tests on the Cervit blanks for the primary and other mirrors, and to the factory of Sir Howard Grubb Parsons at Newcastle-upon-Tyne in England, where the mirrors were ground and polished and the telescope tube was manufactured. He also collaborated with Charles Wynne in England who designed the wide field correctors for the prime focus and with Bill James who made the lenses in Melbourne.

The largest corrector, aided by the slower f/3.3 primary focal ratio, gave the unusually large 1° field of view. This relatively slow focal ratio was later very helpful in allowing the design of a new corrector with a 2° field for multi-fibre spectroscopy with 2dF.

As the telescope neared completion, Ben was appointed Commissioning Astronomer, a post he held from early 1974 until routine observing started in mid-1975. Working closely with the telescope engineers and technical staff, and later with a succession of young British and Australian astronomers who were fortunate to be given time to try out the first instruments – principally prime focus photography, the Wampler-Robinson Image Dissector Scanner and some Cassegrain photometers – Ben took part in many early observations, starting with simple Foucault knife-edge tests on stars and the first photograph, of the globular cluster Ð Centauri, taken in April 1974 with an uncoated primary mirror.

Those early observations very nearly marked the end of Ben's career. He went outside one night to check the weather while Roderick Willstrop was observing at prime focus but failed to return to the Control Room. He was found lying on the wooden floor, within the very limited space between the steel girders of the primary mirror and cell handling trolley. Ben was conscious but very dazed and had obviously fallen some 6 metres from the interior catwalk, and had a badly damaged left arm. His colleagues managed to get him into a car and took him down to the Coonabarabran Hospital.

Peter Gillingham recalls: "As early as I dared in the morning, I phoned the hospital and inquired after Ben's condition. I was greatly relieved when the woman who answered asked if I'd like to speak to Ben. He was in good spirits and I noted that his speech was more fluent than usual. However, a neighbour on the mountain, Julie Lommon, who worked at the hospital as a radiographer, reported that the nurses there were all concerned about the dear professor, who'd had the fall on the mountain that had affected his speech so badly."

Ben's reconstruction of the incident was that he had gone out one of the four doors onto the outside catwalk somewhere near the control room. He closed the outer door and proceeded to walk around the catwalk. He re-entered the dome through what he thought was the same door and avoided lighting his torch because he knew Roderick was exposing a red sensitive plate. He was about 90 degrees off in azimuth. In the dark, he opened the nearest gate (in the outer railing) and stepped forward onto what he thought was the dark stained wooden flooring near the control room. But he stepped into space.

Ben was extremely lucky to have landed between the heavy steel beams of the mirror-handling trolley. Subsequently, he was quite proud of having a stainless pin inserted as part of the repair to his elbow. With the compensation payment, he bought an early HP programmable calculator, about the size of a portable type-writer. Furthermore, his accountant advised him of a tax benefit then in force, with which Ben was able to buy the plotter that mated to the calculator and still come out ahead. He claimed that, if he knew the outcome would be the same, he'd be prepared to do it again! A week or two after his fall, Ben sent a message to Jack Rothwell (Project Office electrical engineer) who'd tripped over a crate when traversing coude west in the dark and chipped a bone, "Anything you can do, I can do better."

A second ring of internal safety railings had already been fabricated and these were immediately installed. The Gascoigne's Leap plaque was made by Paul Lindner, and unveiled by Ben on his last AAT visit before his retirement from MSSSO in 1980.

Despite these distractions, the AAT was to propel Australia and the UK into the front rank of international astronomy for the next 20 years. As the last of a series of four 4-m-class equatorial telescopes built in the 1970s, it was built to be the best telescope the engineers could achieve at the time, and as such it attracted some of the best young astronomers to use it to explore the Universe. It also attracted the best engineers and instrumentalists to keep it at the forefront of technology. That it remains a world-leading telescope to this day says a great deal for the foresight of its designers. It is now used in ways that were undreamt of by its builders, and now observes types of objects that were completely unknown when it was commissioned. Its superb engineering meant that it could be developed and improved to take maximum advantage of new technologies such as infrared detectors, optical fibres and super-sensitive digital cameras as they emerged.

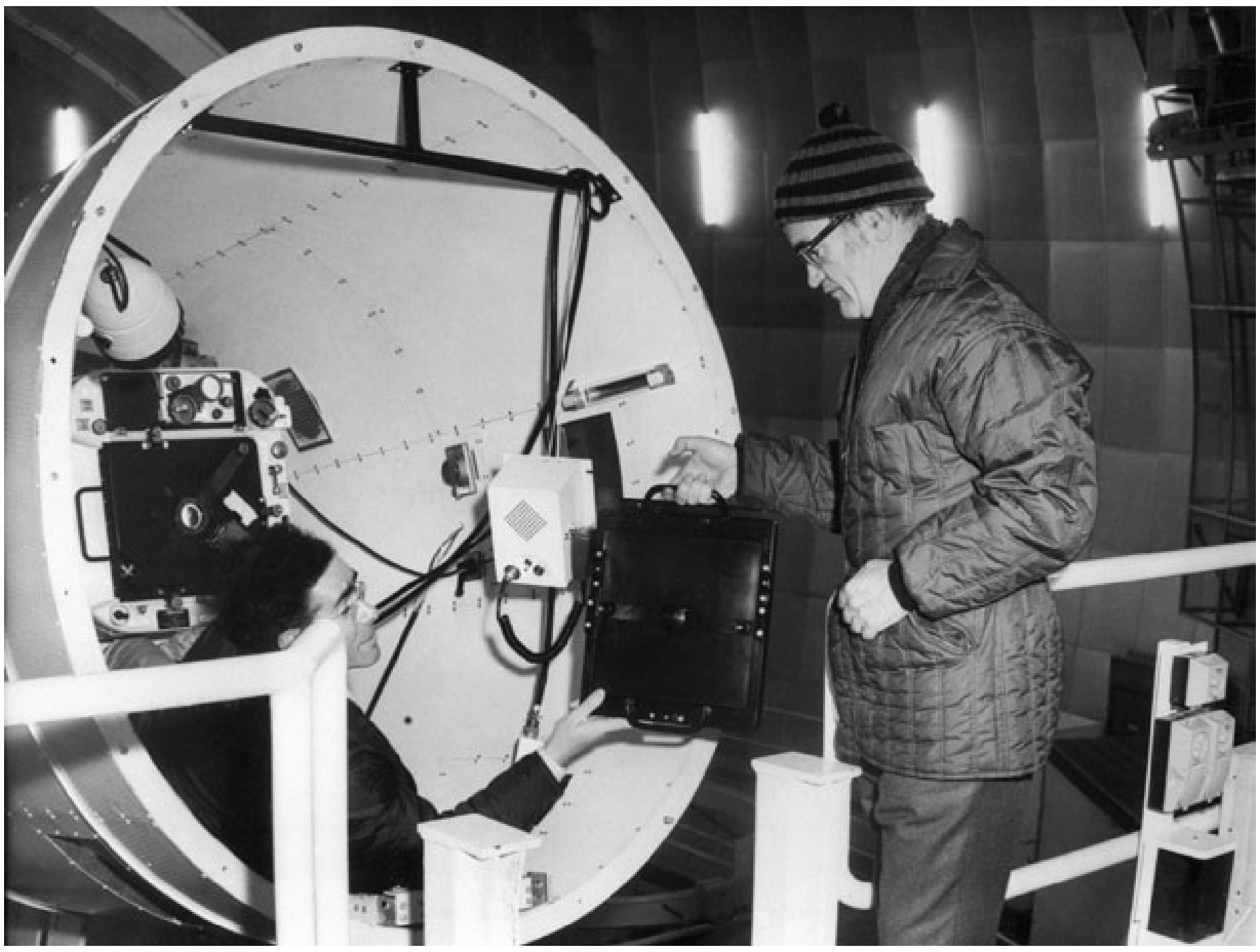

**Figure 4:** Preparing to observe at the AAT prime focus (Roderick Willstrop and Ben Gascoigne, late 1974)

Ben enjoyed using the AAT, pursuing his study of Magellanic Cloud star clusters. However, the days of using single-star photometry to calibrate non-linear photographic plates were numbered. It was already obvious that electronic areal detectors were going to be the way forward, but the devices available then, based on TV cameras, image tubes and electronography, were not sufficiently reliable or uniform, while the computing power was inadequate to analyse the data properly. Bad weather foiled his last attempt to do the job using the IPCS in imaging mode. As Ben wrote "we would have led the field by a decade". The problems were indeed solved within a decade, once CCD detectors had been tamed.

After retiring, Ben capped his astronomical career and returned to his early love of history by writing a number of fascinating papers on the development of astronomy in Australia. Foremost among these must be his lead authorship of the definitive book "The Creation of the AAO", co-authored with Katrina Proust and Mac Robins and published by Cambridge University Press in 1990 (reprinted as a paperback, 2005).

Finally, despite becoming an internationally recognised astronomer and receiving various honours from learned societies in Australia, New Zealand and the UK, and being made an Officer in the Order of Australia in 1996, Ben Gascoigne remained a very modest and approachable man. At the same time, he was well aware of his own abilities and confident in his knowledge, and stood his ground when he knew he was right. Ben also had a good sense of humour, and, despite a persistent stutter, was a fine raconteur who made the occasional cloudy night a great pleasure. In his later years he set an excellent example when he stepped back from his own successful scientific career to allow his wife Rosalie to develop hers as an artist. To those of us who were privileged to know and work with him, and to count him as a friend, he was unforgettable, and we will miss him. Ben died a few weeks before the AAO2010 Symposium and his son Martin spoke briefly and movingly at the meeting about his father's enduring attachment to the Observatory.

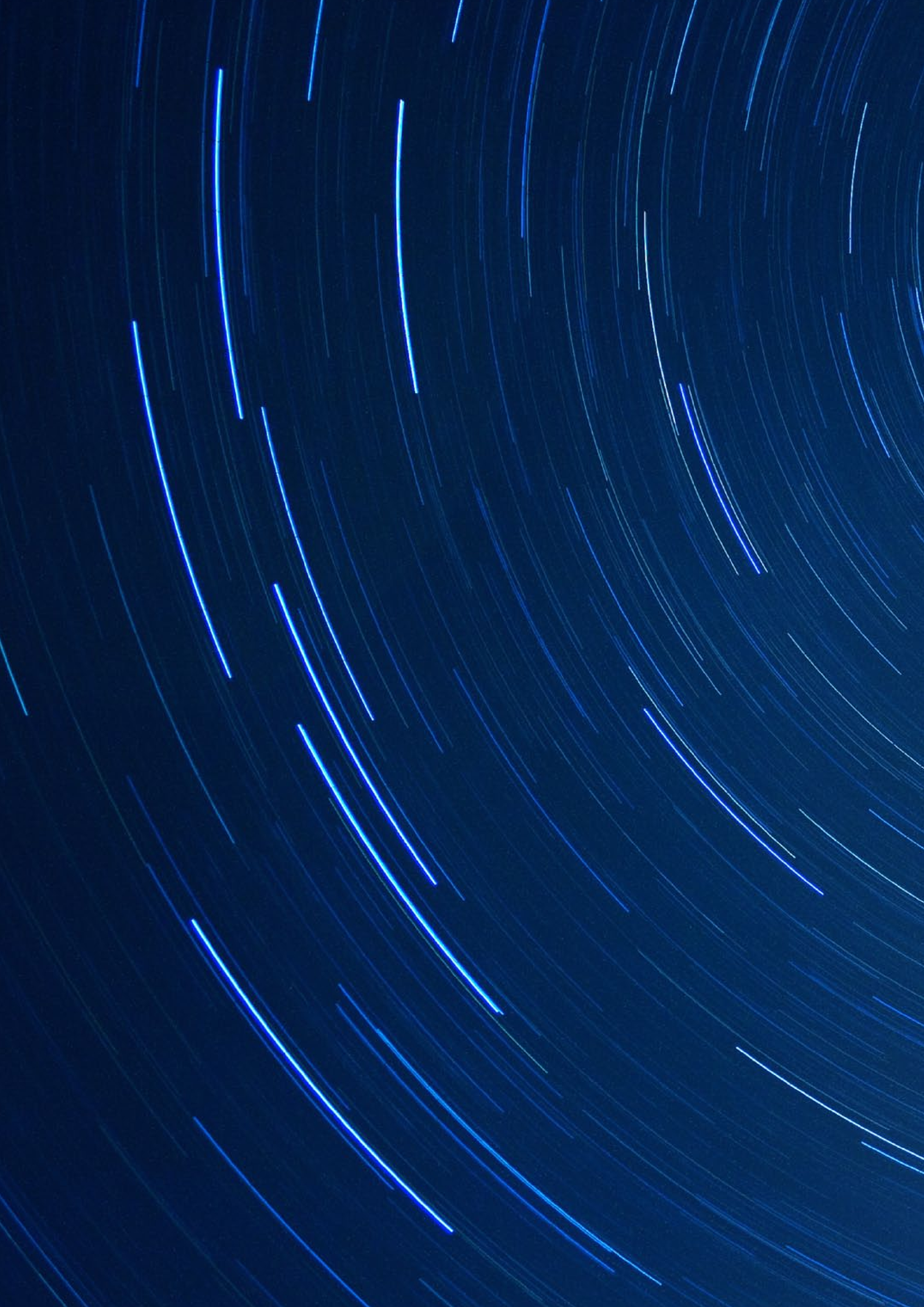

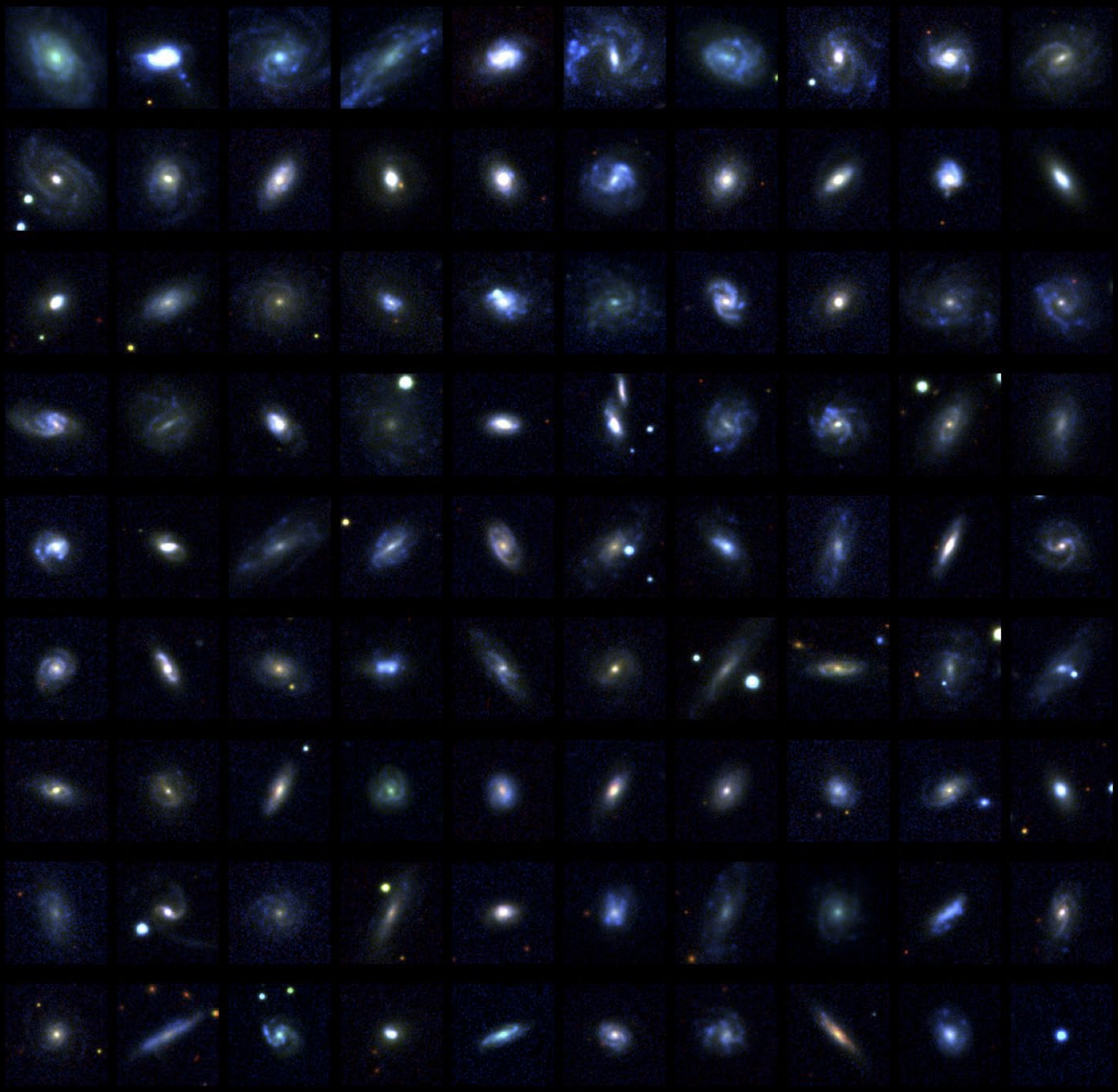

ABOVE: The 90 brightest blue galaxies in the Galaxy and Mass Assembly survey (GAMA; SDSS colour imaging, reprocessed by the GAMA team).

Australian Government
Department of Innovation
Industry, Science and Research

AAO

**AAO Sydney**
Telephone: (02) 9372 4800
Fax: (02) 9372 4880

**AAO Siding Spring**
Telephone: (02) 6842 6291
Fax: (02) 6842 2266

Email: director@aao.gov.au

**www.aao.gov.au**